\UseRawInputEncoding 

\documentclass[twocolumn]{aastex631}

\usepackage{amsmath}
\usepackage{natbib}
\usepackage{graphicx}
\usepackage{color}

\usepackage{soul}
\sethlcolor{yellow}
\usepackage[normalem]{ulem}

\def\kms{{km~s$^{-1}$}}
\def\msun{{${\rm M_\odot}$}}
\def\deg{{$^\circ$}}

\def\zenodo{\url{https://zenodo.org/record/5593154}}

\shorttitle{The forward and reverse shock dynamics of Cassiopeia A}
\shortauthors{Vink et al.}

\graphicspath{{./Figures/}}

\begin{document}

\title{The forward and reverse shock dynamics of Cassiopeia A}

\author[0000-0002-4708-4219]{Jacco Vink}
\affiliation{
Anton Pannekoek Institute/GRAPPA, University of Amsterdam, Science Park 904,
1098 XH Amsterdam, The Netherlands
}

\author[0000-0002-7507-8115]{Daniel J. Patnaude}

\author[0000-0002-0394-3173]{Daniel Castro}
\affiliation{
Harvard-Smithsonian Center for Astrophysics, Cambridge, MA 02138, USA
}

\begin{abstract}
We report on proper motion measurements of the forward-  and reverse-shock regions of the supernova remnant Cassiopeia A (Cas A),
including  deceleration/acceleration measurements of the forward shock.
The measurements combine 19 years of observations with the Chandra X-ray Observatory, using the 4.2-6 keV continuum band,
preferentially targeting X-ray synchrotron radiation.
The average expansion rate is $0.218 \pm 0.029$\%yr$^{-1}$ for the forward shock, corresponding to a  velocity of  $\approx 5800$~\kms.
The time derivative of the proper motions indicates deceleration in the east, and an acceleration  up to   $1.1\times 10^{-4}$~yr$^{-2}$ in the western part.
The reverse shock moves outward in the East, but in the West it moves toward the center 
with an expansion rate of $-0.0225\pm 0.0007$ \%yr$^{-1}$, corresponding to $-1884\pm 17$~\kms.
In the West the reverse shock velocity in the ejecta frame is $\gtrsim 3000$~\kms, peaking at $\sim 8000$~\kms, explaining the presence of X-ray synchrotron emitting
filaments there. 
The backward motion of the reverse shock can be explained by either a scenario in which the forward shock encountered a partial, dense, wind shell, or one in which
the  shock transgressed initially through a lopsided cavity, created during a  brief Wolf-Rayet star phase.
Both scenarios  are consistent with the local acceleration of the forward shock.
Finally we report on the proper motion of the northeastern jet, using both the X-ray continuum band, and the  Si XIII K-line emission band.
We find expansion rates of respectively 0.21\%yr$^{-1}$ and 0.24\%yr$^{-1}$, corresponding to velocities at the tip of the X-ray jet of 7830--9200~\kms.
\end{abstract}


\section{Introduction}

The early dynamical evolution of supernova remnants (SNRs) is  as much determined by the density structure of the
progenitor's circumstellar medium (CSM) as by the velocity distribution of the supernova (SN) ejecta. 
The CSM properties and SN explosion energy determine the expansion velocity of the blast wave, whereas the structure of the SN ejecta determine
at what rate energy from the freely expanding, cold ejecta, is added to the hot shell after the ejecta are shocked by the reverse shock.
The  dynamical evolution of young SNRs is, therefore, of interest as
it reveals information about  both the structure of the CSM, determined by the SN progenitor mass-loss history, as well as the  structure of the  ejecta, which provides insights
into the SN explosion properties.

The presence of a reverse shock in young SNRs was first pointed out by \citet{mckee74}. The hydrodynamics of young SNRs was further
explored by \citet{chevalier82}, who showed that self-similar solutions for the expansion and relative locations of the forward and reverse shocks
with respect to the contact continuity exist if one assumes that the freely expanding SN ejecta density has a velocity profile of the form $\rho_{\rm ej}\propto v^{-n}$,
with $5<n\lesssim 12$.
Under these conditions the forward and reverse shock radii evolve as  $R_{\rm fs}\propto R_{\rm rs} \propto t^m$, with $m=(n-3)/(n-s)$ the expansion parameter. 
The parameter $s$ here indicates the density structure of the
CSM, $\rho_{\rm csm}\propto r^{-s}$, with $s=0$ a uniform density profile, and $s=2$ corresponding to the density
profile of a steady SN progenitor wind. Of course, the CSM structure may be more complicated than described by
an $s=0$ or $s=2$ model.

According to Chevalier's model, valid for the earliest SNR phase,  the reverse shock expands outward in the observer frame.
This analytical model is no longer valid once the reverse shock has reached the core of the ejecta density distribution, where the density is assumed to
be more or less uniform---but decreasing with time as $t^{-3}$. Once most of energy of the SN ejecta has been processed by the reverse shock,
and is contained in the hot SNR shell, the SNR has entered the Sedov-Taylor phase, for which  the forward shock is expected to
expand as $R_{\rm fs}\propto t^{2/(5-s)}$.

The \citet{truelove99} model describes the transition
in the expansion from the early phase  to the Sedov-Taylor phase. For the $s=0$ models the reverse shock eventually reverses its course and moves backward also
in the frame of the observer. For SNRs expanding in a wind steady wind---i.e. $s=2$---semi-analytical models  \citep{hwang12,micelotta16,tang17}, but also hydrodynmical
simulations \citep{orlando16,orlando21}, show that reverse shock keeps expanding outward for 2000--3000 yr with the reverse shock velocity in the frame of the ejecta
being more or less constant.

The most important example of a young core-collapse SNR, and one that is most likely expanding in a dense wind, is Cassiopeia A (Cas A).
Its forward shock has reach a radius of $\approx$2.8\arcmin, conveniently corresponding to 2.8~pc at the distance of 3.4~kpc \citep{reed95}.
Cas A is the result of a Type IIb SN \citep{krause08,rest11} that occurred around the year 1672 \citep{thorstensen01}, but went unnoticed at the time. 
The SN ejecta contains very little hydrogen \citep{fesen91}, and the total ejecta mass is estimated to be around 2--4~\msun\ \citep{vink96,willingale03,laming20}.
The thermal X-ray emission from Cas A is dominated by the shocked ejecta, with prominent emission lines from intermediate mass elements (IMEs: Si, S, Ar) and
iron.  The IMEs and Fe have distinct spatial distributions and Doppler shift profiles \citep{willingale01}. 
In the optical, ejecta knots  can be separated based on spectra rich in [S II] or in [OIII] line emission,
with optical knots dominated by  [S II] lines being preferentially found in the northeast to southwest direction.

One important feature of the ejecta distribution is the presence of two oppositely placed "jet-like" structures, which are visible in the optical \citep{vandenbergh70,fesen16}, 
X-rays \citep{vink04a,hwang04}, and infrared \citep{hines04}. The jets stand out clearly, because they happen to move almost perpendicular to our line of sight \citep{fesen06}. X-ray and optical spectroscopy
reveal that they are rich in IMEs, particular in silicon and sulfur. 
Another peculiar morphological feature is the presence of ring-like features in three-dimensional space (sky coordinates + Doppler velocity), as revealed by optical Doppler velocity mapping
\citep{milisavljevic13}. These rings are likely the result of the inflation of radio-active $^{56}$NI/$^{56}$Co bubbles,  pushing aside the non-radioactive
IME-ejecta \citep{blondin01c}.

The observed deviations from spherical symmetry  mostly relate to asymmetries in the explosion itself \citep[e.g.][]{rest11}.
The outer shock wave, as marked by the thin rims of X-ray synchrotron
radiation, appears nearly circular, and suggests a relatively spherically symmetric CSM density distribution.
However, in the optical the CSM is traced by some high density knots, the so-called quasi-stationary flocculi \citep[QSFs,][]{minkowski57}. These knots have typical velocities of $\sim 150$~\kms\
\citep{kamper76},
and are nitrogen-rich, while still containing hydrogen. These QSFs appear to be the products of a pre-SN mass loss phase, and their distribution is not spherically symmetric 
\citep[e.g.][]{lawrence95,fesen01b,alarie14,koo18}, with a large concentration in the northern part, and a protruding arc of QSFs in the Southwest. \citet{weil20} recently reported the detection of faint optical emission from clumpy CSM outside the current shock radius.

Here we report on an expansion study of Cas A with archival Chandra X-ray Observatory  \citep[Chandra,][]{chandra03} data , spanning 19 years of observations. 
The goal is to both understand the dynamics of the forward
shock and the reverse shock, and the northeastern jet. 
We limit this study to the X-ray continuum band of 4.2-6 keV, which for the forward shock, but likely also for part of the reverse shock \citep{hughes99,helder08,uchiyama08}, is dominated
by X-ray synchrotron emission from plasma immediately downstream of the shock. An alternative hypothesis for the hard X-ray emission from (part of) the reverse shock 
region is non-thermal bremsstrahlung caused by suprathermal electrons, see for example \citet{asvarov90,laming01a,vink03a}, and the discussions in \citet{vink08a,grefenstette15}.

This study is a follow up of many previous X-ray expansion measurements focussing on different
regions and using different measurements, starting with the Einstein HRI and ROSAT  HRI observations \citep{vink98a,koralesky98}, and various measurements with  Chandra \citep{delaney03,delaney04,patnaude09,sato18}. For the forward shock these studies indicate an average expansion rate of  typically 0.2~\%yr$^{-1}$, faster than the radio expansion rates
based on proper motion of bright knots, $\sim 0.1$~\%yr$^{-1}$, but with a large intrinsic scatter \citep{tuffs86,anderson95}. A radio expansion measurement based on the overall fitting of the expansion in the UV-plane
at 151 MHz listed a much faster expansion rate of 0.19--0.24~\%yr$^{-1}$ \citep{agueros99}.

The main results reported here consist of direct proper motion measurements in azimuthal sectors, with annuli
encompassing the forward- and reverse-shock regions, and a separate measurement of the northeastern jet. 
The results for the reverse shock show a wide variety in expansion properties, which does not agree with the theoretical expectations.

We also report on a measurement of the acceleration/deceleration of the expansion, albeit with some caution given the
sensitivity of these measurements to systematic errors.

\section{Method and data}
\label{sec:method}

For certain periods of time the expansion of an SNR can be described by a self-similar solution model of the form
\begin{equation}\label{eq:selfsimilar}
R=R_0 \left(\frac{t}{t_0}\right)^m,
\end{equation}
with $R$ a certain radius---e.g. the forward shock radius, $R_{\rm fs}$---$R_0$ the reference radius, corresponding to age $t_0$, and
with $m$ the expansion parameter. For example, for a point explosion into a uniform density medium
one expects the evolution to be described by $m=2/5$, corresponding to the Sedov-Taylor model. Previous X-ray expansion measurements of
Cas A indicate that $m\approx 0.7$ \citep{vink98a,koralesky98,delaney03,patnaude09}, close to  $m=2/3$, expected for an SNR
evolving inside  a steady progenitor wind with $\rho(r)\propto r^{-2}$---i.e. the $s=2$ case.

The  method used here measures the expansion projected onto the plane of the sky  by comparing X-ray images based on observations in different years.
One reference image is stretched or shrunk by the expansion factor $f$ until it provides a good fit to an image based on a later, respectively an earlier observation.
We can approximate the development of the expansion factor by a Taylor series around a reference time $t_0$
\begin{equation}\label{eq:expansion}
f(t)= \frac{R(t)}{R_0} =  1 + a(t-t_0) + \frac{1}{2}b(t-t_0)^2 + ....,
\end{equation}
with $t_0$ being the age of the SNR at time the reference image ("the model" image) was taken.
For a self-similar evolution we expect 
\begin{align}\label{eq:exp_rate_pars}
a = &\left(\frac{1}{R}\frac{dR}{dt}\right)_{t=t_0} = \frac{m}{t_0}, \\\nonumber
b = &\left( \frac{1}{R} \frac{d^2R}{dt^2}\right)_{t=t_0} =  \frac{m(m-1)}{t_0^2}.
\end{align}
We will refer to $a$ as the expansion rate, and will use the units $\% {\rm yr}^{-1}$. Its inverse is the expansion time scale
\begin{equation}
\tau_{\rm exp}\equiv a^{-1},
\end{equation}
which gives the age of the SNR if no deceleration has  taken place over its lifetime, i.e. $m=1$. This shows that the expansion parameter can
be estimated as $m=t_0/\tau_{\rm exp}$. But strictly speaking this is only valid if the SNR shock radius evolved as (\ref{eq:selfsimilar})
for the entire life of the SNR. In other words, a mismatch between $m$ as derived from $a$ or $\tau_{\rm exp}$, and $b$ is an indication that the 
the self-similar model may not be a valid approximation.

The parameter $b$ is the second derivative and can be labeled the acceleration parameter. However, as indicated in (\ref{eq:exp_rate_pars})
for $m<1$ one expects the value of $b$ to be negative. So we define the deceleration rate as $-b$. 
The deceleration rate for Cas A,
at an age of $t_0=331$~yr (in 2004) and $m\approx 0.7$ is expected be  around $-b\approx 1.9\times 10^{-6}$~yr$^{-2}$.

\begin{table}
\centering
\caption{Summary of the observations used.\label{tab:obs}}
\begin{tabular}{lrccr}\hline\hline\noalign{\smallskip}
Epoch & ObsID &  MJD & $\Delta t$ & Exposure\\
 & &  &  {[yr]} & {[ks]}\\\noalign{\smallskip}\hline\noalign{\smallskip}
"Model" & 4634--4639 & 53119.2 & 0 & 845.6\\
Epoch 1 & 114 & 51573.8 & -4.231 & 50.6\\
Epoch 2 & 1952 & 52311.6 & -2.211 & 50.3\\
Epoch 3 & 9117 & 54440.1 & 3.616 & 25.2\\
Epoch 4 & 9773 & 54442.7 & 3.623 & 25.2\\
Epoch 5 & 10935 & 55138.1 & 5.527 & 23.6\\
Epoch 6 & 12020 & 55139.1 & 5.530 & 22.7\\
Epoch 7 & 10936 & 55500.4 & 6.519 & 32.7\\
Epoch 8 & 13177 & 55502.2 & 6.524 & 17.5\\
Epoch 9 & 14229 & 56062.7 & 8.059 & 49.8\\
Epoch 10 & 14480 & 56432.9 & 9.072 & 49.4\\
Epoch 11 & 14481 & 56789.4 & 10.048 & 50.1\\
Epoch 12 & 14482 & 57142.8 & 11.016 & 50.1\\
Epoch 13 & 18344 & 57682.9 & 12.495 & 26.1\\
Epoch 14 & 19903 & 57681.3 & 12.490 & 25.0\\
Epoch 15 & 19604 & 57890.0 & 13.062 & 50.2\\
Epoch 16 & 19605 & 58254.0 & 14.058 & 50.1\\
Epoch 17 & 19606 & 58616.8 & 15.052 & 50.1\\
\noalign{\smallskip}\hline
\end{tabular}
\end{table}

\subsection{The data}

We used  Chandra archival observations of Cas A for which the ACIS-S detector \citep{garmire03} without gratings  was used, and for which the exposure time
was $t_{\rm obs}\gtrsim 20$~ks.  For the reference ("model") image 
we used a combination of the very-large program (VLP) observations with observation IDs 4634 to 4639,
made from May till July 2004. We left out here observations with short observations, and observations with different role-angles, resulting in a net exposure for the reference image of 845~ks. 
The average modified Julian date (MJD) of the VLP  observations used is 53119. This epoch is referred to $t_0$, in connection to Equation~(\ref{eq:expansion}).
We also use $t_0$ to indicate the age of Cas A: $t_0\approx 332$~yr, for an assumed explosion in 1672 \citep{thorstensen01}.
All  observations used were made between January 2000, and May 2019. Table~\ref{tab:obs} lists all the ObsIDs, effective exposure times, and $\Delta t= (t-t_0)$.

As stated in the introduction, we focussed the expansion measurements on the 4.2--6 keV continuum band, as this band is dominated at the forward shock region---but likely also for part of
the reverse shock region---by X-ray synchrotron radiation, which is a tracer for the presence of a fast shock.
We extracted images in this band  from the level-2 event list with proprietary code, which extracts images with identical world coordinates  systems for  all images,
with the center of the image corresponding to the expansion center of Cas A. For the expansion center we used the
measurement by \citet{thorstensen01}, based on measurements of the  fastest moving optical knots: 
RA$_{J2000}=23{\rm h} 23\arcmin 27.77\arcsec$, DEC$_{J2000}=58\degr 48\arcmin  49.4\arcsec$. At this stage no pointing corrections are applied.
The images of $1024\times1024$ pixels were extracted using the nominal Chandra ACIS pixel size of 0.492\arcsec, 
with the exception for the VLP. For the VLP data the image was extracted using an oversampling of factor four,  with a four times larger image size, for reasons explained below.

\subsection{The expansion measurement method}
There are several ways to measure expansions from images, for example extracting radial profiles and comparing them \citep{delaney03,patnaude09}---fitting in one dimension---or by stretching/shrinking one
image (the "model" image) with respect to an image of another epoch, and measuring what stretching factor provides a best fit \citep{vink98a,koralesky98}. This means that the
fitting is done in two dimensions. We apply the latter method, using our self-developed C$^{++}$ code, which we  named {\em multisectorexpansion}.\footnote{The source code and data products generated by it can be downloaded on \zenodo. 
} 
The input for the code consists of a text file containing a list of all necessary input files (images and mask-files, see below) and optional input parameters. 
After the fitting procedure, a new input file is automatically
generated for iterative purposes, which includes also best-fit parameters.
Other output files are postscript files with expansion plots \citep[generated using the pgplot library,][]{pgplot}, and text and \LaTeX-formatted files with the fitting results.
The core of the C$^{++}$ code goes back to one of the earliest X-ray expansion measurements \citep{vink98a}, but has been substantially updated, first
for an expansion measurements of Kepler's SNR \citep{vink08b}, and even more substantially for the current measurements. The most signficant change is that instead of comparing one observation
set with a model for a given epoch, it makes a joint fit to  observations from different epochs simultaneously, greatly increasing the sensitivity.

Like in  previously published proper motion measurements, using earlier versions of the code, 
the fitting is done by stretching (regridding) the "model" image (or part of it) by a factor $f$ (Equation~\ref{eq:expansion}) and comparing
them to another image using the Cash statistic \citep{cash79}, which is the log-likelihood function for Poisson distributions:
\begin{align}
L_k \equiv &-2\ln \mathcal{L} = -2\ln \Pi_{i,j} P_{i,j,k} \\\nonumber
= & -\sum_{i,j} \left[
n_{ijk} \ln \tilde{n}_{ijk} - \tilde{n}_{ijk} - \ln(n_{ijk}!)
\right].
\end{align}
Here $k$ refers to the epoch number, $i,j$ refers to the pixel coordinates, $n_{ijk}$ is the count number in pixel $ij$ of epoch $k$, and $\tilde{n}_{ijk}$  is the predicted pixel  photon count for a given epoch, as derived from the model-image. 
The best-fit model is the one that minimizes $L_k$. As the term  $\ln(n_{ijk}!)$ is  based on the count image alone, which does not depend on model parameters, the term is omitted when comparing different models.

Apart from the model and reference images a set of "mask"-images are provide consisting of pixels with values 0 or 1, which is used to select only those parts
of the images for which the model is evaluated. This was already part of the earlier version of the code.

The novelty of our approach is
that instead of comparing the model image with one other image, the new code compares all epochs simultaneously, by optimising the
sum of the log-likelihoods for all model-image comparisons.
That is, we obtain the best fit model by minimizing 
\begin{equation}\label{eq:lik}
L_{\rm tot} = \sum_k L_k,
\end{equation}
with $k$ the epoch number, and with for each $k$ using a model image $k$ characterised by the expansion rate parameter, $a$ ---and for some regions $b$---and time difference $\Delta t$,
according to Equation~\ref{eq:expansion}.
This greatly decreases the error of the measurement, as now the photon counts in 17 images are combined to give a single log-likelihood value.
The model images are based on the VLP observations. For the center of expansion we use the 

expansion center measured  by \citet{thorstensen01}, based on optical proper motion studies of the fastest moving knots, which are probably least decelerated.
The input model images are four times oversampled. So after the applying the stretching factor the model images are rebinned by a factor four, to bring the pixel size to that of the comparison images.
This reduces interpolation artefacts caused by shifting pixels within a subpixel scale. After this procedure the normalisation has to be adapted to match that of the comparison images.
This is done by normalising the sum of the pixel values for the model images to that of the comparison images within each region of interest, using the mask-images.

The Cash statistic cannot be used for model  input values of zero---due to the $\tilde{n}_{ijk}$ term in the statistic. 
To avoid zero value pixels, we first produced a master mask, which was closely
cropped around Cas A, as not to compare regions that lie far outside Cas A and for which the count statistic is low. Moreover,  we smoothed the image slightly with a gaussian of
$\sigma=0.08$\arcsec, and we added a small number to each pixel of 0.01 to avoid pixels with zero values \citep[see also][]{vink08b}.

\begin{table}						
\centering						
\caption{\label{tab:dxdy}						
Coordinate corrections based on the iterative method (see text).						
$\Delta x,\Delta y$ are in Chandra ACIS-S pixel units (0.492\arcsec) with						
respect to the reference VLP image.						
}						
\begin{tabular}{lrr}\hline\hline\noalign{\smallskip}						
epoch & $\Delta x$  & $\Delta y$ \\\noalign{\smallskip}\hline\noalign{\smallskip}						
1	&	0.52	&	-0.2	\\	
2	&	0.1	&	0.12	\\	
3	&	-0.54	&	0.48	\\	
4	&	-0.08	&	0.72	\\	
5	&	-0.15	&	0.48	\\	
6	&	-0.76	&	0.07	\\	
7	&	-0.38	&	-0.02	\\	
8	&	-0.55	&	0.18	\\	
9	&	-0.19	&	-1.03	\\	
10	&	0.3	&	-0.15	\\	
11	&	0.37	&	0.36	\\	
12	&	-0.51	&	-0.68	\\	
13	&	0.08	&	0.8	\\	
14	&	0.39	&	-0.3	\\	
15	&	-0.3	&	-0.38	\\	
16	&	-0.15	&	-0.2	\\	
17	&	-0.13	&	-0.79	\\\noalign{\smallskip}\hline	
\end{tabular}						
\end{table}

\subsection{Correcting for pointing errors}
\label{sec:pointing}
Apart from stretching the model image, the  code also applies a  displacement correction to the model image for each epoch ($\Delta x_k,\Delta y_k$) in order to correct for  pointing errors.
These can be of the order of 0.6\arcsec\ for Chandra \citep[90\% containment radius][]{chandra03}.  In fact, blinking images with the ds9 viewer \citep{ds9} reveals pointing errors by eye.
The field of Cas A does not contain stable point sources that can be reliably used to realign images, with the exception of the central point source in Cas A, the putative
neutron star left behind by the SN explosion. This point source is off center, and likely has its own proper motion of a few hundred km\,s$^{-1}$ \citep{mayer21}.

We experimented with several solutions for co-aligning the images, including using centroiding of the central point source. In the end we settled on an iterative procedure:
we fitted the expansion parameters ($a$ and $b$)  for all sectors (masks) associated with the forward shock, without using a pointing correction. 
We used the forward-shock masks, as the  errors in the pointing are small compared to the displacement of the shock.
Then, in a separate procedure, we 
fitted all sectors combined to find the optimal  values for $\Delta x_k,\Delta y_k$ for each epoch, using the best-fit expansion parameters. These best fit $\Delta x_k,\Delta y_k$ 
were then used for finding a new optimal expansion solution for individual sectors. Three iterations provided a stable fit for all $\Delta x_k,\Delta y_k$, 
 which are listed in Table~\ref{tab:dxdy}. The Table shows that  typically the corrections
are less than 0.55 pixels  (0.27\arcsec), which is better than the expected Chandra performance of 0.6\arcsec.
Note that this is a correction with respect to the reference image, whereas the 0.6\arcsec\
Chandra error refers to the absolute celestial coordinate reconstruction.

\begin{figure*}
  \centerline{
    \includegraphics[trim=65 160 220 300,clip=true,width=0.45\textwidth]{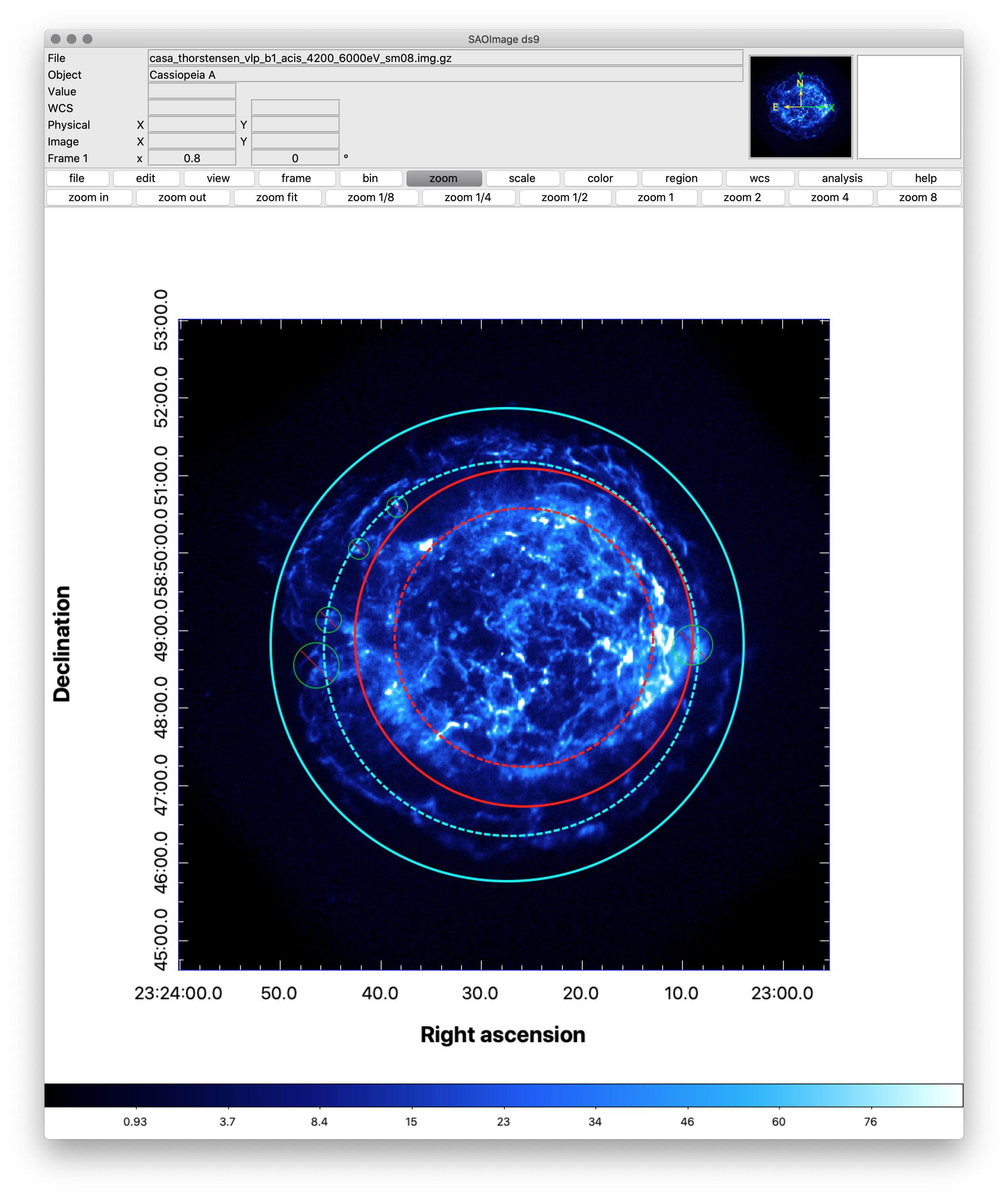}
    \includegraphics[trim=65 160 220 300,clip=true,width=0.45\textwidth]{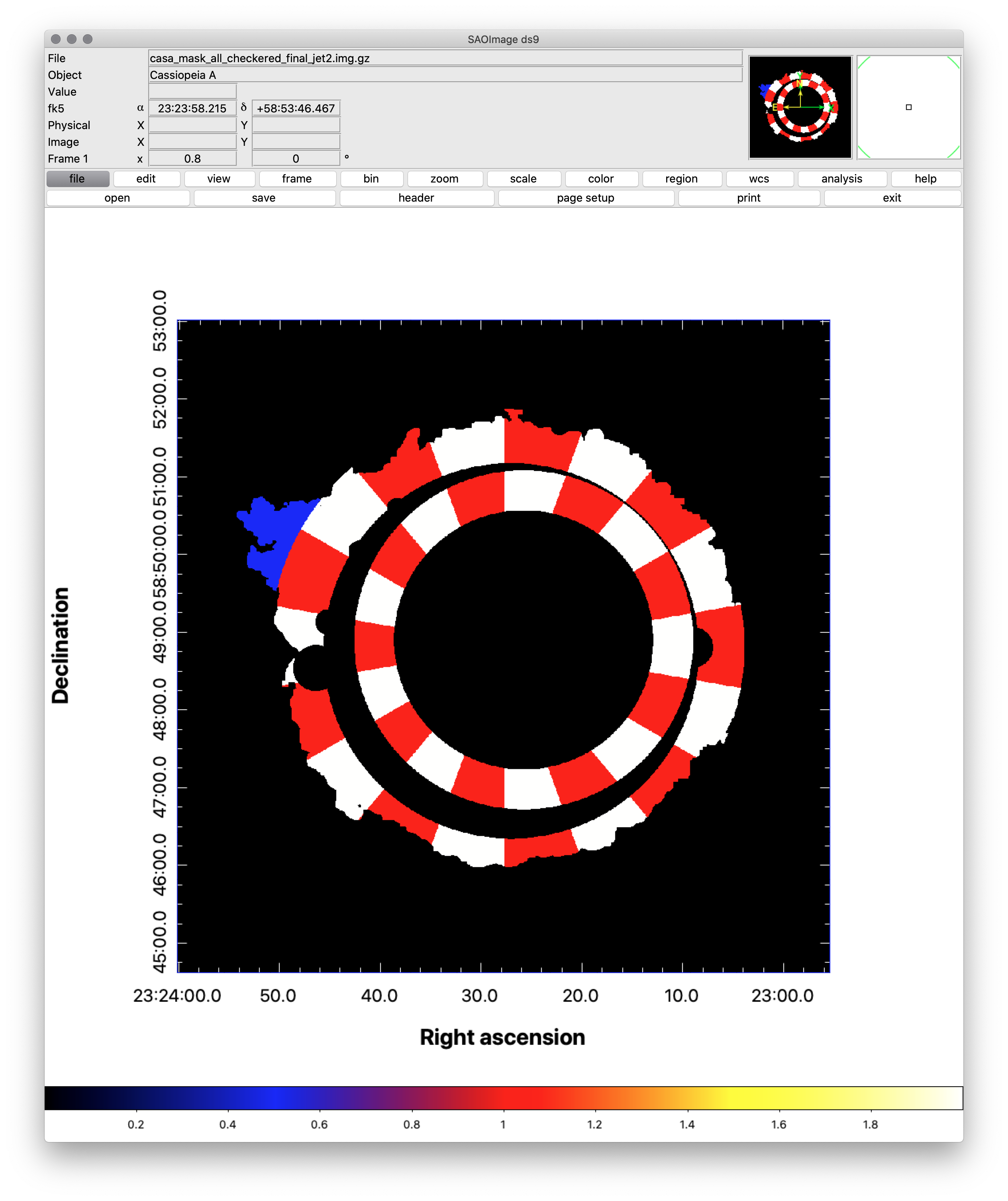}
    }
  \caption{\label{fig:regions}
Left: The Chandra VLP (year 2004) continuum image (4.2-6 keV) with the annuli depicted that were used for selecting the forward (cyan) and reverse shock regions (red).
The small green circles with a red line indicate regions excluded
from the analysis, as for these regions it is ambiguous whether the continuum emission is associated with the reverse or forward shock.
For example in the Southeast they exclude regions with known Fe-rich ejecta knots.
Right: The masks associated with the expansion measurements, in segments of 20\deg, are depicted together, using a festive circus-tent colouring scheme to separate  the different segments.
It includes, in blue, the mask used for  measuring the expansion of the northeastern jet.
  }
\end{figure*}

\subsection{Selected regions}
The main aim of this study  is the dynamics of both the forward and reverse shocks. The Chandra VLP X-ray image in the continuum band shows that the nonthermal filaments associated with
the forward shock roughly lie on a circle
with radius of 2.8\arcmin (2.8~pc); Fig.~\ref{fig:regions}. For an explosion date around 1672 
this amounts to an average velocity of $\approx 8000$~\kms. 
For the expansion study of the forward shock we created the "master" mask by smoothing the  Chandra VLP continuum image (4.2-6 keV) with a gaussian of  2~pixels, and selecting all pixels with a value of more than 1.0. Then a dilation algorithm was applied to add 4 pixels around the periphery of the SNR, plugging some isolated holes in the mask, and removing pixel clusters not connected to the SNR.  Having created the master mask, the forward shock region was selected from it by overlaying an annulus with an inner radius of 2.4\arcmin\ and an outer radius of 3\arcmin. 
Inspecting the continuum image,
some small circular regions were removed, as they appeared to contain structure associated with ejecta knots, or the reverse shock region. 
The selected annuli can be seen in Figure~\ref{fig:regions} (left) and
the final mask in Figure~\ref{fig:regions} (right). The forward shock mask was then further divided in segments with opening angles of 20\deg, as indicated by the color coding in the map on the right.

The reverse shock location is less easily identified. \citet{helder08} used the location of nonthermal filaments and a deprojection method to infer that the reverse shock has a radius of 1.9\arcmin, but shifted
to the West with respect to the explosion center of Cas A; see  \citet{gotthelf01a} for an earlier estimate of 1.6\arcmin.  The disadvantage of this X-ray determination is that the 4-6 keV continuum band is a mix of synchrotron radiation and thermal bremsstrahlung. More recently, \citet{arias18} used free-free absorption in the radio below 100~MHz to map the unshocked ejecta. 
This confirmed that the reverse shock has a  radius of 1.9\arcmin, but with a center shifted 14\arcsec\ to the West of the explosion center as estimated by \citet{thorstensen01}. 
The free-free absorption map shows a more complete map of the outline of the reverse shock than the results of \citet{helder08} in that it includes the eastern part. 
The corresponding free expansion velocities for the ejecta reaching the reverse shock corresponds to $\approx 4400$~\kms in the East and  $\approx 6600$~\kms\ in the West.
Interestingly, a similar discrepancy in velocity is seen for the optical knots in Doppler velocity, with the backside velocity being best fit with velocities $\sim 6000$~\kms, and the front side
with velocities of $\sim 4000$~\kms\ \citep[e.g.][]{reed95,lawrence95,milisavljevic13}.

Note that most of
the optical fast moving knots are located within this inferred reverse shock radius. 
The reason is that these optical knots flare up for tens of years after having encountered the 
reverse shock---so they mark the location of the reverse shock in three dimensions----but many are seen in projection inward of the projected reverse shock radius. 
Only the outer most ejecta knots should be close to the projected radius of the reverse shock.
This can be verified by applying the conversion of projected radius to motion in the plane of the sky as inferred by \citet{milisavljevic13}, $S=0.^{\prime\prime}022~{\rm km^{-s}s}$, 
to their best fit outer most velocities of $\approx $4000--6000\kms, giving a radius ranging from $\approx$1.5\arcmin\ to 2.2\arcmin.

To select the region of the reverse shock we use an annulus of 1.67\arcmin\ to 2.18\arcmin, encompassing the reverse shock location obtained by \citet{arias18}, as indicated in Figure~\ref{fig:regions} in red.
The red/white colored  sectors in  Figure~\ref{fig:regions} (right) show the 18 segments for the reverse shock regions. Like for the forward shock each segment  has an
opening angle of 20\deg, centred on the explosion center,  with the central position angle (PA) starting at $PA=10$\deg, i.e. the first segment starts due North and ends at PA$=20$\deg.

\subsection{Optimisation methods}
For finding the optimal solution for  expansion  measurements per region, the code  searches for the lowest value of $L_{\rm tot}$ (Equation~\ref{eq:lik}) in each region (i.e. for each mask).
For the reverse shock region and the jet region we only optimised for the best value of the expansion rate, $a$. For the forward shock region we  optimised jointly for $a$ and the second derivative, 
the acceleration parameter $b$.

The code has the options to simply scan in fixed steps the parameter space from $a_{\rm min}$ to $a_{\rm max}$, or by a more
efficient way,  but with the drawback that a local rather than a global minimum is found. The latter method is the default option after having tested both options. 
The efficient way consists of first determining  $L_{\rm tot}$ for six values of
$a$, and  selecting the values of $a$ corresponding to the lowest values of $L_{\rm tot}$. These are then used to determine a parabola through these three points ($a$,$L$),
which is then used to guess the minimum of the parabola. The actual value of $L_{\rm tot}$ for this inferred mininum is then determined, and compared to the other three values of ($a$,$L$),
and the worst one is discarded. The remaining three values are then used for a next iteration, until the improvement in  $L_{\rm tot}$ is $\Delta L<4$.
It can be that the parabola is inverted or no valid solutions are found, in which case a random point is chosen and the procedure is repeated. If this happens repeatedly, a scan of the parameter
space is used. 

For finding the parameter $b$ for the forward shock sectors, we used initially a similar efficient procedure as for determining $a$, in an iterative procedure: determine $a$, and fix it to the best-fit
value, and then determine $b$, and then repeat this procedure. But here we did find that sometimes the absolute minimum is not found,
due to a strong statistical correlation between parameters $a$ and $b$. So instead we scanned the parameter space in $a$ and $b$, but optimised computing time by using a mesh-refinement method:
we first scan $L$ on a very course grid  of $5\times5$ values in $a$ and $b$. We then refine the grid by  a factor $3\times3$ and continue the scan of the parameter space for those grid values for which $\Delta L=L-L_{\rm min}$ is above a given threshold. A refinement to a parameter grid of $135\times 135$ values was found to be sufficient, but in a few cases (see Figure~\ref{fig:likmaps_hires} in the appendix)
we also performed an additional refinement to $405\times 405$ grid points.
The starting values of $a$ was found using the efficient method described above. The initial threshold for a refinement of the likelihood search was $\Delta L=2048$, 
the final threshold was set to $\Delta L=16$.  The result of the mesh refinement procedure, as well as the statistical correlation between $a$ and $b$  can be judged from the
examples in Figure~\ref{fig:likmaps_hires}.
The resulting grid values of  $\Delta L$ are also used to determine the measurement error, with $\Delta L=1,4,9$ corresponding to respectively $1\sigma$,2$\sigma$ and 3$\sigma$
confidence intervals. See also the log-likelihood curves in Figure~\ref{fig:logLfs}.

For the separate optimisation of the misalignment correction of the images (Section~\ref{sec:pointing}) a simple scan of ($\Delta x, \Delta y$) was used,  iteratively using first a course grid and then zooming
in on the best value with a smaller grid size.

\begin{figure}
  \centerline{
  \includegraphics[trim=70 50 60 50,clip=true,width=0.5\textwidth]{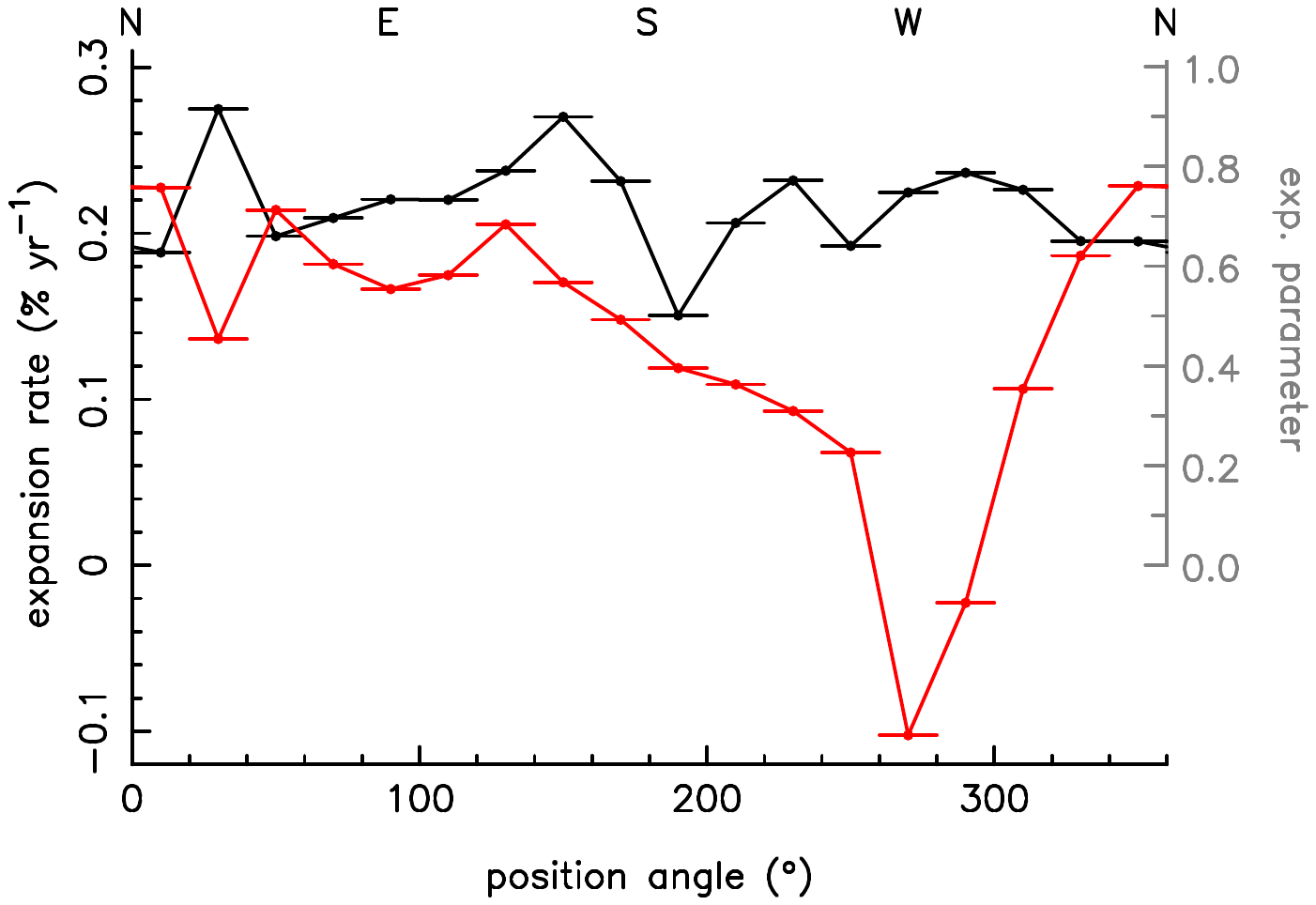}
  }
  \caption{\label{fig:expansion}
 The expansion rates for the forward and reverse shock regions. 
 The grey scale  shows the corresponding expansion parameter, defined here as $m\equiv  t_0/\tau_{\rm exp}$, determined using $t_0=332$~yr.
     North corresponds to PA$=0$\degr, and position angle is defined counter clockwise.}
\end{figure}

\begin{figure}
  \centerline{
    \includegraphics[width=\columnwidth]{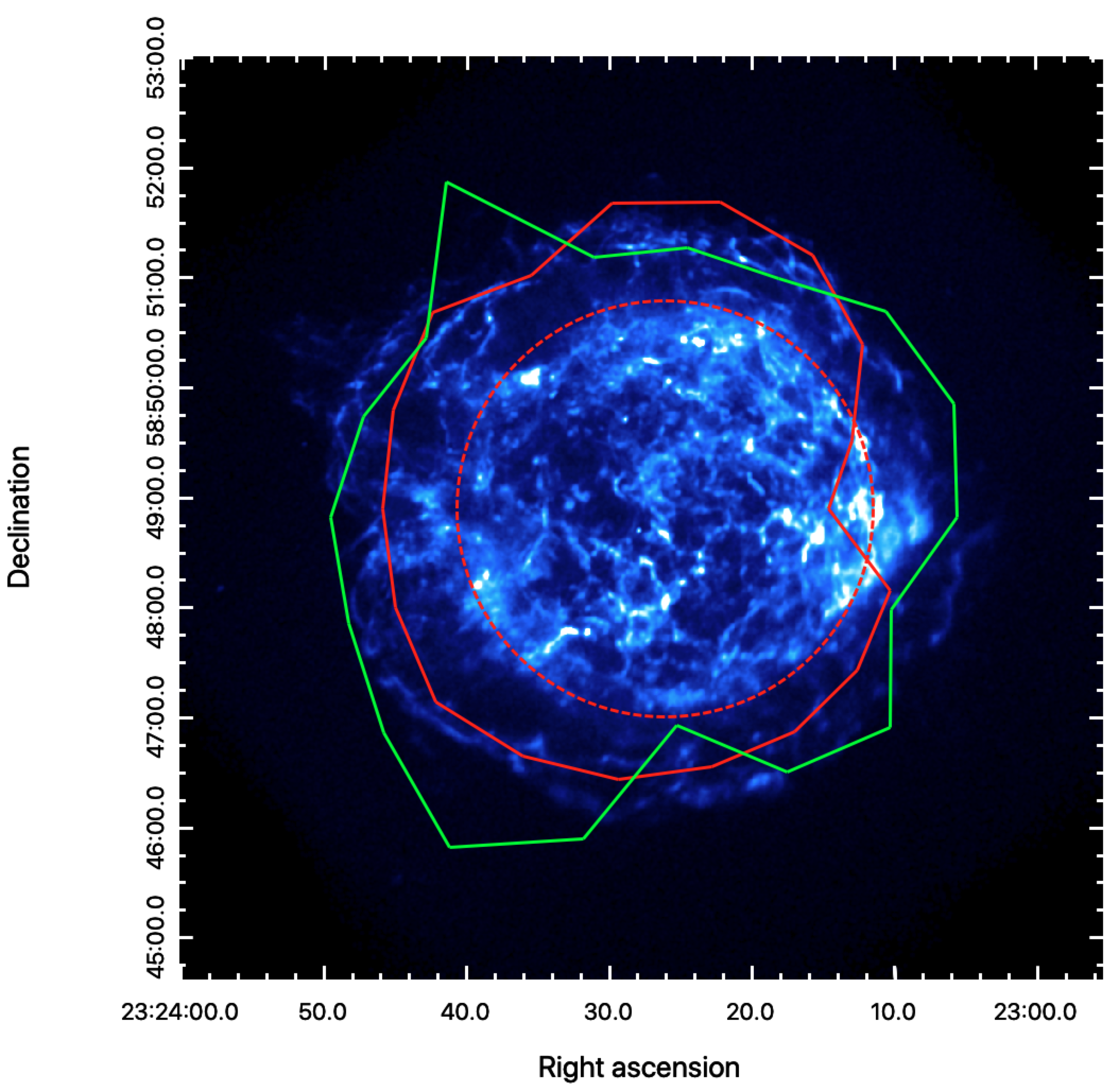}
  }
  \caption{\label{fig:spider}
    Same image as Fig.~\ref{fig:regions} but now with a spider diagram
    overlayed that visualises the expansion rate as a function of position angle.
    For the forward shock (green)
    the radial extent of the spider diagram is linearly proportional to the
    expansion rate. For the reverse shock (red) the radial coordinate provides
    the expansion relative to the dashed circle--inside the circle indicates a motion toward the interior.
  }
\end{figure}

\section{Results}

\begin{figure}
  \centerline{
    \includegraphics[trim=100 80 100 263,clip=true,width=0.25\textwidth]{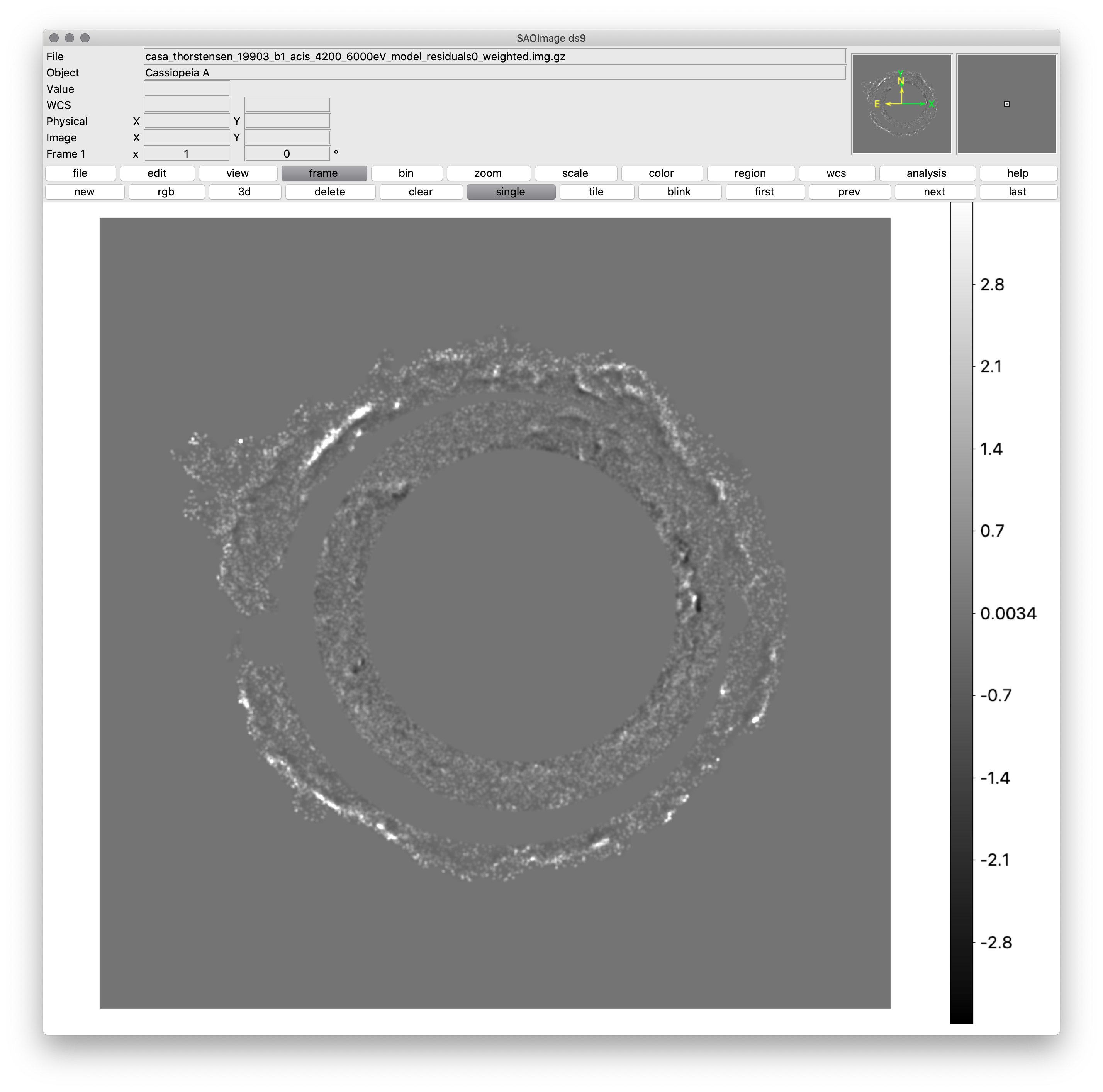}        
    \includegraphics[trim=100 80 100 263,clip=true,width=0.25\textwidth]{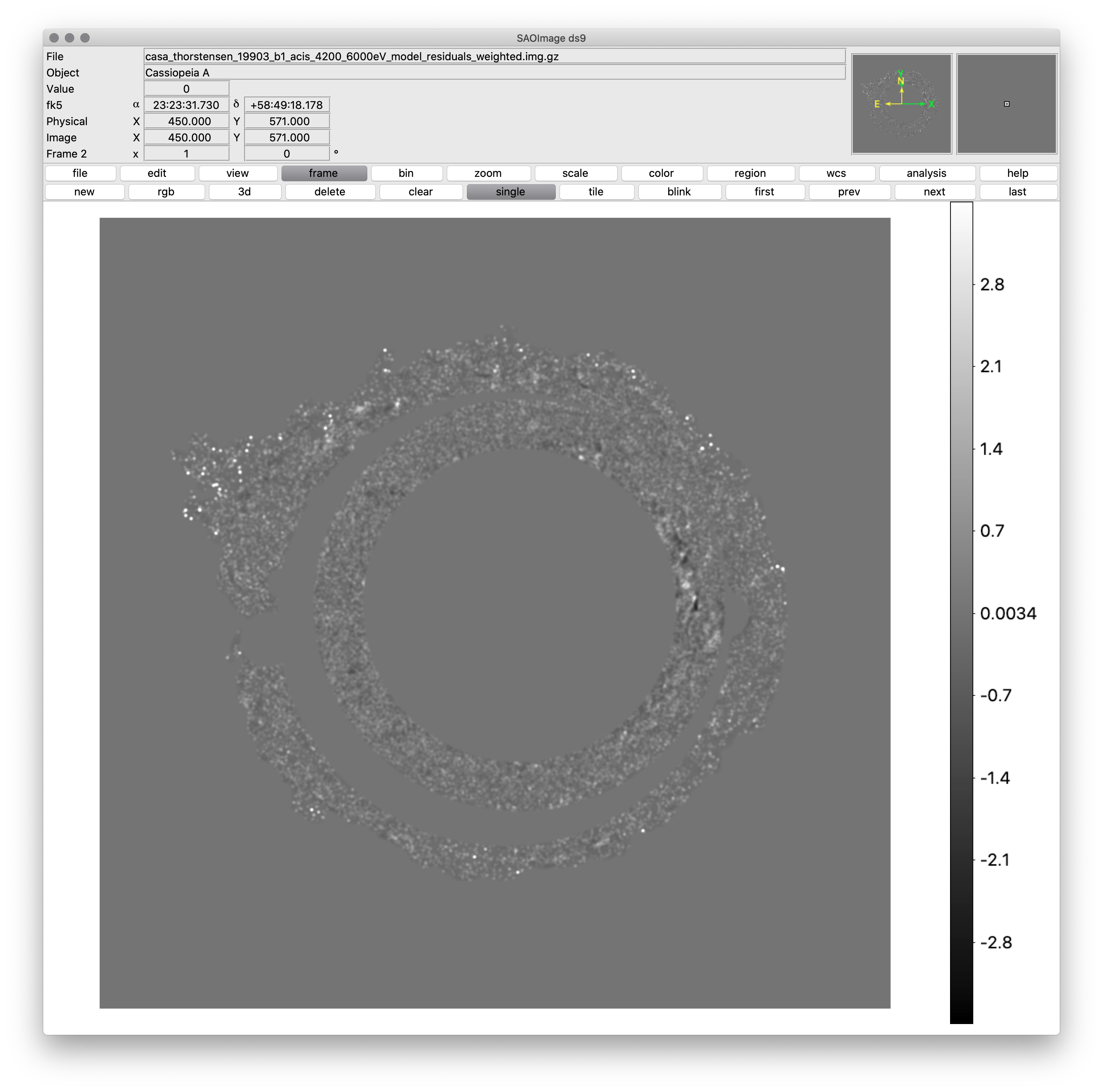}
  }
  \caption{
    Residual images for ObsID 19903, which consists of the masked data image with subtracted the
    combined model image, and then divided by the squareroot values of the
    model image (the expected error per pixel).
    The residual images were smoothed with a gaussian with $\sigma=3$ pixels to bring out features.
    On the left: the model is uncorrected for expansion, i.e. using the uncorrected
    VLP image as a model. On the right: the model was a composite of all
    individual models for each sector.
    The gray scales for both residuals images are identical.
\label{fig:ripples}
  }
\end{figure}

We discuss the results of our X-ray expansion study in the 4.2-6 keV band  for the forward shock and reverse shock regions separately. The results for both shock regions are combined in Figure~\ref{fig:expansion} as a function of PA. Fig.~\ref{fig:spider} provides a more visual illustration of the expansion as function of position in Cas A.
As an illustration for how well the best-fit expansion models fit the data we  show in Figure~\ref{fig:ripples} two residual maps, of the model with the 2019 observation.
On the left, the model is simply the unaltered 2004 VLP observation, whereas in the right-hand image the model consists of the sum of the best-fit models for all 37  sectors.

By comparing the models with individual observations one can indeed see brightness changes over time \citep[see also][]{patnaude07,uchiyama08,patnaude11,sato17},
and occasionally also filaments that change in shape, perhaps due to variable expansion rates along the filaments. These type of changes are a source systematic errors,
and cause most of the residuals seen in Figure~\ref{fig:ripples} (right).

In the appendix the profiles of $\Delta L$ are shown as a function of $a$--- and for the forward  shock $b$---which shows that the best-fit values are well-defined; although some local minima
indicate that there are some ambiguities, possibly due to substructure in some sectors.

\begin{table*}
\centering
\caption{Expansion measurements for the forward shock region. For $v_{\rm s,obs}$ and $m$ a radius of 2.8\arcmin\ and an SNR age of 333~yr are assumed.\label{tab:forward}}
\begin{tabular}{rcccccr}\hline\hline\noalign{\smallskip}
PA & $a$ & $b$ & $\tau_{\rm exp}$ & $m$ & $v_{\rm s,obs}$ & $L$\\ 
{[$^{\circ}$]} & {[\% yr$^{-1}$]} & $[10^{-5}$ yr$^{-2}$] & {[yr]} & & {[km\,s$^{-1}$]}\\\noalign{\smallskip}\hline\noalign{\smallskip}
10 & 0.1900 $\pm$ 0.0012 & -1.67 $\pm$ 0.09 & $526.4^{+3.5}_{-3.4}$ & $0.6326\pm 0.0041$ & 5052 $\pm$ 33 & 126861.2\\
30 & 0.2760 $\pm$ 0.0009 & -6.74 $\pm$ 0.12 & $362.3^{+1.2}_{-1.2}$ & $0.9190\pm 0.0030$ & 7339 $\pm$ 24 & 102128.6\\
50 & 0.1990 $\pm$ 0.0012 & -1.02 $\pm$ 0.03 & $502.4^{+3.0}_{-3.0}$ & $0.6628\pm 0.0039$ & 5293 $\pm$ 31 & 165853.7\\
70 & 0.2098 $\pm$ 0.0011 & -2.07 $\pm$ 0.21 & $476.7^{+2.5}_{-2.5}$ & $0.6986\pm 0.0036$ & 5579 $\pm$ 29 & 162899.0\\
90 & 0.2217 $\pm$ 0.0016 & -6.44 $\pm$ 0.19 & $451.1^{+3.3}_{-3.3}$ & $0.7382\pm 0.0054$ & 5895 $\pm$ 43 & 60783.0\\
110 & 0.2212 $\pm$ 0.0016 & -3.82 $\pm$ 0.15 & $452.1^{+3.3}_{-3.2}$ & $0.7366\pm 0.0053$ & 5882 $\pm$ 42 & 95685.0\\
130 & 0.2384 $\pm$ 0.0008 & 0.67 $\pm$ 0.06 & $419.5^{+1.5}_{-1.5}$ & $0.7938\pm 0.0028$ & 6339 $\pm$ 22 & 114881.9\\
150 & 0.2707 $\pm$ 0.0020 & 1.17 $\pm$ 0.31 & $369.4^{+2.8}_{-2.7}$ & $0.9014\pm 0.0067$ & 7199 $\pm$ 53 & 64903.1\\
170 & 0.2329 $\pm$ 0.0024 & 4.17 $\pm$ 0.07 & $429.4^{+4.5}_{-4.4}$ & $0.7756\pm 0.0081$ & 6194 $\pm$ 65 & 52860.5\\
190 & 0.1518 $\pm$ 0.0014 & 11.75 $\pm$ 0.07 & $658.6^{+5.9}_{-5.8}$ & $0.5056\pm 0.0045$ & 4038 $\pm$ 36 & 63778.8 \\
210 & 0.2076 $\pm$ 0.0030 & -3.39 $\pm$ 0.12 & $481.7^{+7.1}_{-6.9}$ & $0.6913\pm 0.0100$ & 5521 $\pm$ 80 & 67648.2\\
230 & 0.2331 $\pm$ 0.0017 & -1.87 $\pm$ 0.15 & $429.1^{+3.1}_{-3.1}$ & $0.7761\pm 0.0056$ & 6198 $\pm$ 44 & 53429.5 \\
250 & 0.1934 $\pm$ 0.0013 & 11.09 $\pm$ 0.04 & $517.1^{+3.4}_{-3.4}$ & $0.6439\pm 0.0042$ & 5142 $\pm$ 34 & 97039.3\\
270 & 0.2251 $\pm$ 0.0009 & 0.33 $\pm$ 0.06 & $444.3^{+1.8}_{-1.8}$ & $0.7494\pm 0.0030$ & 5985 $\pm$ 24 & 114108.3\\
290 & 0.2376 $\pm$ 0.0010 & 4.33 $\pm$ 0.05 & $420.9^{+1.8}_{-1.8}$ & $0.7911\pm 0.0034$ & 6318 $\pm$ 27 & 104277.1 \\
310 & 0.2274 $\pm$ 0.0008 & -2.67 $\pm$ 0.15 & $439.7^{+1.6}_{-1.5}$ & $0.7573\pm 0.0027$ & 6048 $\pm$ 21 & 113542.6\\
330 & 0.1963 $\pm$ 0.0016 & -2.24 $\pm$ 0.30 & $509.5^{+4.3}_{-4.2}$ & $0.6536\pm 0.0055$ & 5220 $\pm$ 44 & 156104.4 \\
350 & 0.1958 $\pm$ 0.0020 & 2.21 $\pm$ 0.10 & $510.8^{+5.4}_{-5.3}$ & $0.6519\pm 0.0068$ & 5206 $\pm$ 54 & 142819.6\\
\noalign{\smallskip}\hline\noalign{\smallskip}
Mean & 0.218 $\pm$ 0.029 & 0.21 $\pm$ 4.94 & & 0.727 $\pm$ 0.096 & 5803 $\pm$ 763\\
\noalign{\smallskip}\hline
\end{tabular}
\end{table*}

\subsection{The forward shock region}

For determining the proper motion of the forward shock expansion we used  two methods, one with solving also for the deceleration of the shock---$b$ in Equation~\ref{eq:exp_rate_pars}---and one
with assuming $b=0$. The best-fit parameters including the measurement of the deceleration rate  are summarised in Table~\ref{tab:forward}.

For the best-fit values of $a$, fitting also for $b$ does not make much of a difference for the expansion rates. 
For example,  solving for $b$ gives an average expansion rate of $0.218\pm 0.029$\% yr$^{-1}$,
compared to  $0.219\pm 0.030$\% yr$^{-1}$ when keeping $b=0$---the errors indicate the rms of the variations. 
These values correspond to  expansion time scales of $\tau_{\rm exp}=457\pm 60$~yr, or $m=t_0/\tau_{\rm exp}=0.73\pm 0.10$.

There is quite some variation in the expansion parameter as a function of PA, with for a PA  of 30\deg\   an expansion parameter of $m=0.919\pm 0.003$, which 
approaches free expansion ($m=1$). The slowest expansion is measured for a PA  of 190\deg (South) with $m=0.506 \pm 0.05$.
As illustrated in Figure~\ref{fig:comparison} (see appendix), the expansion rates are relatively robust to the choice of pointing corrections.

\begin{figure}
    \centerline{
\includegraphics[trim=50 50 100 100,clip=true,width=0.5\textwidth]{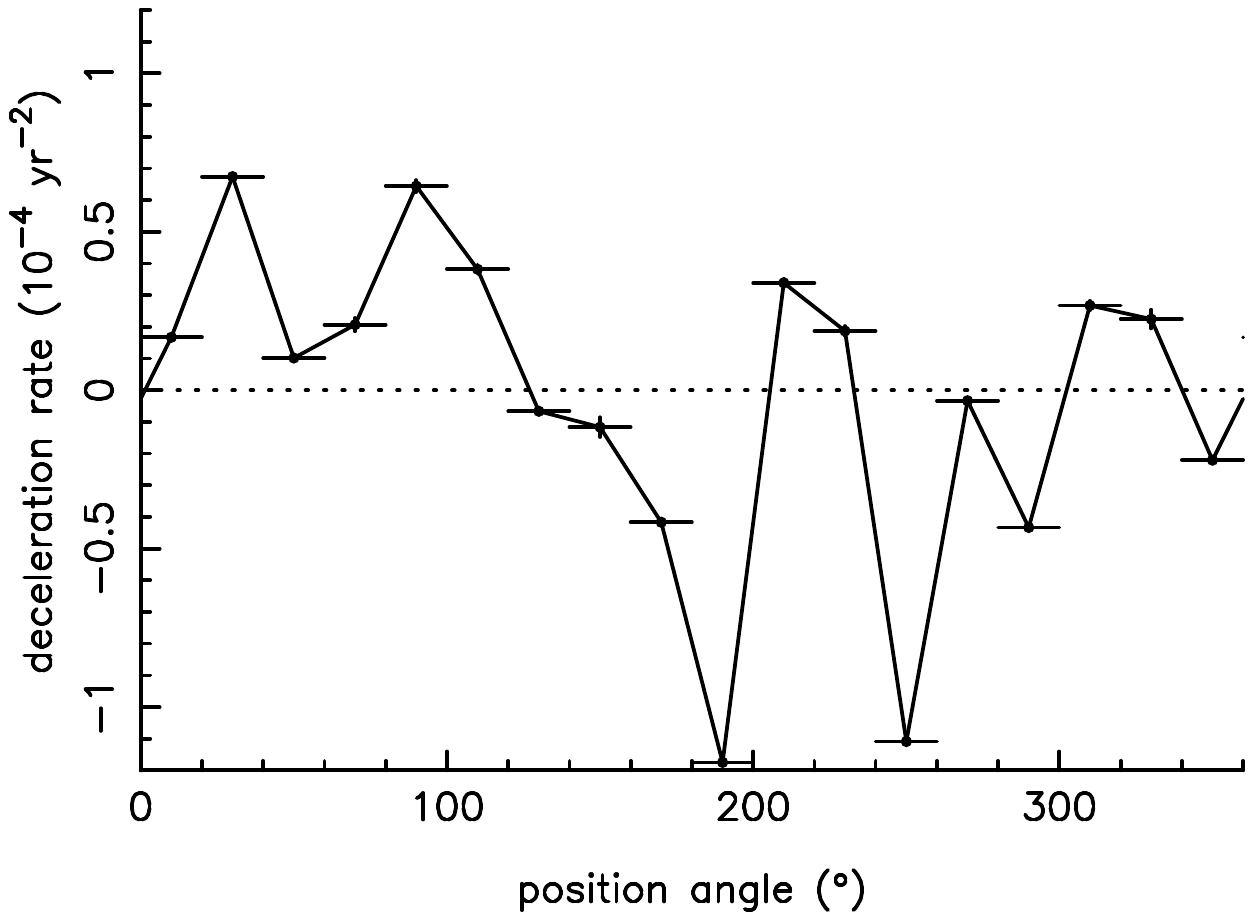}
    }
  \caption{\label{fig:deceleration}
    The measured deceleration rates (defined here as deceleration$\equiv -b$) as
    a function of position angle.
  }
\end{figure}

Fitting for $b$ does  significantly improve the  fitting results  for each PA, with the exception of $PA=270$\deg for which the best fit value is anyway $b\approx 0$.
For the other PAs the improvement ranges from 
$\Delta L=-4.7$ for  PA$=50$\deg\ to $\Delta L=-400$  for PA$=250$\deg. For the sum of $L$ over all 18 sectors of the forward shock region, we have 
 $\Delta \sum_{\rm PA}L=-1155$.

 The average  of the best-fit values for the expansion rate derivative is $b=(-0.21\pm 4.94)\times 10^{-5}$~yr$^{-2}$, which shows there is a larger spread in $b$ than the average
 value.  Contrary to expectations, there appear to be portions of the forward shock  that are accelerating---in particular around PA$=190$\deg, and 250\deg.
 Moreover, the spread is about an order of magnitude larger than the expected
 value of $|b|\approx 2\times 10^{-6}$~yr$^{-2}$ (see Section~\ref{sec:method}).
 For each individual PA, however, the detection is determined with an
 error of the order of (1--2)$\times 10^{-6}$~yr$^{-2}$. So for several individual PAs we can report a statistically significant measurement of a deceleration or an acceleration.
 
 There are some reason to be cautious  about the values of the deceleration rates. Firstly, because the measured values are larger than the theoretical expectations, and therefore
 require more scrutiny. More importantly, the
 measured deceleration rates are more sensitive to systematic errors than the expansion rates. Potential systematic errors are (1) within each shock region multiple filaments exists with potentially different 
  dynamics; (2) filaments brighten and dim, and new filaments may appear over the time span of interest (2000--2019), so one should be aware that the shock front is not uniquely
  defined during the whole observation period; and, last but not least, (3) unlike the expansion rate $a$ the deceleration rate is more  sensitive to misalignments of images of
  different epochs. We illustrate this by showing  in the appendix also the measured deceleration rates without solving for ($\Delta x, \Delta y$) for each individual
  epoch (see Section~\ref{sec:pointing}). 
  
Note that $\Delta x=0, \Delta y=0$ is a very conservative approach to test systematic errors. In fact,
setting $\Delta x=0, \Delta y=0$ leads to an overall deterioriation of  the test statistic of $\Delta \sum_{\rm PA}L=3802$.

So we think that, indeed, the western part of the forward shock is accelerating. But it important
to confirm this with more observations, preferentially by another deep Chandra deep observation, which could be used as an alternative reference image.

\begin{table*}
\centering
\caption{Expansion measurements for the reverse shock region. For $v_{\rm s,obs}$ and $m$ a radius of 1.9\arcmin---but not centered on the explosion center---and an SNR age of 333~yr are assumed.\label{tab:reverse}}
\begin{tabular}{rccccr}\hline\hline\noalign{\smallskip}
PA & $f$ & $\tau_{\rm exp}$ & $m$ & $v_{\rm s,obs}$ & $L$\\ 
{[$^{\circ}$]} & {[\% yr$^{-1}$]} & {[yr]} & & {[km\,s$^{-1}$]}\\\noalign{\smallskip}\hline\noalign{\smallskip}
10 & 0.2275 $\pm$ 0.0005 & $439.6^{+1.0}_{-1.0}$ & $0.7574\pm 0.0018$ & 4179 $\pm$ 10 & 132400.8\\
30 & 0.1364 $\pm$ 0.0011 & $733.3^{+5.8}_{-5.7}$ & $0.4541\pm 0.0036$ & 2506 $\pm$ 20 & 72990.0\\
50 & 0.2140 $\pm$ 0.0008 & $467.3^{+1.7}_{-1.7}$ & $0.7127\pm 0.0026$ & 3932 $\pm$ 14 & 125390.4\\
70 & 0.1814 $\pm$ 0.0023 & $551.2^{+7.0}_{-6.8}$ & $0.6041\pm 0.0075$ & 3333 $\pm$ 42 & 119088.9\\
90 & 0.1663 $\pm$ 0.0018 & $601.2^{+6.4}_{-6.3}$ & $0.5539\pm 0.0058$ & 3056 $\pm$ 32 & 122083.8\\
110 & 0.1748 $\pm$ 0.0013 & $572.2^{+4.2}_{-4.2}$ & $0.5819\pm 0.0043$ & 3211 $\pm$ 23 & 105591.0\\
130 & 0.2053 $\pm$ 0.0016 & $487.1^{+3.9}_{-3.8}$ & $0.6837\pm 0.0054$ & 3772 $\pm$ 30 & 127765.6\\
150 & 0.1704 $\pm$ 0.0012 & $586.7^{+4.2}_{-4.1}$ & $0.5675\pm 0.0040$ & 3131 $\pm$ 22 & 110526.2\\
170 & 0.1479 $\pm$ 0.0009 & $675.9^{+4.2}_{-4.2}$ & $0.4926\pm 0.0030$ & 2718 $\pm$ 17 & 101548.7\\
190 & 0.1188 $\pm$ 0.0012 & $841.8^{+8.5}_{-8.3}$ & $0.3956\pm 0.0039$ & 2183 $\pm$ 22 & 110024.0\\
210 & 0.1090 $\pm$ 0.0008 & $917.7^{+6.6}_{-6.5}$ & $0.3628\pm 0.0026$ & 2002 $\pm$ 14 & 111771.7\\
230 & 0.0929 $\pm$ 0.0013 & $1076.4^{+15.7}_{-15.3}$ & $0.3094\pm 0.0044$ & 1707 $\pm$ 25 & 127131.4\\
250 & 0.0679 $\pm$ 0.0014 & $1471.8^{+31.8}_{-30.5}$ & $0.2262\pm 0.0048$ & 1248 $\pm$ 26 & 88390.1\\
270 & -0.1023 $\pm$ 0.0012 & $-977.5^{+11.1}_{-11.3}$ & $-0.3407\pm 0.0039$ & -1880 $\pm$ 22 & 55628.8\\
290 & -0.0226 $\pm$ 0.0007 & $-4425.0^{+124.0}_{-131.3}$ & $-0.0753\pm 0.0022$ & -415 $\pm$ 12 & 109937.4\\
310 & 0.1062 $\pm$ 0.0008 & $941.4^{+7.3}_{-7.2}$ & $0.3537\pm 0.0027$ & 1952 $\pm$ 15 & 149720.9\\
330 & 0.1865 $\pm$ 0.0012 & $536.1^{+3.5}_{-3.5}$ & $0.6211\pm 0.0040$ & 3427 $\pm$ 22 & 130213.0\\
350 & 0.2285 $\pm$ 0.0004 & $437.7^{+0.7}_{-0.7}$ & $0.7608\pm 0.0012$ & 4198 $\pm$ 7 & 142704.4\\
\noalign{\smallskip}\hline\noalign{\smallskip}
Mean & 0.134 $\pm$ 0.084 & & 0.446 $\pm$ 0.279 & 2459 $\pm$ 1540 \\
\noalign{\smallskip}\hline
\end{tabular}
\end{table*}

\subsection{The reverse shock region}

For the reverse shock region we did not attempt to measure the deceleration, given the systematic errors involved, but also because the expansion rates
as a function of PA are varying substantially, making the expansion rate already an interesting parameter from a dynamical point of view. The results for the reverse shock expansion rates are listed in Table~\ref{tab:reverse} and depicted in red in Figure~\ref{fig:expansion}.
The most remarkable feature is that the reverse shock has a negative value---indicating that the reverse shock moves toward the center---for PAs between
260\deg and 300\deg, corresponding to the southwestern and western region.

\begin{table}																		
\centering																	
\caption{\label{tab:jet}	Expansion measurements for the northeastern jet. 																
}																	
{\scriptsize																	
\begin{tabular}{lcccc}\hline\hline\noalign{\smallskip}																	
	&	Energies	&	$a$			& 	$\tau_{\rm exp}$			&	$m$			\\		
	&	[keV]	&	$[\% {\rm yr}^{-1}]$			&	[yr]			&				\\\noalign{\smallskip}\hline\noalign{\smallskip}		
Cont.	&	4.2--6.0	&	0.2041	$\pm$	0.0004	&	489.9	$\pm$	0.88	&	0.680	$\pm$	0.001	\\		
Si-K	&	1.75--1.97	&	0.2401	$\pm$	0.0003	&	416.5	$\pm$	0.45	&	0.800	$\pm$	0.001	\\\noalign{\smallskip}\hline\noalign{\smallskip}		
\end{tabular}																	
}																	
\end{table}

\subsection{The northeastern jet}
The measurement of the northeastern jet stands  apart from the other expansion measurements reported here,
 in that one cannot claim to measure the expansion of a shock front.
 Instead the measured  proper motions concern the movement of plasma in the jet, mostly caused by the movements of
 bright features within the overall jet envelope. In this sense the rational for using the 4.2-6 keV energy band is not valid, also because, as
 far as we know, the continuum is thermal rather than nonthermal radiation. In the continuum band the northeastern
 jet is relatively faint, but the jet is rich in IMEs. 
 For that reason we report here both an expansion in the continuum band, to be consistent with the measurements of the forward and
 reverse shock regions, and a measurement in the 1.75--1.97 keV band, which is dominated by K-shell  emission from Si XIII.
 The results are summarised in Table~\ref{tab:jet}.
The measurements in the two band are not consistent with each other, with $a=0.2041$\%yr$^{-1}$ for the continuum band,
and $a=0.2401$\%yr$^{-1}$ for the Si XIII band. But the expansion rates are within a range similar to that of the forward shock.

\begin{figure}
  \includegraphics[trim=50 50 50 50,clip=true,width=\columnwidth]{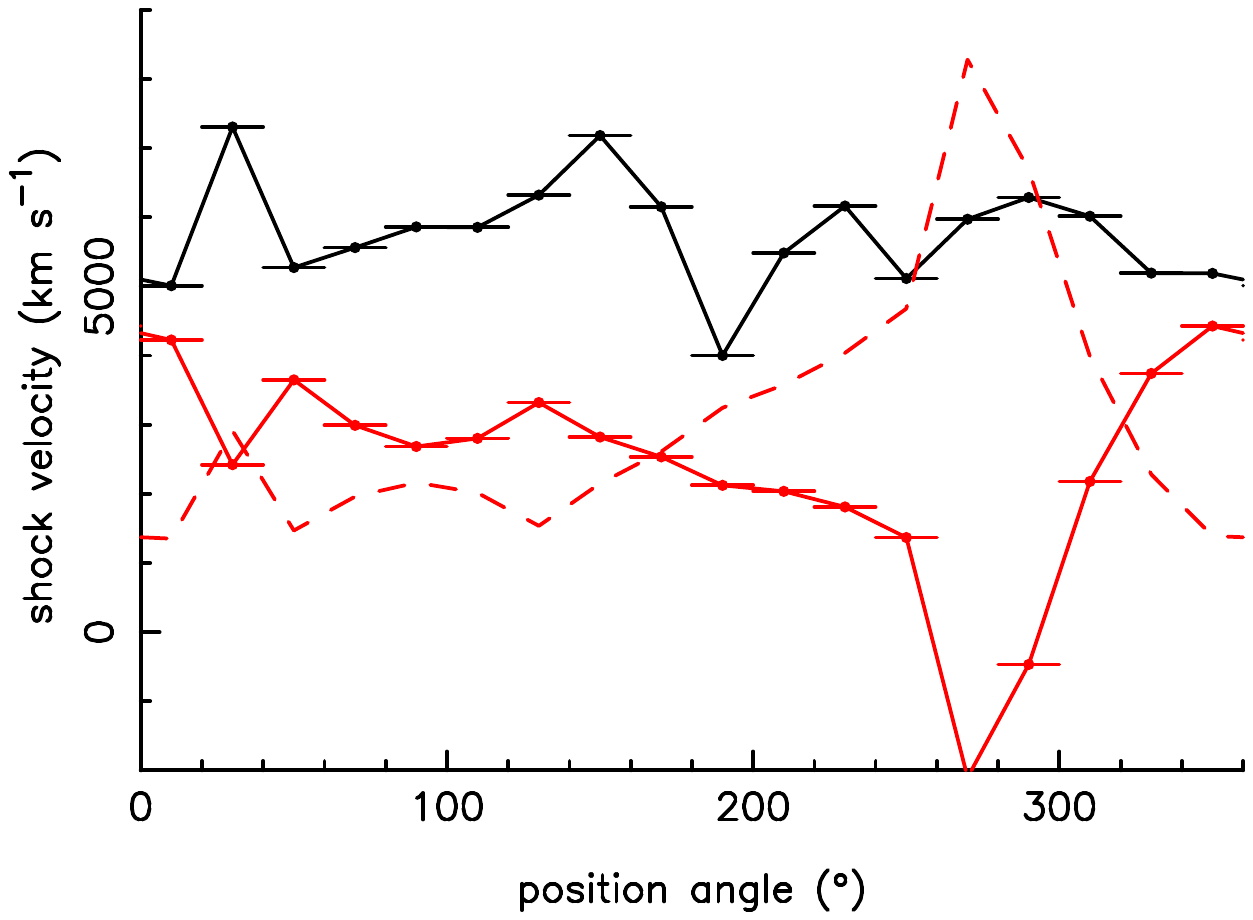}
  \caption{\label{fig:velocities}
    The implied velocities of the forward- (black) and reverse-shock regions (red), based on the approximate angular distances of the shock with respect
    to the expansion center \citep{thorstensen01},
    taking into account the shift of the reverse
    shock with respect to this center \citep{arias18}.
    The solid lines are the velocities in the frame of the observer. The
    red dashed line shows the reverse shock velocity in the frame of the
    freely expanding ejecta.
  }
\end{figure}

\section{Discussion}

We reported here X-ray proper motion measurements for
 the forward and reverse shock region, as well as for the northeastern jet. 
The two shock regions are each divided in 18 annular segments with opening angles of 20\deg.  This is different from various previous studies \citep{delaney03,patnaude09,sato18},
 most of which focus on individual filaments. We reported here the proper motions in terms of the expansion rate $a$ (\%yr$^{-1}$). For the forward shock we also measured the deceleration/acceleration rate,
and the expansion of the northeastern jet. Both the deceleration/acceleration measurements, as well as the jet expansion in X-rays, are here reported for the first time, as far as we know.

The forward shock proper motions reported here generally agree with previous measurements. They are comparable with those measured by \citet{delaney03}, who
reported an average expansion rate of  $0.20\pm0.07$\%yr$^{-1}$ ($m=0.66$), and they are somewhat larger than those reported by  \citet{patnaude09}, but for fewer filaments.
The forward shock expansion is also consistent with the 151~MHz radio expansion measurement by \citet{agueros99}, reporting a dynamical time scale of 400--500~yr, which corresponds
to 0.25--0.20\%yr$^{-1}$.

The   proper motions near the reverse shock reported here show that the reverse shock is moving outward in most regions, but is negative for position angles between 260\deg\ and  300\deg.
This is not the first time that a  reverse shock velocity directed toward the center is reported. In X-rays the first hints for much lower velocities in the West were reported in\citet{vink98a},
but negative velocities of filaments associated with the reverse shock were reported by \citet{delaney04} and more recently by \citet{sato18}. \cite{sato18}  reported projected
velocities of 2100~\kms--3800~\kms, toward the center.
These X-ray measured proper motions find their radio-band counterpart in the proper motion measurements of individual knots \citep{bell77,tuffs86,anderson95}, which provide a much slower expansion rate
than the overal expansion of the SNR in the radio as reported by  \citet{agueros99}. In fact, \citet{anderson95} even reported proper motions toward the center in the West.
With this in mind, it is clear that these radio knots, and the overall bright radio ring they are embedded in, contains relativistic electrons accelerated by the reverse shock, a conclusion already obtained
by \citet{helder08,uchiyama08}.

For the discussion on the implications of our reverse shock measurements, 
we assume that in the western part the X-ray continuum from the reverse shock region 
is dominated by synchrotron radiation, rather 
than nonthermal bremsstrahlung \citep{laming01a,vink03a}. 
X-ray synchrotron radiation from the western reverse shock region is currently the more prevalent hypothesis \citep[e.g.][]{helder08,vink08a,uchiyama08,grefenstette15,sato18},
but a definitive proof  is lacking.
In the near future  X-ray polarisation measurements with the Imaging X-ray Polarimetry Explorer \citep[IXPE,][]{weisskopf21} may be able to provide a definitive proof for
a synchrotron radiation nature by detecting polarised X-ray radiation from the
western region, provided the local magnetic-field turbulence is not too high \citep{bykov20}.

In the case of X-ray synchrotron radiation the measured proper motions most likely reflect the proper motion of the
reverse shock itself, as X-ray synchrotron radiation requires a strong shock, and the highly relativistic electrons need to be close to the shock on account of the rapid synchrotron cooling rate.
For  nonthermal bremsstrahlung it is plausible that they may be powered by internal, secondary shocks as well, allowing for the possibility that the measured proper motions are not directly linked to
the reverse shock proper motion. 
 However, as we will discuss below, there are also hints from optical observations of a reverse shock that moves toward the center---but puzzlingly more extensively so than found in our study. 
 Finally, the nonthermal X-ray filaments in the western part coincide with the expected location of the reverse shock as determined by the radio absorption measurements \citep{arias18}.

The expansion rates can be converted into a (projected) velocity using the distance to Cas A of 3.4~kpc \citep{reed95}, and using the shock radii.
For the forward shock  we use a radius of 2.8\arcmin, which conveniently corresponds to 2.8~pc. Note that often 2.5~pc is used, but 2.8~pc is closer to
the location of the filaments in the Chandra observations. 

For the reverse shock we used the estimate of its location
obtained from radio-absorption measurements of unshocked ejecta gas by \citet{arias18} for the conversion of expansion rate to 
velocity. 
The location of the reverse shock is
approximated by a 1.9~pc radius, but centered on RA$=$23h23m26s, DEC$=$58\deg48\arcmin54\arcsec, which is $\sim14$\arcsec\ toward the (north)west of the explosion center derived by
 \citet{thorstensen01}. One assumption is that the velocities measured by us are largely confined to the plane of the sky. Although this will not be the case everywhere, by measuring in annuli targeting the outer
 radius, and the inferred projected reverse shock position, we hope to have minimized the chance for large velocity components along the line of sight.

\begin{figure}
  \centerline{
 \includegraphics[trim=100 80 100 263,clip=true,width=0.25\textwidth]{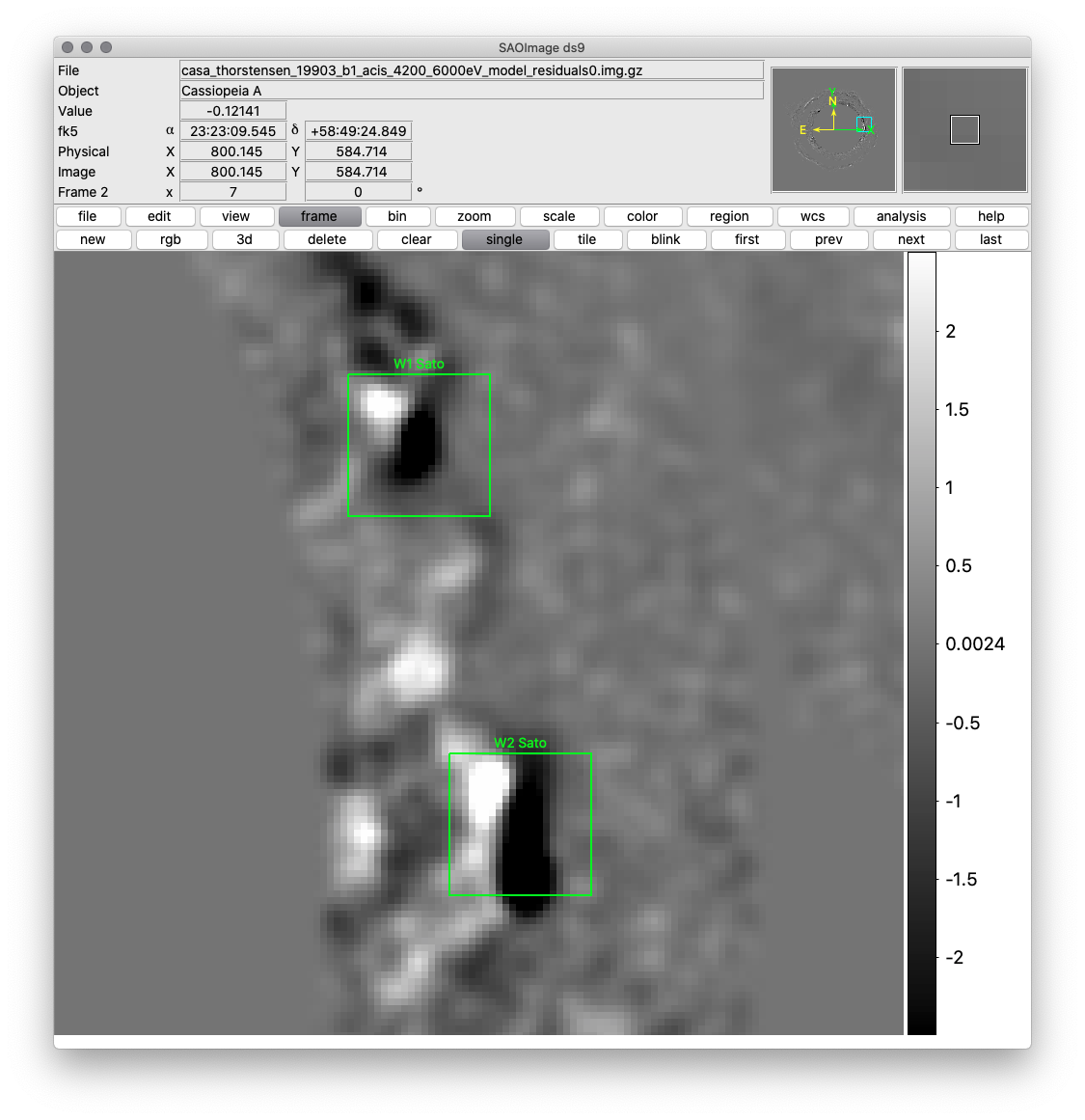}
    \includegraphics[trim=100 80 100 263,clip=true,width=0.25\textwidth]{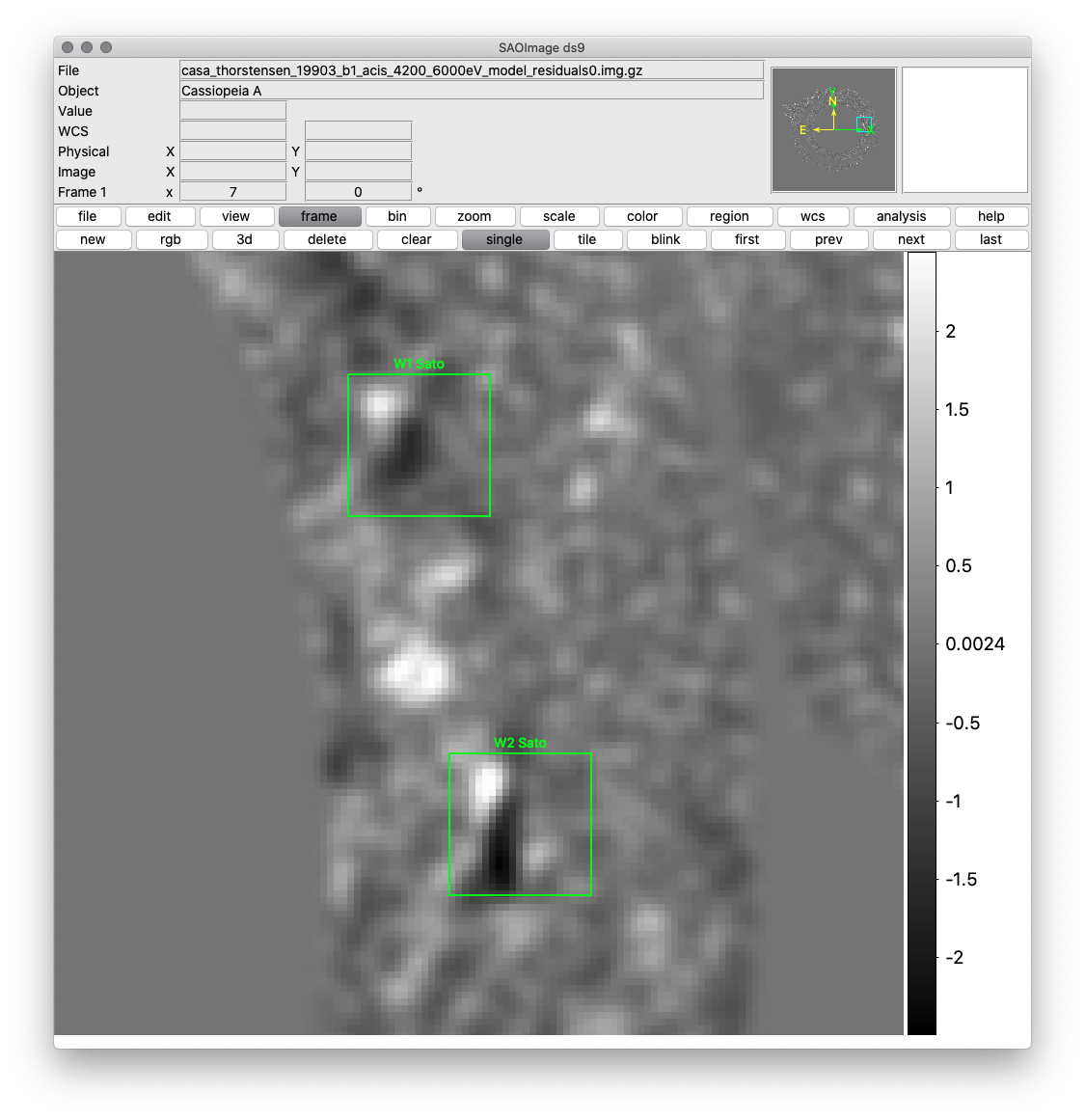}
    }
  \caption{\label{fig:ripples_detail}
    Detail of Fig.~\ref{fig:ripples}, focussing on two filaments, indicated
    by the squares, with
    proper motion measurements by \citet{sato18}.
  }
\end{figure}

The inferred  velocities are listed in Table~\ref{tab:forward}  and Table~\ref{tab:reverse}, and  plotted in Figure~\ref{fig:velocities}. For the forward shock the velocities range from 4040~\kms\ to
7340~\kms, the latter at a PA of 30\deg, with an expansion parameter of $m=0.91$, which is close to free expansion ($m=1$). 
The maximum observed proper motion for the reverse shock region corresponds to 4200~\kms\ in the North, but at a PA of 270\deg\ the projected reverse shock velocity in the observer frame is -1880~\kms,
which is somewhat smaller than the values reported in this region by \cite{sato18}. However, \citet{sato18} handpicked the filament, and the value reported by us correspond to a larger region.
Indeed if we focus on this region, as done in Figure~\ref{fig:ripples_detail} we see that the residuals are still large, even after expansion correction, 
around the regions picked by \citet{sato18} (their W1 and W2). We note that the residuals around W1 are also reflecting brightness changes and  changes in the shape of the filaments.

For the forward shock the inferred velocities of the X-ray synchrotron filaments correspond to the shock velocity $V_{\rm fs}$, but for the reverse shock velocity 
we have to make a distinction between the velocity in the frame of the observer (red solid line in Figure~\ref{fig:velocities}) and the  velocity difference between the shock front
and the unshocked ejecta. The latter the defines the thermodynamical and particle acceleration properties of the shock.

For the reverse shock the unshocked gas close to the shock front moves with the free expansion velocity $v_{\rm ej}=R_{\rm rs}/t$,
whereas the reverse shock in the frame of the observer can be expressed as $v_{\rm rs,obs}=dR_{\rm rs}/dt$. This gives for the reverse shock velocity in the frame of the ejecta:
\begin{equation}\label{eq:vs_rs}
\left|V_{\rm rs}\right|= \left| \frac{R_{\rm rs}}{t} - \frac{dR_{\rm rs}}{dt}\right|,
\end{equation}
with $t$ the age of the SNR. In Figure~\ref{fig:velocities} $V_{\rm rs}$ is shown as a red dashed line. 
For the western part the free expansion velocity of the ejecta close to the reverse shock is $R_{\rm rs}/t_0 \approx 6300$~kms, whereas for the eastern part it is  $R_{\rm rs}/t_0 \approx 4900$~kms.
 The difference is due to the different projected distances of the reverse shock location with respect to the explosion center \citep{arias18}.

\subsection{The reverse shock velocity and X-ray synchrotron radiation}

X-ray synchrotron radiation from the vicinity of SNR shocks requires fast acceleration, and hence fast shocks, as the electrons responsible for the radiation---typically with energies $\gtrsim 10$~TeV---have a relatively short loss time scale \citep[e.g.][]{ginzburg65,vinkbook}:
\begin{equation}
\tau_{\rm syn,loss}\approx 20\left(\frac{E}{10~{\rm TeV}}\right)^{-1}\left(\frac{B}{250~{\rm \mu G}}\right)^{-2}~{\rm yr},
\end{equation}
with $B\approx 250~{\rm\mu G}$ the typical postshock magnetic field for Cas A, and $E$ the electron energy.  This sets the typical time scale for brightness variations in X-ray synchrotron, but it also 
means that the acceleration rate needs to be faster than the synchrotron loss rate. Using the theory of diffusive shock acceleration, while incorporating radiative energy losses, one can show that
the typical photon energy, $h\nu$, relates to the shock velocity as \citep[e.g.][]{aharonian99,zirakashvili07,vinkbook}
\begin{equation}
h\nu\lesssim 3\eta^{-1}\left(\frac{V_{\rm s}}{3000~{\rm km\,s^{-1}}}\right)^2~{\rm keV},
\end{equation}
with $\eta\gtrsim 1$ a parameter that is inversely proportional to the magnetic-field turbulence on length scales of the electron gyroradius.
So X-ray synchrotron radiation in the 4.2-6 keV band requires shock velocities that are above 3000~\kms,  in conjunction with a highly turbulent magnetic field.

The forward shock velocity reported here, and in previous studies, is everywhere well above this limit. However, for the reverse shock region this is not the case.
Note that $V_{\rm s}$ for the reverse shock refers to $V_{\rm rs}$ in Equation~\ref{eq:vs_rs}, and is depicted as a dashed line in Figure~\ref{fig:velocities}.
Apart from the difference in reverse shock motion, $dR_{\rm rs}/dt$, also the difference in location with respect to the explosion center adds to the difference in shock velocity
of about 1400~\kms\ between the eastern and western parts.

It is clear that  $V_{\rm rs}$ is only sufficiently high for X-ray synchrotron emission for  PA $\sim 180$\deg\ to 310\deg, 
which explains why indeed only in the western part of Cas A the reverse shock
X-ray continuum emission is dominated by synchrotron radiation \citep[see][Figure~6]{helder08}, and even why both X-ray and radio synchrotron radiation is
very bright around a PA of 270\deg: the shock velocity there is $V_{\rm s} \approx 8000$~\kms, exceeding even the forward shock velocity.

However, there are two caveats, which are related to each other. First of all, if X-ray synchrotron emission is only originating from the reverse shock regions between PA $\sim 180$\deg\ to 310\deg\ than one
rational for selecting the 4.2-6 keV band---namely that we wanted to focus on X-ray synchrotron radiation as it traces the shock front---is not entirely valid. 
The other caveat is that  studies of the optical knots near the reverse shock \citep[Fesen et al, in preparation, see also][]{fesen19}, shows evidence that all around the SNR the reverse
shock is either at a standstill or returning toward the center, clearly at odds with our study where the reverse shock in the observer frame has a proper motion corresponding
to 2000--4000~\kms\ outside PAs of 180\deg--310\deg. Note that the optical emission from the reverse shock is caused by ejecta clumps, entering the reverse shock and lighting up for 
a few years. So the motion of the reverse shock in the optical refers to the pattern of the locations where the knots start are appearing, rather than relying on direct proper motion measurements. 

So what is the potential effect of having a region dominated by thermal bremsstrahlung rather than synchrotron radiation in the continuum band?
Due to the short radiative loss time scale, synchrotron radiation is confined to a region close to the shock front, and moves along with the shock front.
For thermal bremsstrahlung the shock front itself will be the onset of the emission, but the shocked plasma itself remains visible for a long time. So likely the proper motion is a combination
of the proper motion of the shock front, and of the plasma motion behind it.
The plasma motion with respect to the shock front is given by $\Delta v=V_{\rm rs}/\chi$, with $\chi=4$ the shock compression ratio---assuming that the cosmic-ray acceleration is not very efficient,
in which case $\chi > 4$\ \citep[e.g.][]{berezhko99,vinkbook}.  This means that the shock plasma in the frame of the observer will move with
\begin{equation}
v_{\rm plasma,obs}=\frac{dR_{\rm s}}{dt} + \frac{V_{\rm rs}}{\chi}.
\end{equation}

We can conservatively estimate the magnitude of the plasma velocities on the proper motions of the eastern part by assuming that the reverse shock is stationary, i.e. $dR_{\rm rs}/dt=0$ and 
 $V_{\rm rs}=|R_{\rm rs}/t|$,
and that the proper motion reflects solely the  plasma velocity.
In that case
$v_{\rm plasma,obs} = R_{\rm rs}/(\chi t)\approx 1400$~\kms, for $R_{\rm s}=1.9$~pc and $\chi=4$. Clearly, the measured velocities in the eastern part are larger than 1400~\kms, with velocities as high as
4200~\kms, and an average shock velocity of $2460$~\kms. 
So, although we may have overestimated the  reverse shock velocity outside the western part in our frame, the reverse shock cannot be stationary or returning backward
for PAs 0\deg\ to 220\deg, or from 310\deg to 360\deg.

An alternative explanation for the difference between the optically determined reverse shock proper motion, and the proper motions reported here, may be that using the optical ejecta knots for
determining the reverse shock velocity biases locally
toward stationary or returning reverse shocks.
Another observational bias may be that most of the optical knots are projected toward the inside of the projected reverse shock outer radius. For the X-ray proper motions we selected the movements
near the edge of the reverse shock, whereas for optical measurements there may be a bias to select knots lying well inside the inner radius of the reverse shock region used
by us.

There may well be more regions of the reverse shock going backward missed in our study, as we concentrated on the projected reverse shock region. 
For example, optical studies show that the backside of the optical shell seems to move faster than the front side \citep{reed95,milisavljevic13},
which likely has a similar origin as the displacement of the reverse shock location in the plane of the sky.

It could well be that the backside indeed also has a reverse shock moving inward. This may also cause X-ray synchrotron emission from the backside of the shell.
But since we observe the narrow filaments  on the backside 
face on, rather than edge on, the X-ray synchrotron may be less easily identified. Figure~6 in \citet{helder08}, which is based on purely spectral analysis does give a hint of more extended
X-ray synchrotron emission projected to the inside. 
But their figure still indicates a lack of diffuse X-ray synchrotron emission from the eastern part.

\subsection{Why did the reverse shock reverse its direction?}

The standard scenario for the evolution of Cas A is that of  a remnant of  a core-collapse SN of Type IIb \citep{krause08,rest11}, with a low ejecta mass of 2-4~\msun\ evolving
in the clumpy, dense wind of its progenitor \citep{vink96,willingale03,chevalier03,vink04a,hwang12,weil20}. The observed deviations from spherical symmetry, like the jet and ejecta rings, 
are usually attributed to asymmetries in the explosion itself, for which light echos from different directions indeed provide evidence \citep{rest11}.
The pre-shock density at the forward shock is  $n \approx 1$~cm$^{-3}$ \citep{lee14}, corresponding to $\rho_{\rm csm}\approx 2\times 10^{-24}~{\rm g\,cm^{-3}}$.
The density is expected to fall off as $\rho(r)=\dot{M}/4\pi r^2v_{\rm w}$, implying
a mass loss rate of $\dot{M}\approx 2.5\times 10^{-5}$~\msun\,yr$^{-1}$ for an assumed wind velocity of $v_{\rm w}=10$~\kms.

For such a $1/r^2$ CSM density profile, the reverse shock is not expected to move back to the center at the current age of Cas A, but rather  when its
age is 2000--3000 yr; see Figure~3 in  \citet{micelotta16} for a semi-analytical model, and Figure~4 in \citet{orlando21} for the hydrodynamical simulation of Cas A. So the question arises
why the reverse shock in the western part of Cas A does not agree with well-understood evolutionary expansion models.

One can obtain some intuitive understanding of the reverse shock dynamics 
 by starting with the well known Rankine-Hugoniot relation $P_1+\rho_1v_1^2=P_0 + \rho_0v_0^2$, expressing momentum flux conservation,
with subscript 0 referring to the unshocked gas, and subscript 1 to the postshock gas. For a strong shock $P_0$ can be neglected, and $v_0=V_{\rm s}$. 
Using the shock compression ratio $\chi=\rho_1/\rho_0=v_0/v_1$, and using for the density the ejecta density at the reverse shock ($\rho_0=\rho_{\rm ej,0}$) we can transform this into
 \citep[c.f.][]{vinkbook}:
\begin{equation}
P_{\rm rs,1}= \left(1 - \frac{1}{\chi}\right)\rho_{\rm ej,0} V_{\rm rs}^2.
\end{equation}
$P_1$ refers to the pressure in the shocked  gas, which is of the same order, but somewhat lower \citep{chevalier82},  than  the pressure in the shell just behind the forward shock 
$P_{\rm fs,1}=(1-1/\chi)\rho_{\rm csm,0}V_{\rm fs}^2$.
Writing $P_{\rm rs,1}=\xi P_{\rm fs,1}$, with $\xi\approx 0.5$, and assuming the compression ratios are similar, we see that 
\begin{equation}
\rho_{\rm ej,0} V_{\rm rs}^2=  \xi \rho_{\rm csm,0} V_{\rm fs}^2.
\end{equation}
For $ V_{\rm rs}$ we can insert Equation~\ref{eq:vs_rs} and  $V_{\rm fs}=mR_{\rm fs}/t$. 
The turning around of the reverse shock happens when 
$dR_{\rm rs}/dt=0$. 
We conclude, therefore, that the reverse 
shock is moving inward when \citep{vinkbook}
\begin{equation}\label{eq:rho_csm_ej}
\rho_{\rm ej,0} < \xi\rho_{\rm csm,0}  m^2 \left(\frac{R_{\rm fs}}{R_{\rm rs}}\right)^2\approx  1.1\xi \rho_{\rm csm,0},
\end{equation}
for which we used $m=0.7$, $R_{\rm rs}=1.9$~pc and $R_{\rm fs}=2.8$~pc.

For the average ejecta density we can  write
\begin{align}\label{eq:rho_ej}
\overline{\rho_{\rm ej,0}}= &\frac{M_{\rm ej}}{\frac{4\pi}{3}R_{\rm rs}^3} \\\nonumber
\approx& 9.5\times 10^{-24}\left(\frac{M_{\rm ej}}{4~M_\odot}\right)\left(\frac{R_{\rm rs}}{1.9~{\rm pc}}\right)^{-3}{\rm g\,cm^{-3}}.
\end{align}
Combining Equation~\ref{eq:rho_csm_ej} and Equation~\ref{eq:rho_ej}, we see that the ejecta density
needs to be a factor 2--5 lower than the average density 
for a CSM density of $2\times 10^{-24}~{\rm g~cm}^{-3}$,
in order for the reverse shock to move toward the center.

So under the hypothesis that the reverse shock moves inward due to relatively low ejecta density in the western part, the implication is that the
ejecta density is much lower than $2\times 10^{-24}~{\rm g\,cm^{-3}}$,
or, alternatively,
the forward shock is moving through dense material, or has done so recently.

The ejecta  density in the western part is already somewhat  lower, as here the reverse shock is furthest from the explosion center. 
The low density  could be more enhanced, if locally there is a large bubble, inflated by radioactive $^{56}$Ni/$^{56}$Co during the SN phase.
Such bubbles give rise to ringlike structures surrounding the inflated bubbles, which have been identified in the threedimensional layout of the
 optical knots  \citep{milisavljevic13}.
Indeed, there is a high concentration of Fe-rich ejecta close the western part of the reverse shock \citep[e.g. last panel Figure1 in][]{hwang12}.

The hypothesis that the reverse shock in the western part moves toward center due to a large ejecta bubble seems, therefore, plausible.
A problem is, however,
that there are similar large concentrations of Fe-rich ejecta, and also ejecta rings,  in the northern part and in the southeastern part, in regions
for which we do not measure an inward movement of the reverse shock. 
Moreover, the hydrodynamical simulations of Cas A's evolution by \citet{orlando21}---based on a realistic ejecta  distributions that include underdense regions caused by  $^{56}$Ni/$^{56}$Co decay---
show that the reverse shock is 
expected to move outward with 2000--4000~\kms\ everywhere. These velocities are consistent with the measurements reported here outside PAs in the range $\sim$220\deg--320\deg.

Alternatively, the reverse shock in the western part may have  been affected by the local CSM structure. 
There is a two decade old discussion on the possibility of an interaction of the western part of the SNR with a molecular cloud at the western edge of Cas A \citep[e.g.][]{keohane96,kilpatrick14}.
There is indeed a molecular cloud projected toward the western part of Cas A, and the southwestern (counter)jet is much more irregularly shaped than the northeastern jet, which could be the result
of some form of interaction of the southwestern jet with a density enhancement \citep{schure08}.
However, there is no direct evidence that the Cas A SNR shock has a direct interaction with the molecular cloud seen in absorption. Such an interaction should lead to bright thermal X-ray emission 
in the western part due to the high local density, which is not observed. Nor is there evidence for a direct interaction between the forward shock and the molecular cloud as probed by CO molecular
line studies \citep{zhou18}.

In a follow up study   \citet{orlando22} studied the possibility of the interaction of the forward shock with an asymmetric mass-loss shell  with which the forward shock 
interacted in the recent past, based on a preliminary presentation of the work reported here. Their simulations can indeed reproduce the reversal of the reverse shock motion in the current epoch of the SNR evolution. 
The interaction of the forward shock with the shell creates an increase in pressure in the shell, and induces reflected shocks, which enhance the shell pressure at the reverse shock. 

A constraint on when the encounter with the putative shell  could have taken place comes from early radio maps of Cas A:
the radio bright  region in the western part, likely associated with the strong reverse shock encountered there in the frame of the ejecta, was already
present in the first radio synthesis map of Cas A produced in the early 1960s by \citet{ryle65}. So likely the reverse shock has been going back in that region  prior to 1960.

On the other hand, the optical morphology underwent quite some changes over the last 50 yr \citep{patnaude14}, with the northern shell already being prominent in the
1950s, and the southeastern part only appearing in the 1970s. This may provide a window on the history of the reverse shock development and its possible connection to
the the CSM structure. A complication is  that the connection between the evolution of the optical  structures and the structures seen in X-ray or radio is not fully understood.
For example, the emergence of optical knots in certain regions seems to precede the local X-ray brightening of more diffuse emission by several years \citep{patnaude14}.
Also note that the southeastern shell, emerging in the 1970s, already has a radio counterpart in the earliest radio map \citep{ryle65}.

The debris of this overrun shell may be associated with the quasi-stationary flocculi in the western part of Cas A \citep{koo18}. If true, the current reverse shock dynamics may
reveal some information about the mass loss history of the Cas A progenitor, which could be combined with the optical  information about unshocked CSM as reported by \citet{weil20}
to shed light on the nature of the progenitor star.

Another type of  CSM structure that could have affected the evolution of the reverse shock is if the progenitor went through a short Wolf-Rayet star phase. The progenitor was a stripped star---the SN being a SN IIb---
and it is possible that it exploded as a Wolf-Rayet (WR) star. 
In the case of a  late WR phase of the progenitor, the fast (1000--2000~\kms) WR wind creates a low density cavity. After the explosion the forward shock initially moves fast through the
cavity until it reaches the WR/red supergiant wind boundary, and then rapidly decelerates. Once more energy is transferred to the shocked red-supergiant wind, the shock reaccelerates.

The possible CSM configurations,  and the subsequent SNR evolution of Cas A  inside the wind structure, 
was modelled by \citet{vanveelen09}, who concluded that the WR phase must have been absent, or relatively shortlived ($\lesssim 15,000$~yr).
The main argument was that for a longer WR phase the reverse shock would be moving backward, which was thought not to be the case in 2009. 

Now that we know that the reverse shock moves back, at least in the western part, one should reconsider this scenario. Indeed, also in this scenario the forward shock could be still
accelerating, in agreement with our findings \citep[see Fig. 8 in][]{vanveelen09}. The difficulty for this scenario is how to explain that  the WR wind cavity is aspherical, given
the fast wind speed.  One possibility is that the
 progenitor star  was moving through the interstellar medium, creating a lopsided wind-region, which  also affected the shape of the inner cavity \citep[c.f.][]{weaver77,meyer21}.

The effect of a WR phase on the hydrodynamics of the jets in Cas A was investigated by \citet{schure08}, who concluded that the jets would not survive the encounter was the
edge of the cavity for a WR phase that lasted longer than $\approx 2500$~yr. Interestingly, the northeastern jet in Cas A is well-defined in X-rays, whereas the southwestern jet looks broken up.
This is also in agreement with the idea that there was a WR wind cavity, elongated in the southwestern direction. The late encounter of the southwestern
jet partially broke then up the southwestern jet, whereas the northeastern jet encountered the boundary layer early enough and pierced through it unhindered.

The nature of the progenitor of Cas A is still a mystery. In particular the cause of its large mass loss rate is often attributed to a closely interacting binary system. But
there is no evidence for a surviving companion star \citep{kerzendorf19}. Nevertheless, the shock dynamics reported here provide important hints on the late mass-loss history of
the progenitor, be it in the form of a partial, asymmetric shell from episodic mass loss, an aspherical cavity created by a brief WR phase wind, or perhaps even a combination of
both.

\subsection{Deceleration versus acceleration of the forward shock}

Interestingly, the interaction of the forward shock with a dense mass-loss shell  \citep{orlando22} or the edge of a WR wind cavity \citep{vanveelen09} initially leads to a strong deceleration of the forward shock, followed by an acceleration
once the shock has penetrated through the shell. 
Indeed, one of the surprising results of the deceleration/acceleration measurements reported here is that the forward shock appears to be accelerating around PAs of 180\deg\ (South)
and 250\deg\ (West), see Figure~\ref{fig:deceleration}. The latter agrees with the location of a returning reverse shock, and it also the location where several tests  of the deceleration/acceleration
measurements
provide robust results (Figure~\ref{fig:comparison}).

On the other hand, we do find a rather strong deceleration toward the northern part of the SNR, an order of magnitude stronger than expected based on Equation~\ref{eq:exp_rate_pars}.
In the simulations of \citet{orlando22} it is assumed that the mass-loss shell is denser in the West. But the strong deceleration in the North may  potentially indicate that the  shell is not partial, but rather
aspherical, and that in the North the shock is just encountering the shell and decelerating.

Clearly following up the deceleration/acceleration measurements with further monitoring of Cas A with Chandra is important, to make the measurements more precise and confirm these results,
with even longer expansion base lines.

The reverse shock motion itself could be potentially explained by the asymmetries in the ejecta distribution, but with some difficulties. However, the combination of an inward moving reverse shock coinciding
with an acceleration of the forward shock in the western region is more naturally explained by a more complex CSM structure.

\subsection{The proper motion of the northeastern jet}
The measurements of northeastern jet regions reported is the first measurement of the motion of this jet measured in X-rays. The X-ray measurement motions reveal more about the dynamics of the
shock heated gas, which is a combination of shocked ejecta and shocked CSM. Previous measurements of the jet velocities are based on optical measurements, which rely on dense
optical knots, which are expected to have velocities close to the free expansion velocities $R/t$. Indeed, \citet{fesen06,fesen16} report proper motions up to $\sim$15,000~\kms. 
Note that the optical knots  traces  the northeastern jet  out to 5.3\arcmin---further out than the X-ray counter part, which goes out  to 4\arcmin. 
For our measurement we used measurements out to a radius of 3.9\arcmin.

The inclusion of a measurement of the NE jet region stands apart from the rest of the measurements here, which focus on the forward and reverse shock, whereas the X-ray structure
of the jet shows a number of bright "streaks", which are likely associated with ejecta material. For that reason we deviated from the overal setup for the measurements, and included
X-ray images from 1.75-1.94 keV, dominated by Si XIII line  emission. The measurements in the continuum band and Si-K line emission band differ substantially, with expansion rates
of 0.2\%yr$^{-1}$ for the continuum band, and 0.24\%yr$^{-1}$ for the Si-K band. This difference may be real and related to the fact that Si-K is almost exclusively associated with shocked ejecta, whereas the continuum emission is likely a mix of thermal bremsstrahlung from shocked ejecta, plus more diffuse thermal bremsstrahlung from shocked CSM. Either way, both expansion rates
are consistent with an expansion parameter of $m=0.68$--$0.8$ similar to, or even larger than, the average expansion parameter of the forward shock.
The similarity of the expansion parameters suggest that the NE jet is still moving through the red-supergiant wind of the progenitor \citep[c.f.][]{schure08}.

The tip of the jet in X-rays is around 3.9\arcmin, which, combined with the expansion parameter and age of Cas A of 332~yr (in 2004), suggests an expansion velocity of $\approx$7830--9200~\kms.

\section{Conclusions}

We  reported here on the proper motion of the forward- and reverse shock regions of Cas A along the entire projected edges of both shocks using Chandra observations in the
4.2-6 keV continuum band. We used a method in which the
proper motions are measured by directly combining 17 observations and comparing them to a model  based on the Chandra VLP observations obtained in 2004.
The combination of many images, spanning 19 years in observations, greatly increased the sensitivity, which allowed us  to measure the second order derivative of the proper motion of
the forward shock, 
i.e. the acceleration/deceleration rate.

For the forward shock we report an average expansion rate of $0.218$\%yr$^{-1}$, varying from 0.152\%yr$^{-1}$ to 0.276\%yr$^{-1}$, corresponding with a mean expansion parameter of $m=0.73$
with variations from $m=0.51$ to $m=0.91$. The mean is somewhat higher than reported by \citet{patnaude09} and close to what was reported by \citet{delaney04}, but both did not use the entire
edge of the SNR and cannot be compared one to one. 
The second-order derivatives show large variations, and values higher than predicted by a model in which the shock moves through a smooth CSM with a $1/r^2$ density profile.
The average deceleration rate we found is $(0.2\pm 4.9)\times 10^{-5}$yr$^{-2}$. The eastern part of the forward shock appears to be decelerating, whereas in the western part, around position angles
of 190\deg\ and 250\deg, the shock appears to be accelerating.

The reverse shock  is moving outward in the frame of the observer in the eastern part of the SNR, with typical expansion rates of 0.1--0.2\%yr$^{-1}$, but is lower between position
angles of 190\deg\ and 310\deg, with  negative expansion rates for the range of 260\deg--300\deg: here the reverse shock moves toward the center. 
At 270\deg the negative expansion rate translates into a projected velocity of $-1884\pm 17$\kms, with perhaps individual filaments moving even faster \citep{sato18}.
The reverse shock velocity in the frame of the ejecta exceeds the velocity of the forward shock around $PA=$270\deg\, reaching a value as high as 8000~\kms.
For position angles between $\sim 180$\deg\ and 310\deg, the reverse shock velocity in the frame of the ejecta exceeds 3000~\kms, and it is  for these position angles
that Cas A appears to emit X-ray synchrotron radiation.  Indeed, the theory for diffusive shock acceleration predicts that X-ray synchrotron radiation is only expected
for $V_{\rm s}\gtrsim 3000$~\kms. 

The inward motion of the reverse shock in the western part does not agree with the  evolutionary model for a SNR expanding into a dense wind at an age less than 1000~yr.
We discussed two possibilities for what may have caused the inward reverse shock motion in the western part: i) a locally low density of the ejecta in the west, or ii) the encounter of the forward
shock with a density enhancement in the past, created by a mass-loss shell, or by a brief Wolf-Rayet star phase. 
Hypothesis (ii) is preferred, but we note that the expansion of the forward shock is as high as around the rest of the SNR shock.
So a density contrast or shell in the CSM in this region,  created by the  progenitor's mass outflow, needs to have been encountered in the past. 
Currently the forward shock has recovered its velocity, although it appears to be still accelerating.

Finally we reported on the expansion rate of the northeastern jet, for which measured an expansion rate of 0.20-0.24\%yr$^{-1}$,  similar to the forward shock. This implies
that the jet is moving through the same type of density structure as most of the forward shock, and suggest a velocity at the tip of the X-ray emitting jet of  $\approx$7830--9200~\kms.

\begin{acknowledgments}
We thank several colleagues for helpful discussions on Cas A: with Martin Laming on nonthermal bremsstrahlung and  the possible implications of multiple reflected/secondary shocks, 
with Salvatore Orlando on the effects of mass loss shells on  the dynamics of the forward and reverse shocks,
 and with Rob Fesen and Danny Milisavljevic on the optical emission from Cas A.
JV is partially supported by funding from the European UnionÕs Horizon 2020 research and innovation programme under grant agreement No 101004131 (SHARP).
D.J.P and D.C. acknowledge support from NASA contract NAS8-03060.
\end{acknowledgments}

\vspace{5mm}
\facilities{
Chandra X-ray Observatory (CXO).
}
\software{
The software for measuring the expansion can be downloaded from here \zenodo.
We have made used of the FITS-image viewer ds9 \citep{ds9}, and the subroutine libraries
{\em pgplot} \citep{pgplot} and {\em cfitsio} \href{https://heasarc.gsfc.nasa.gov/docs/software/fitsio}{https://heasarc.gsfc.nasa.gov/docs/software/fitsio}. 
}

\bibliographystyle{apj}  

\appendix
\section{Log-likelihood profiles}
\begin{figure*}
\centerline{
\includegraphics[trim=50 20 120 30,clip=true,width=0.15\textwidth]{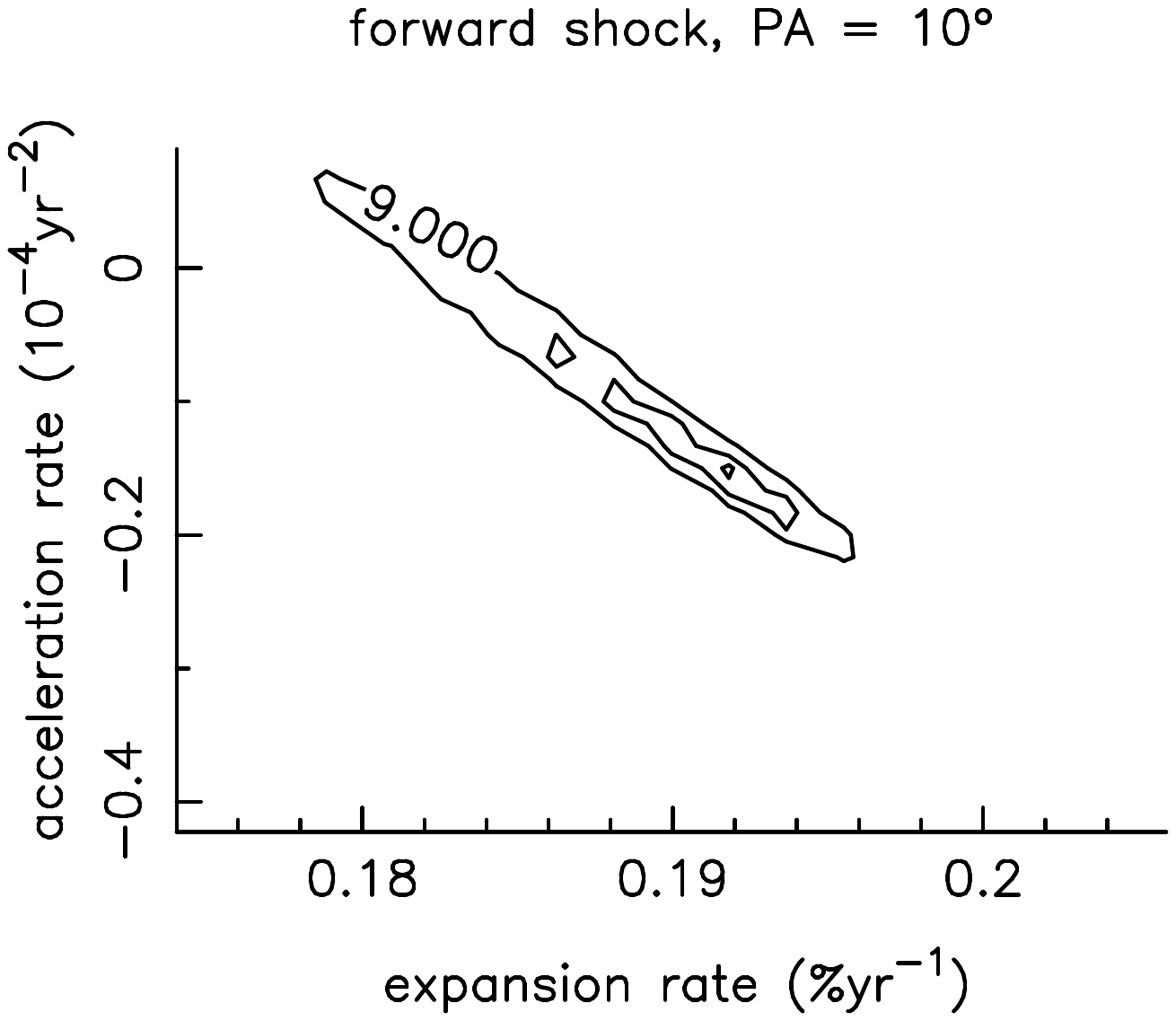}
\includegraphics[trim=50 20 120 30,clip=true,width=0.15\textwidth]{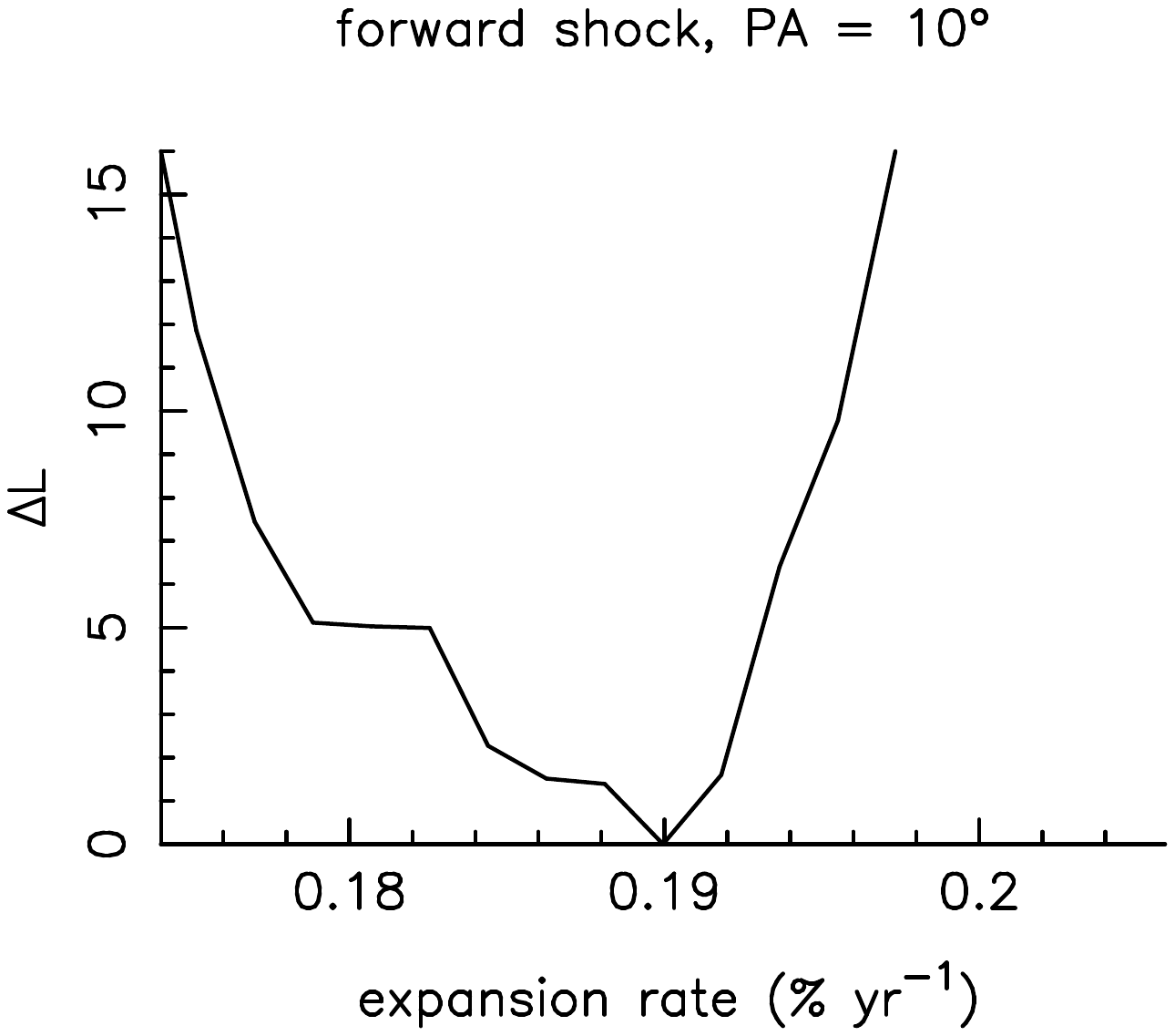}
\includegraphics[trim=50 20 120 30,clip=true,width=0.15\textwidth]{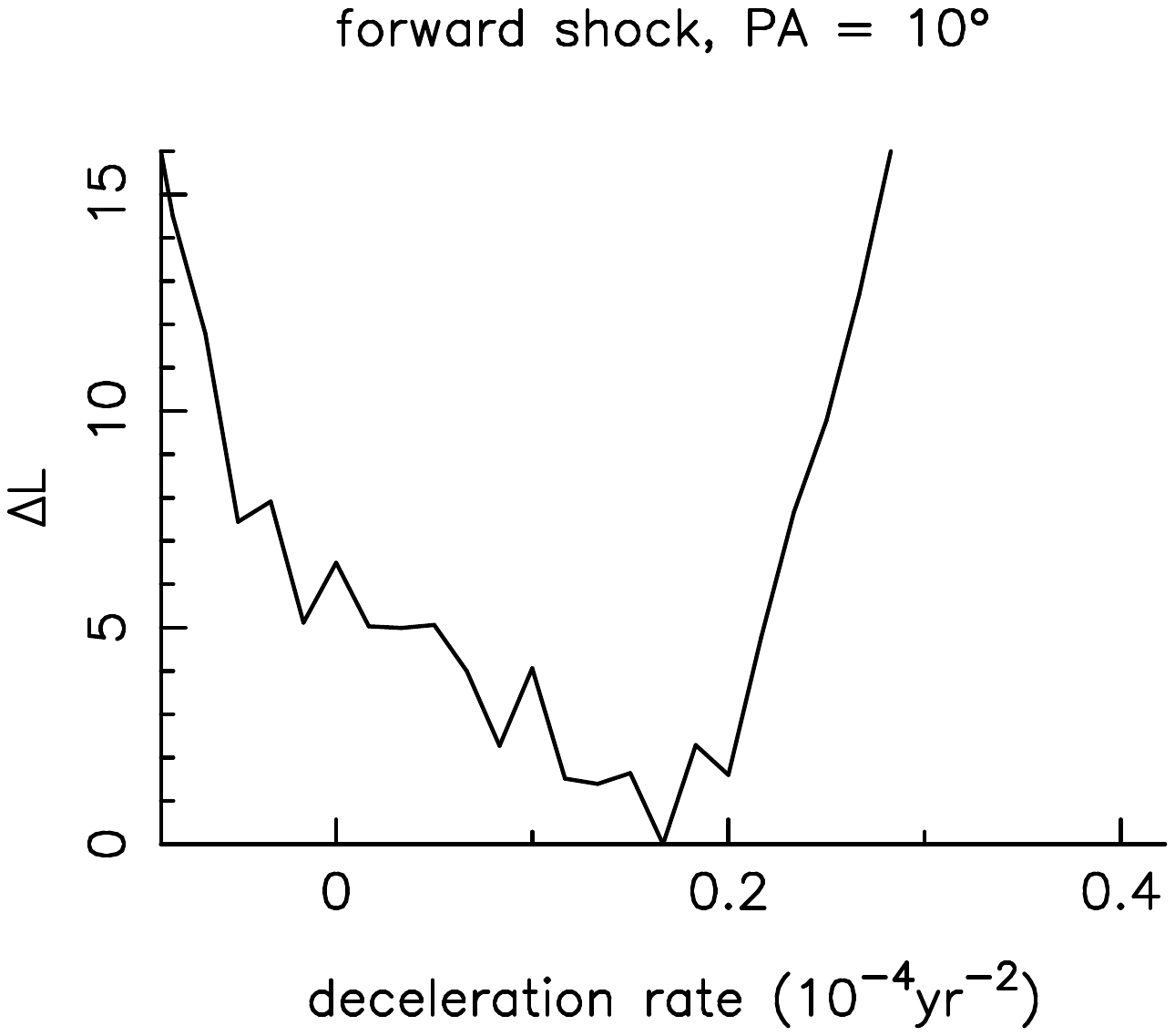}
\includegraphics[trim=50 20 120 30,clip=true,width=0.15\textwidth]{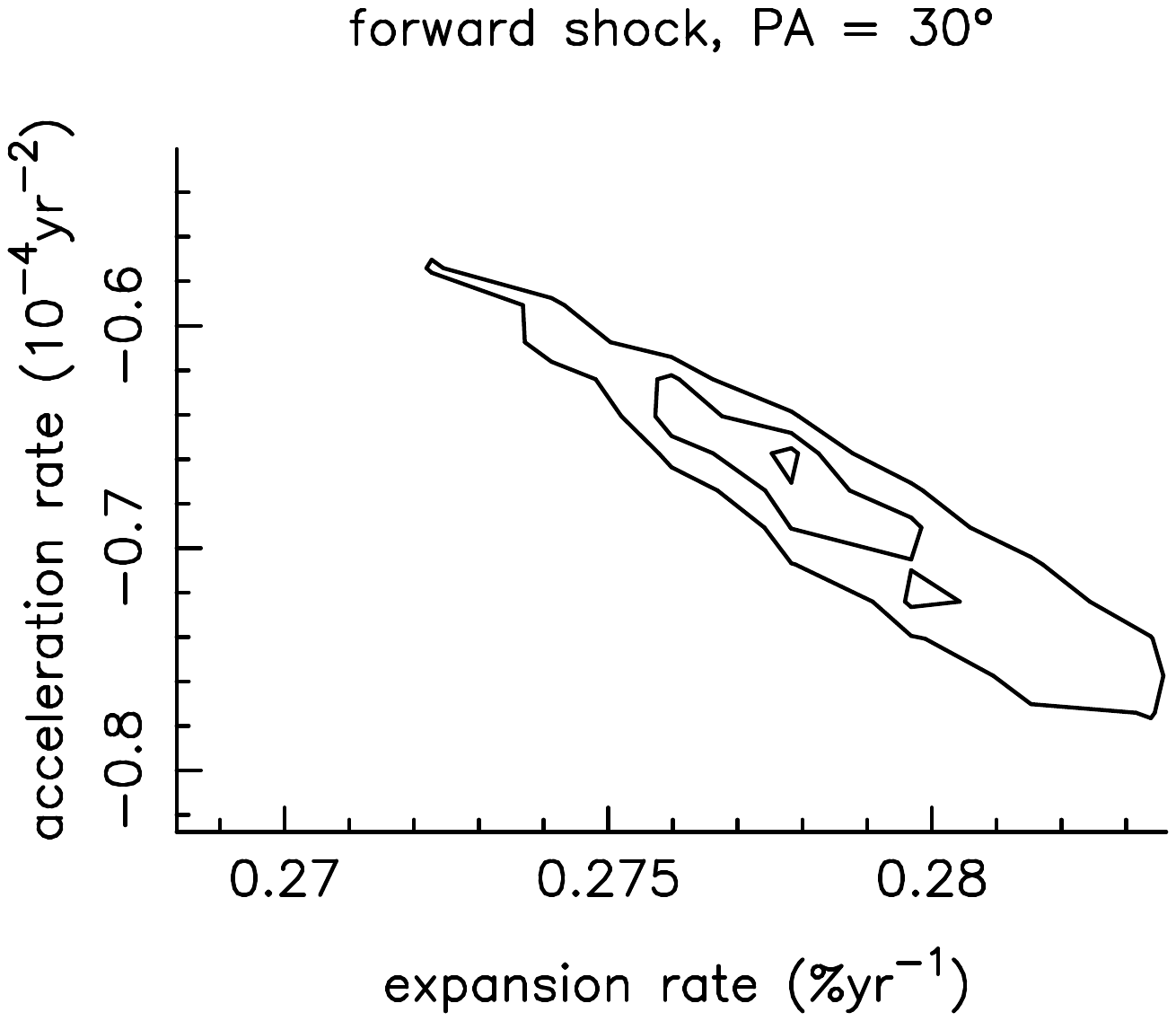}
\includegraphics[trim=50 20 120 30,clip=true,width=0.15\textwidth]{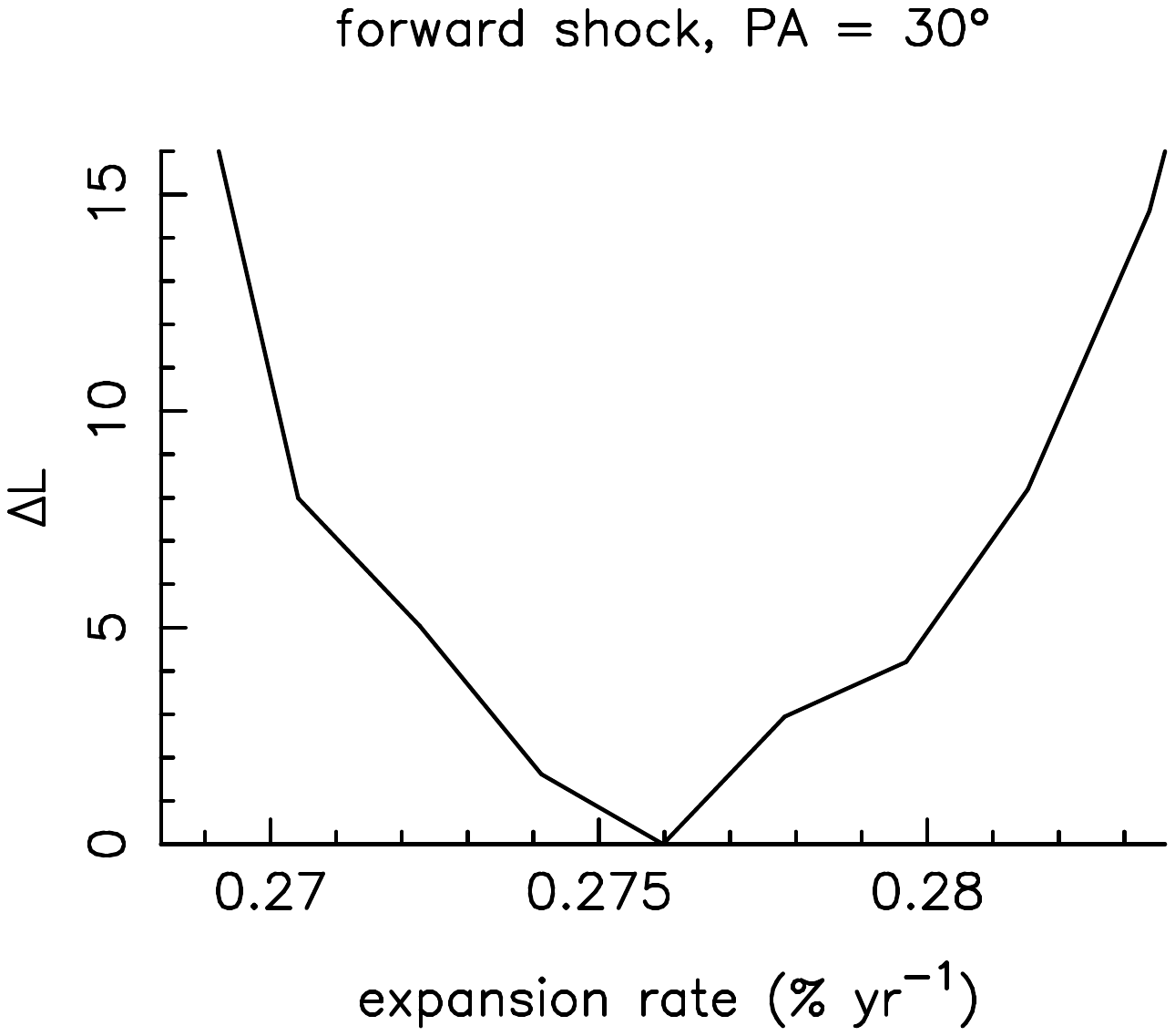}
\includegraphics[trim=50 20 120 30,clip=true,width=0.15\textwidth]{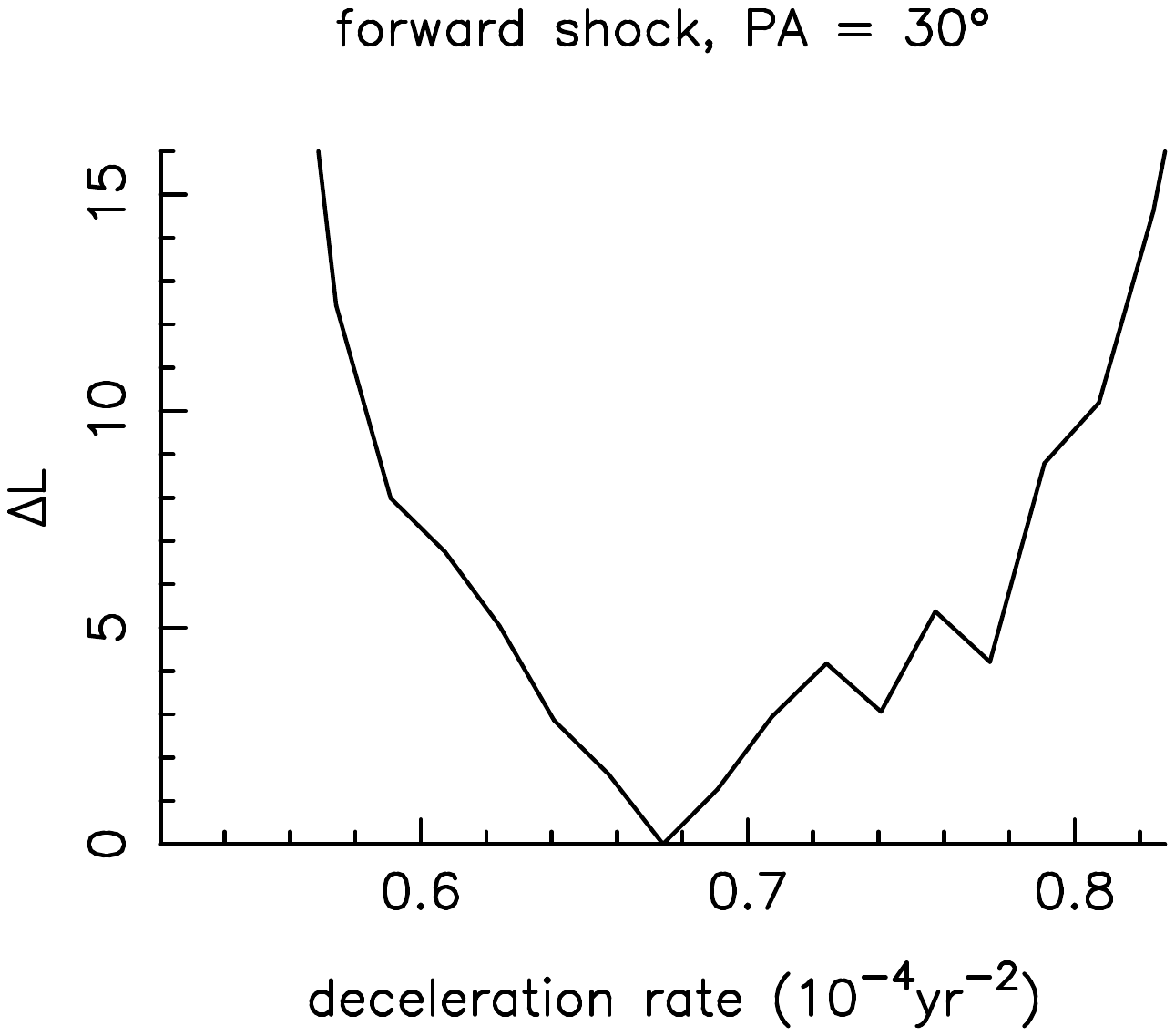}
}

\centerline{
\includegraphics[trim=50 20 120 30,clip=true,width=0.15\textwidth]{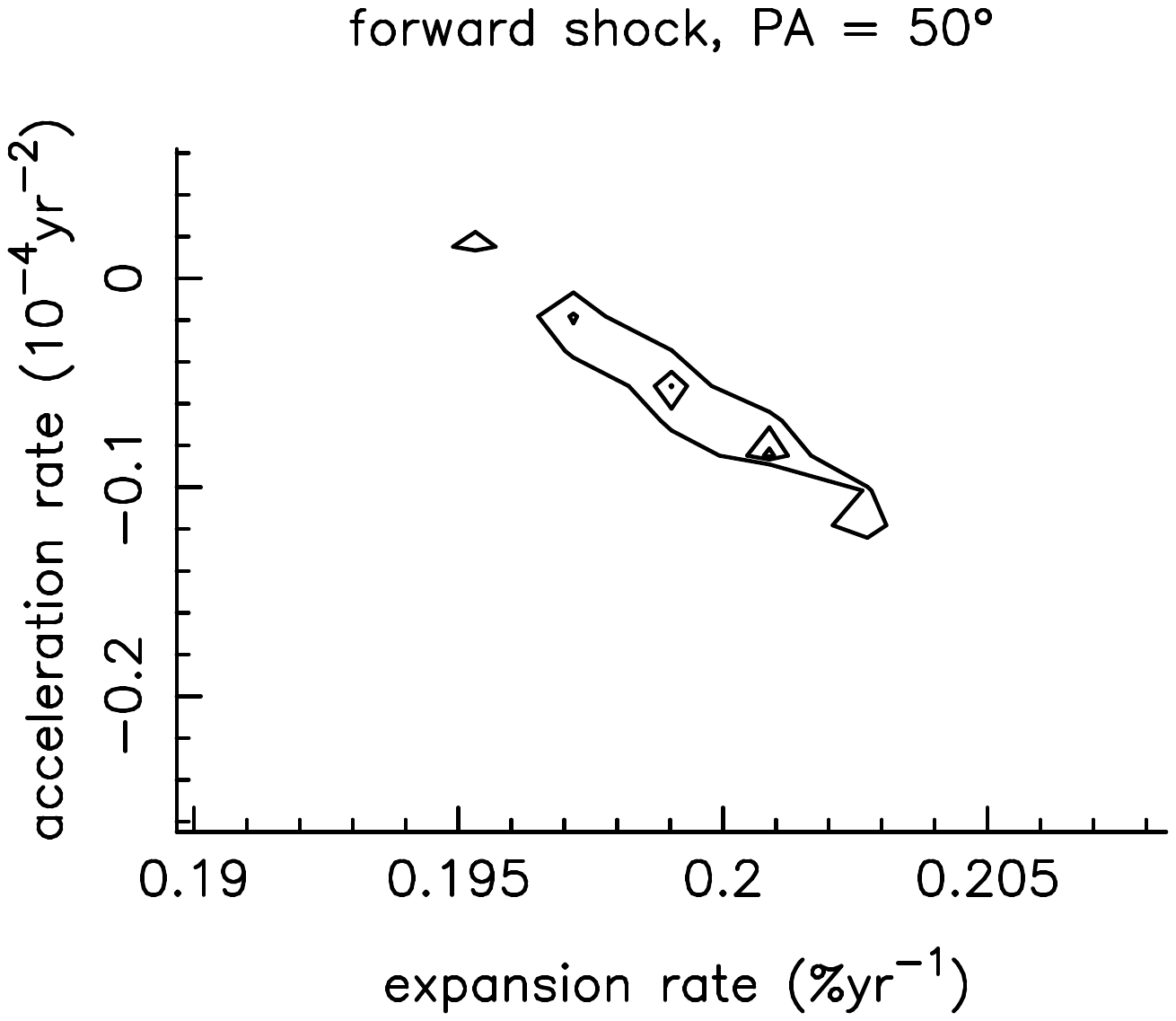}
\includegraphics[trim=50 20 120 30,clip=true,width=0.15\textwidth]{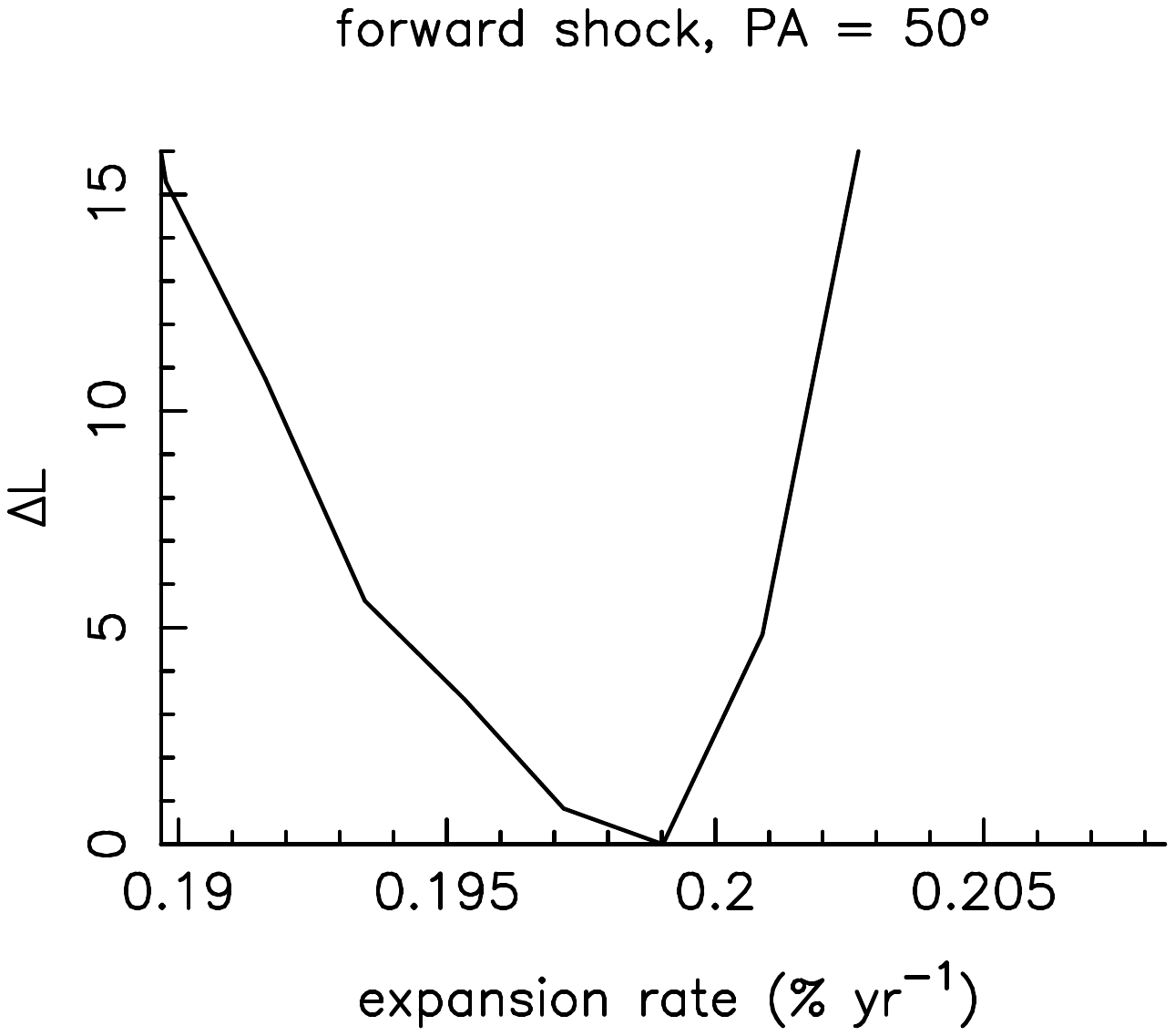}
\includegraphics[trim=50 20 120 30,clip=true,width=0.15\textwidth]{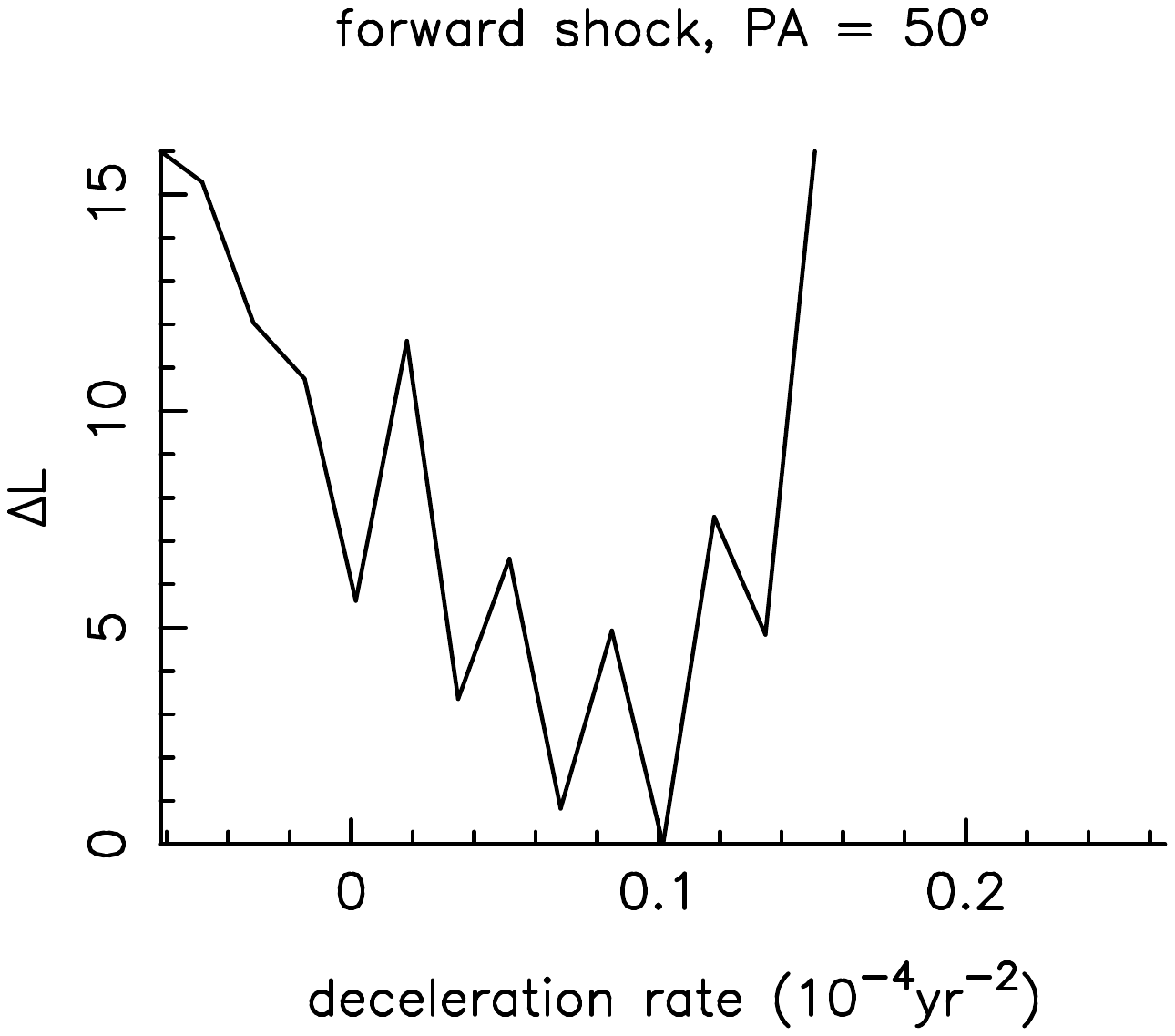}
\includegraphics[trim=50 20 120 30,clip=true,width=0.15\textwidth]{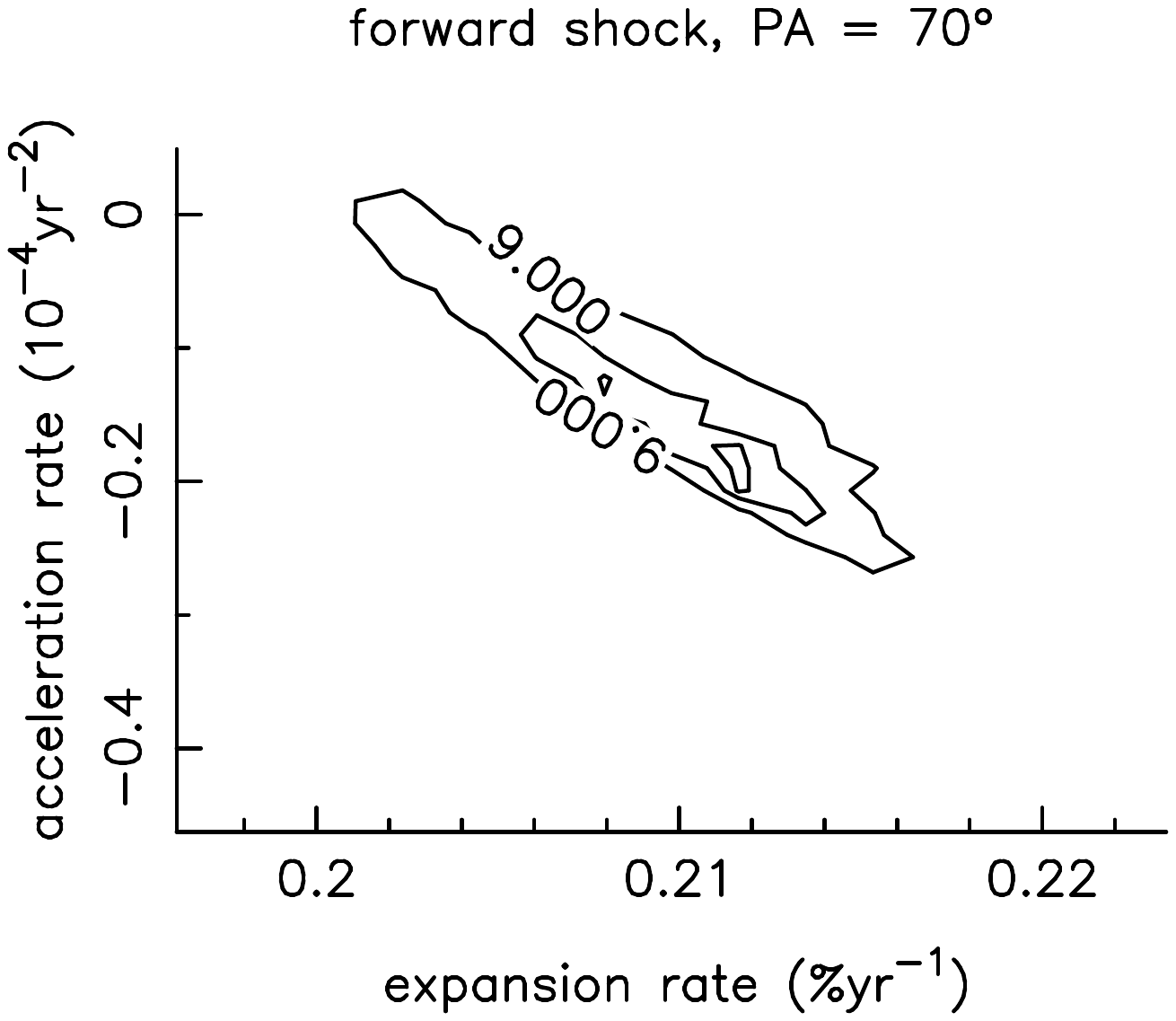}
\includegraphics[trim=50 20 120 30,clip=true,width=0.15\textwidth]{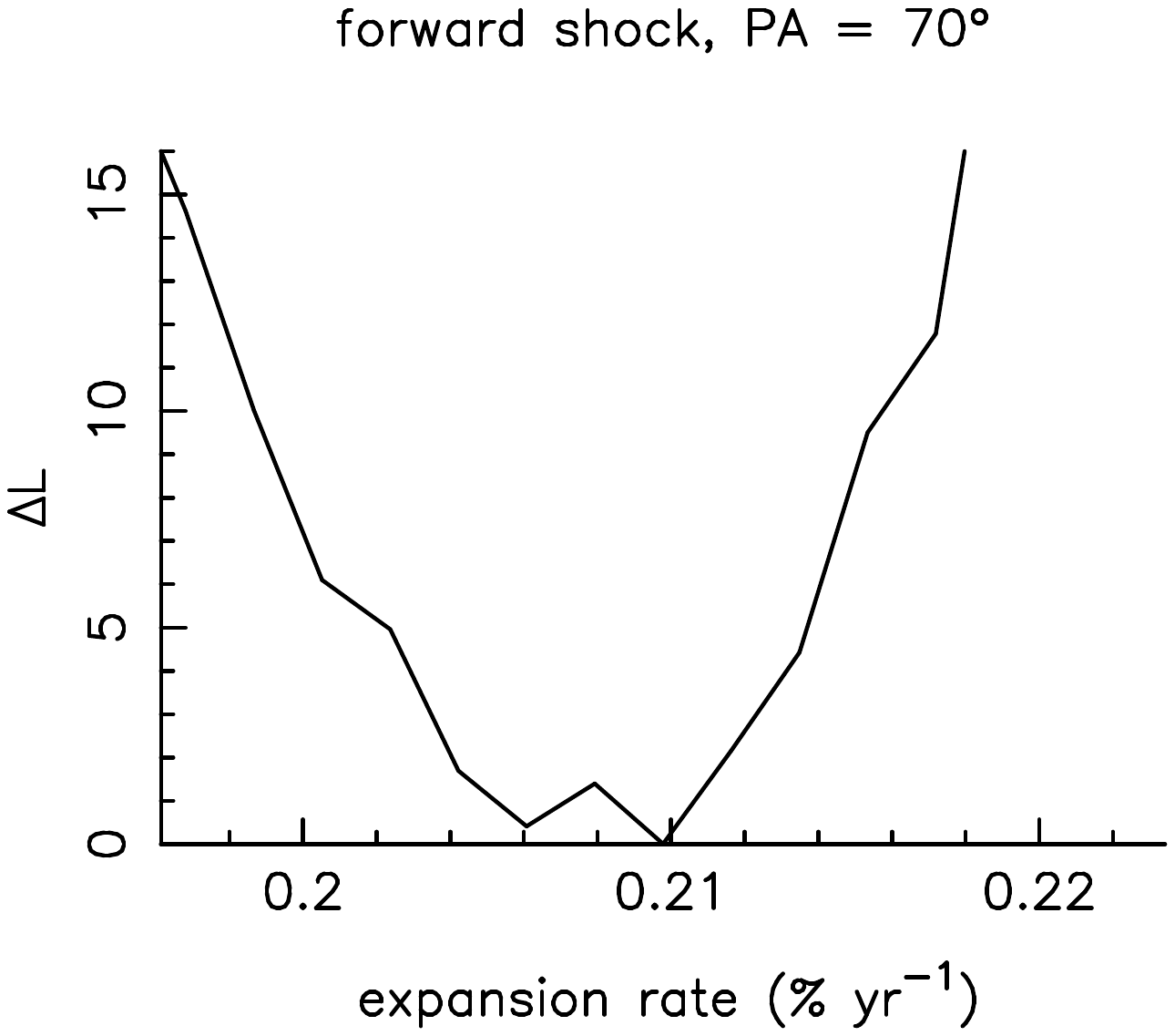}
\includegraphics[trim=50 20 120 30,clip=true,width=0.15\textwidth]{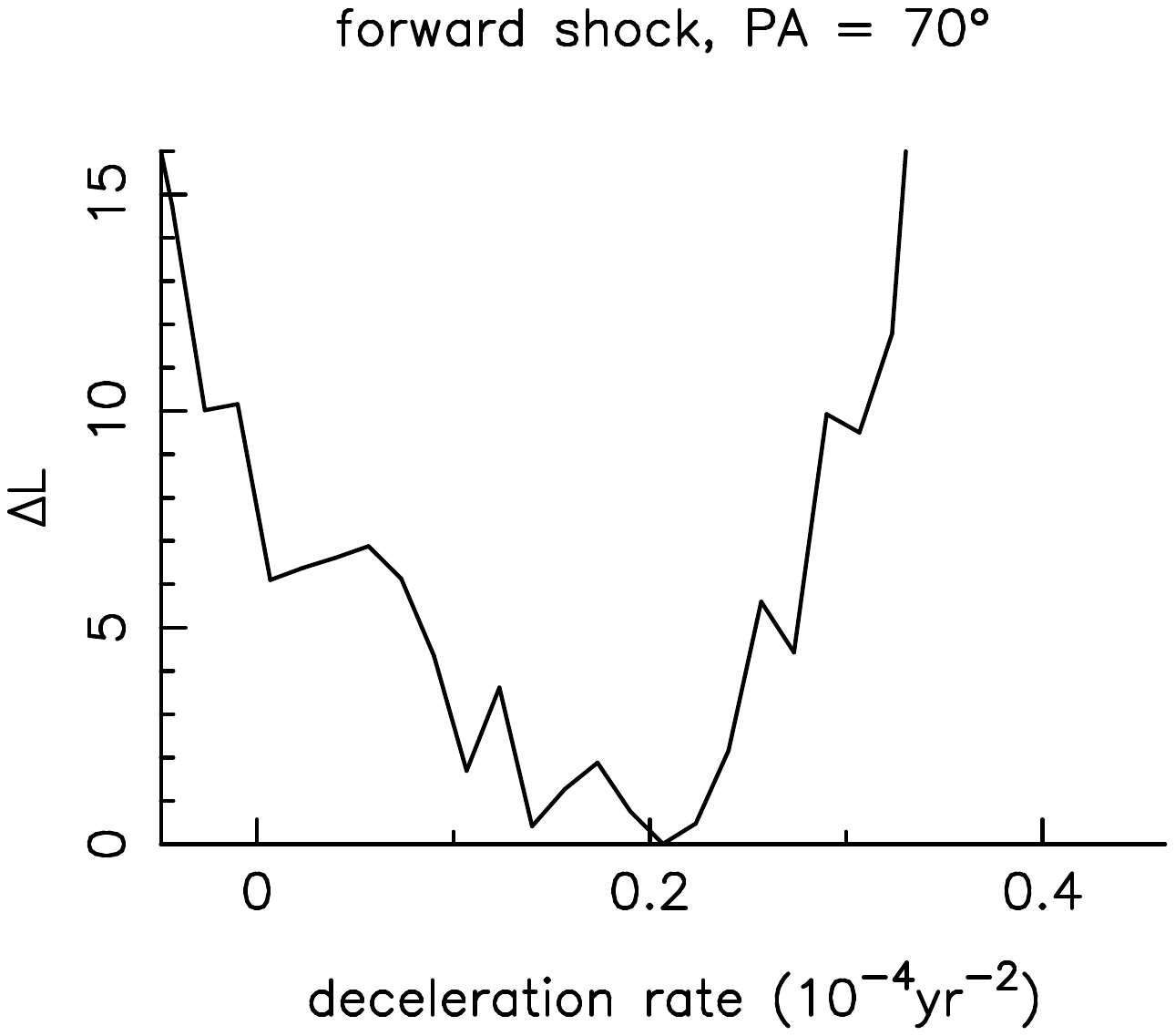}
}

\centerline{
\includegraphics[trim=50 20 120 30,clip=true,width=0.15\textwidth]{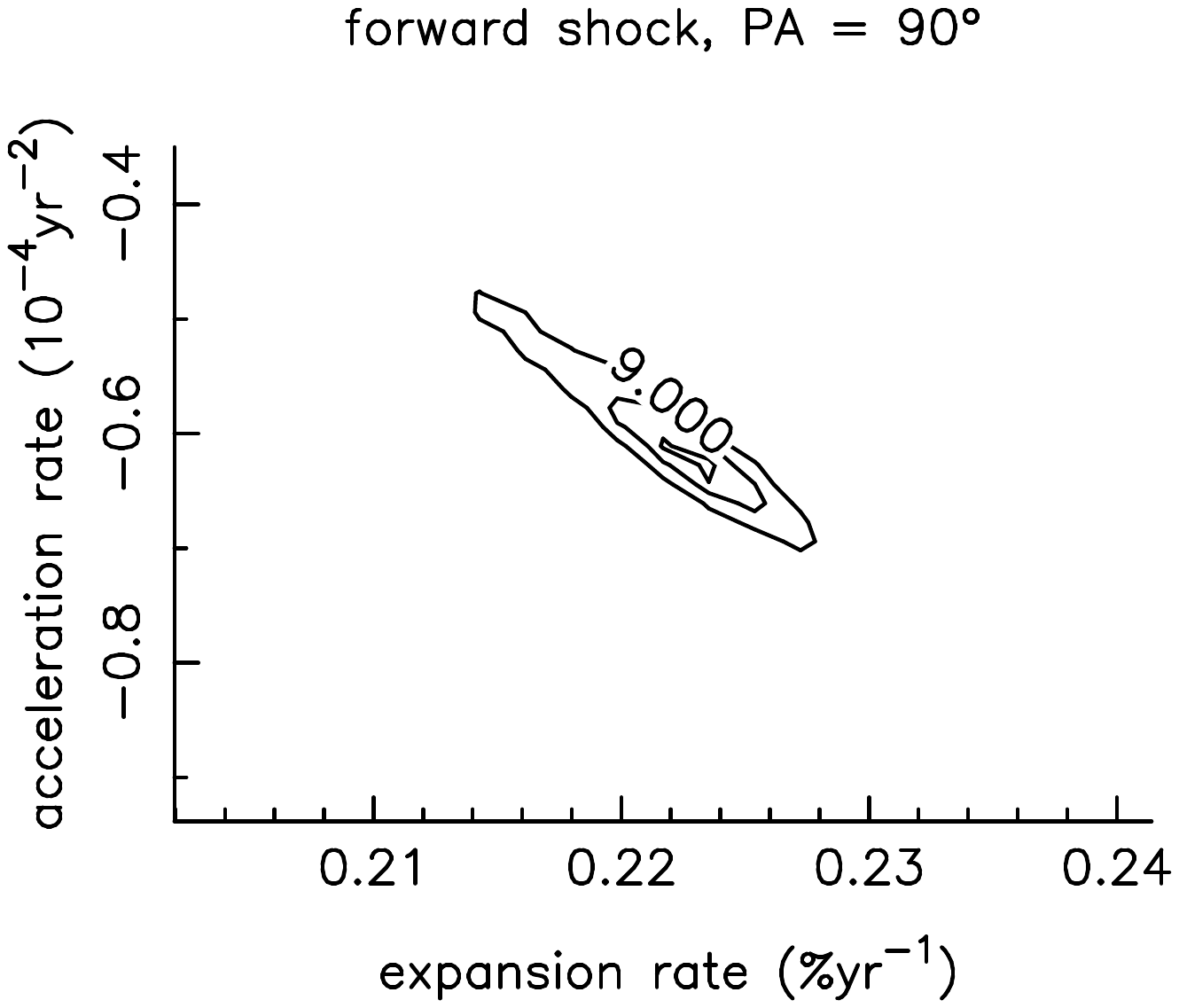}
\includegraphics[trim=50 20 120 30,clip=true,width=0.15\textwidth]{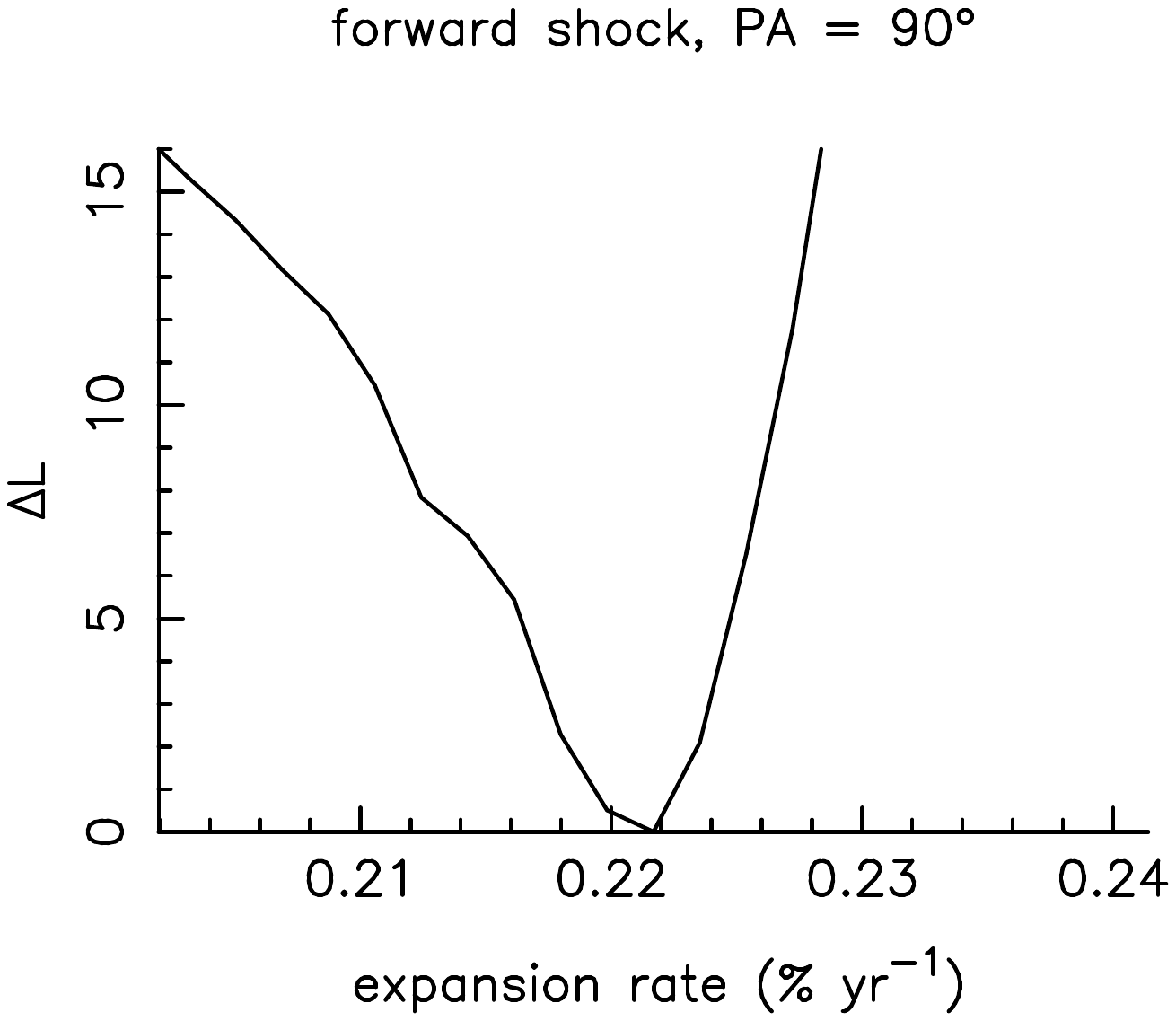}
\includegraphics[trim=50 20 120 30,clip=true,width=0.15\textwidth]{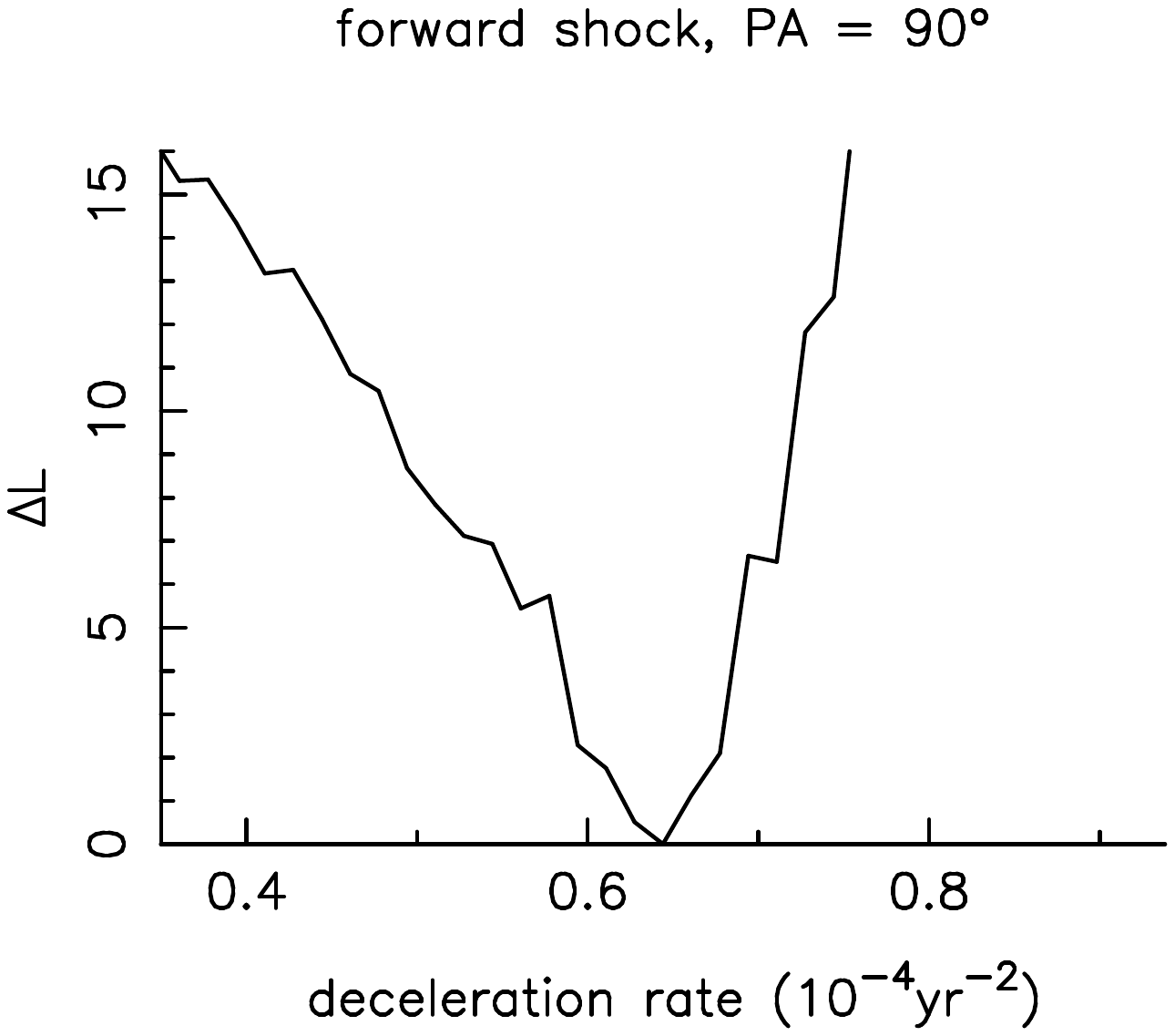}
\includegraphics[trim=50 20 120 30,clip=true,width=0.15\textwidth]{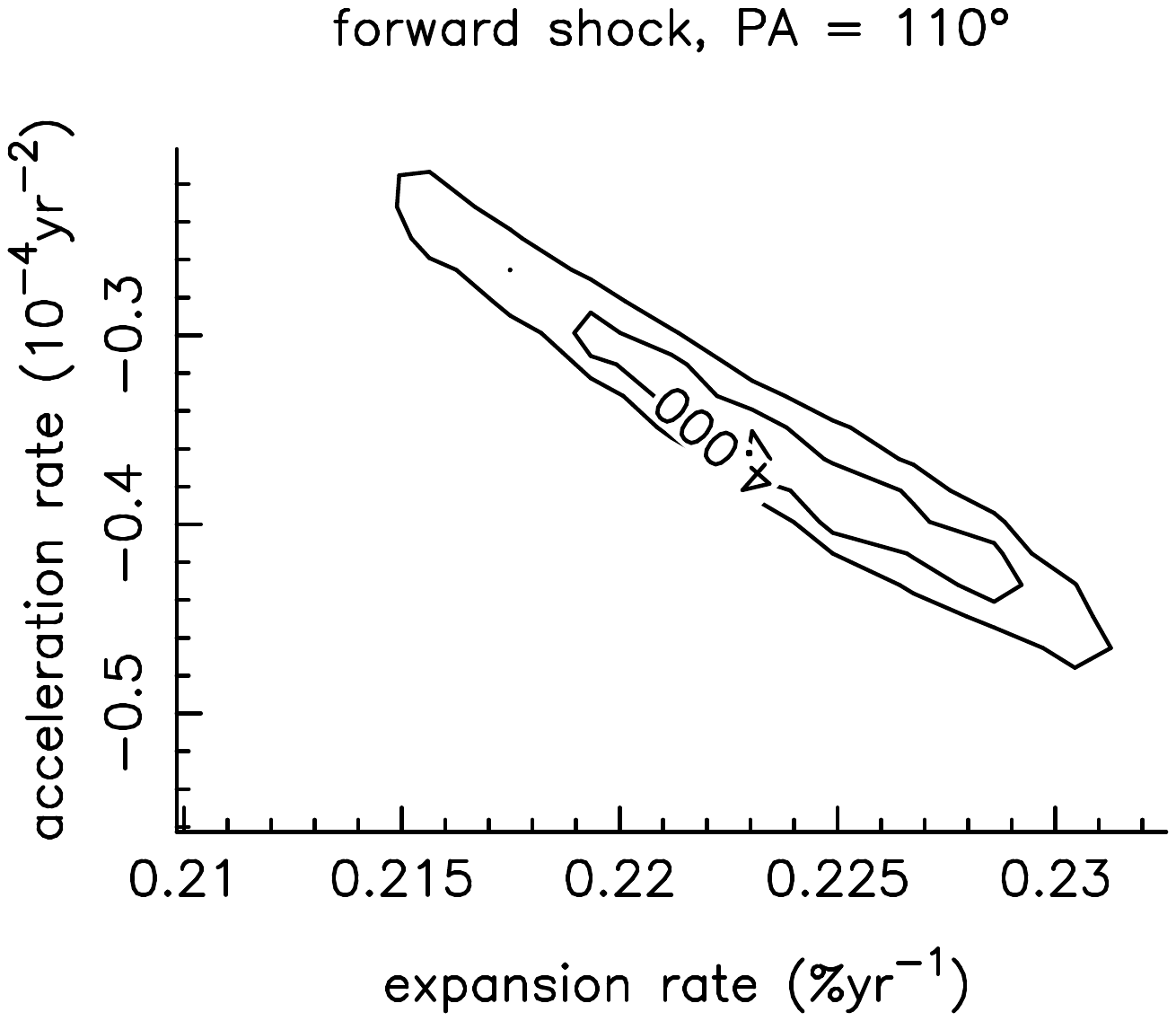}
\includegraphics[trim=50 20 120 30,clip=true,width=0.15\textwidth]{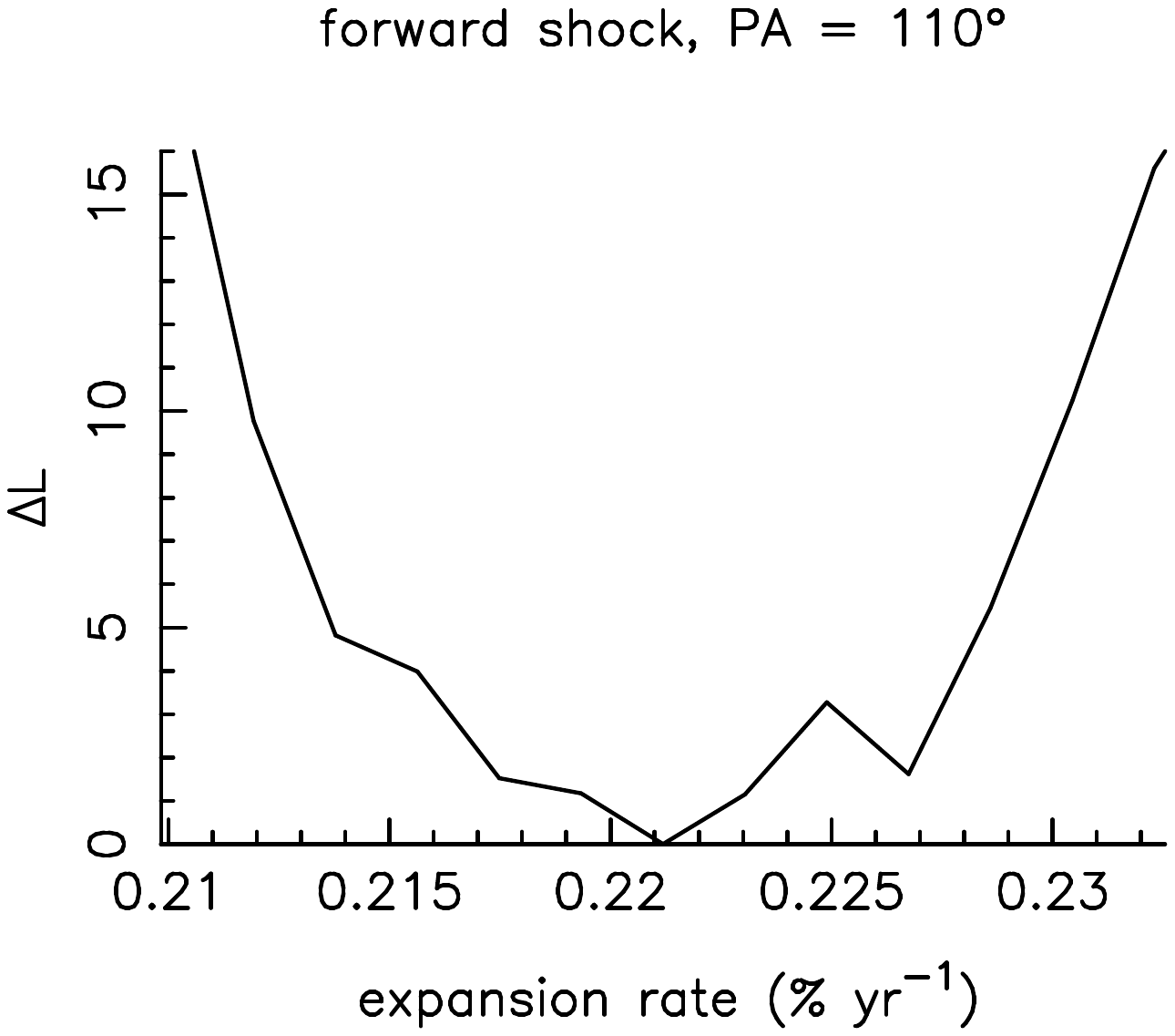}
\includegraphics[trim=50 20 120 30,clip=true,width=0.15\textwidth]{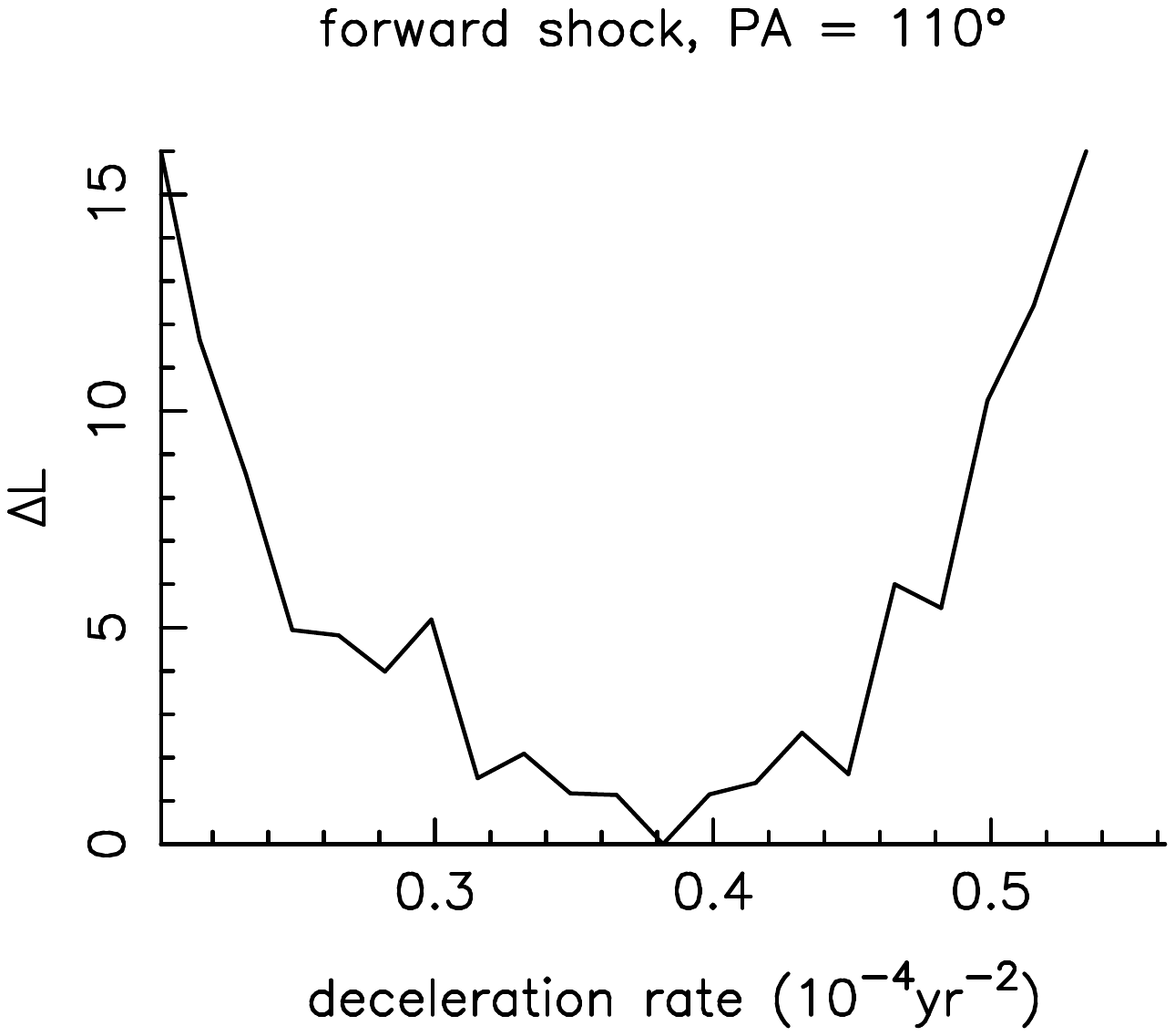}
}

\centerline{
\includegraphics[trim=50 20 120 30,clip=true,width=0.15\textwidth]{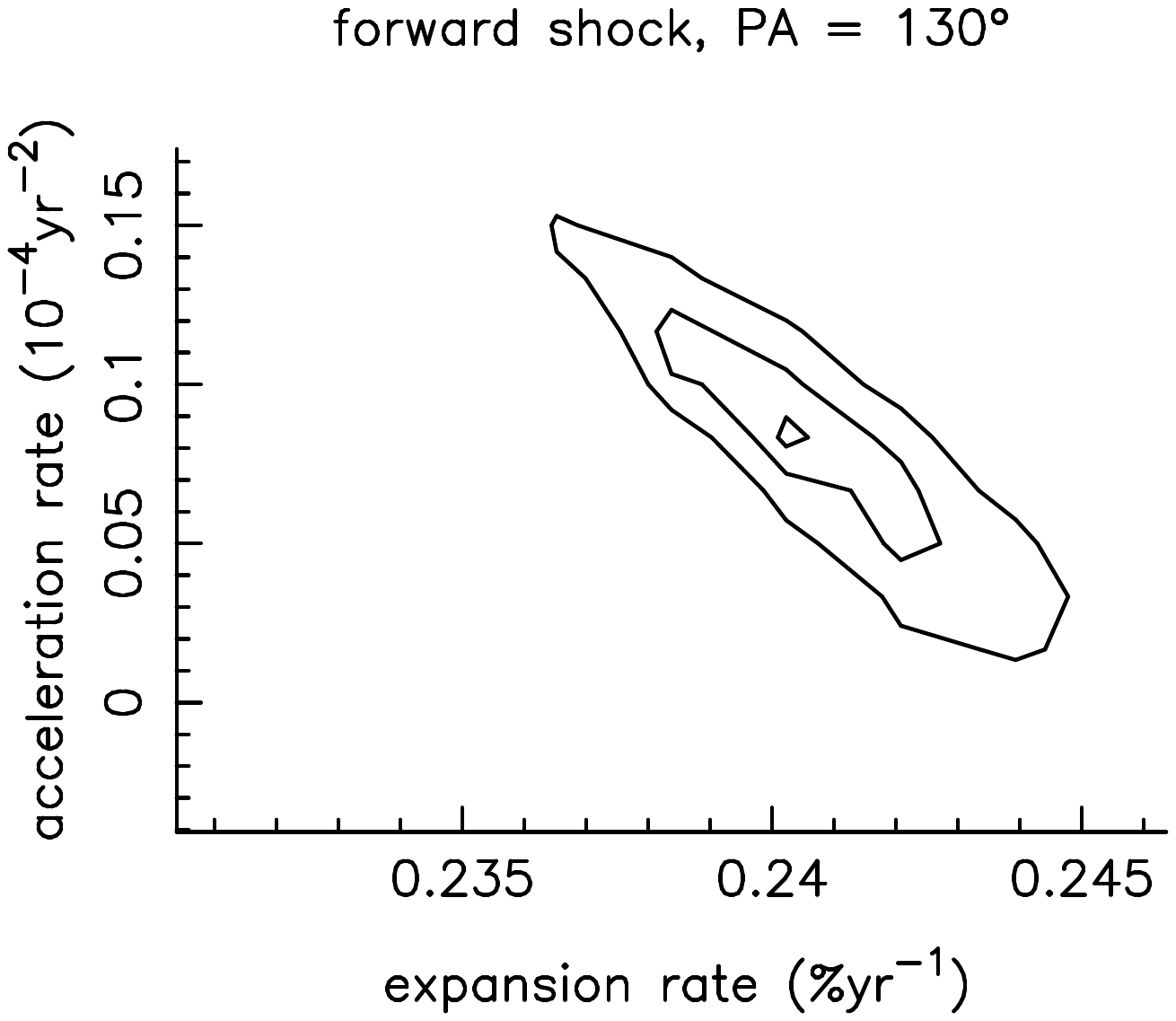}
\includegraphics[trim=50 20 120 30,clip=true,width=0.15\textwidth]{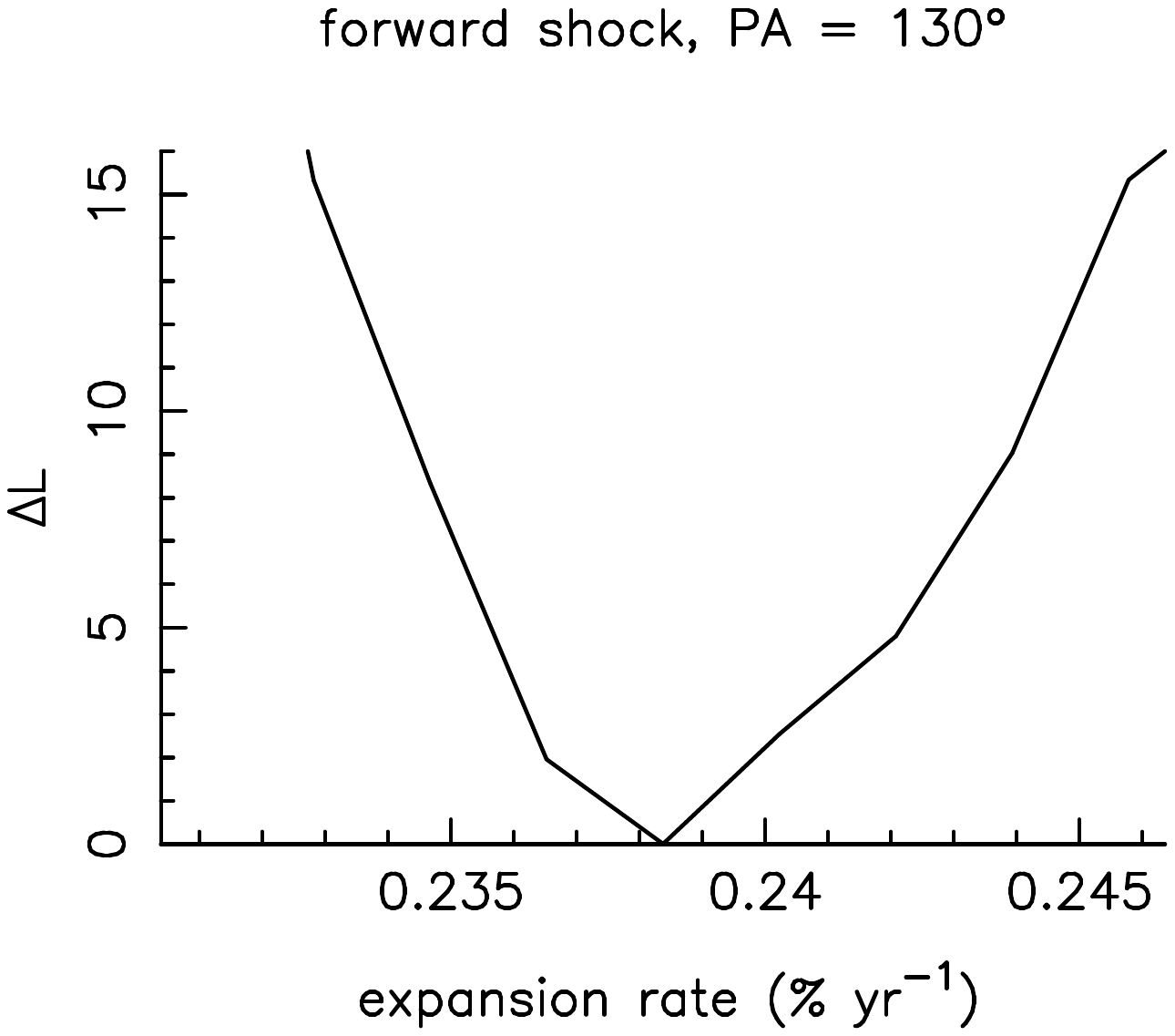}
\includegraphics[trim=50 20 120 30,clip=true,width=0.15\textwidth]{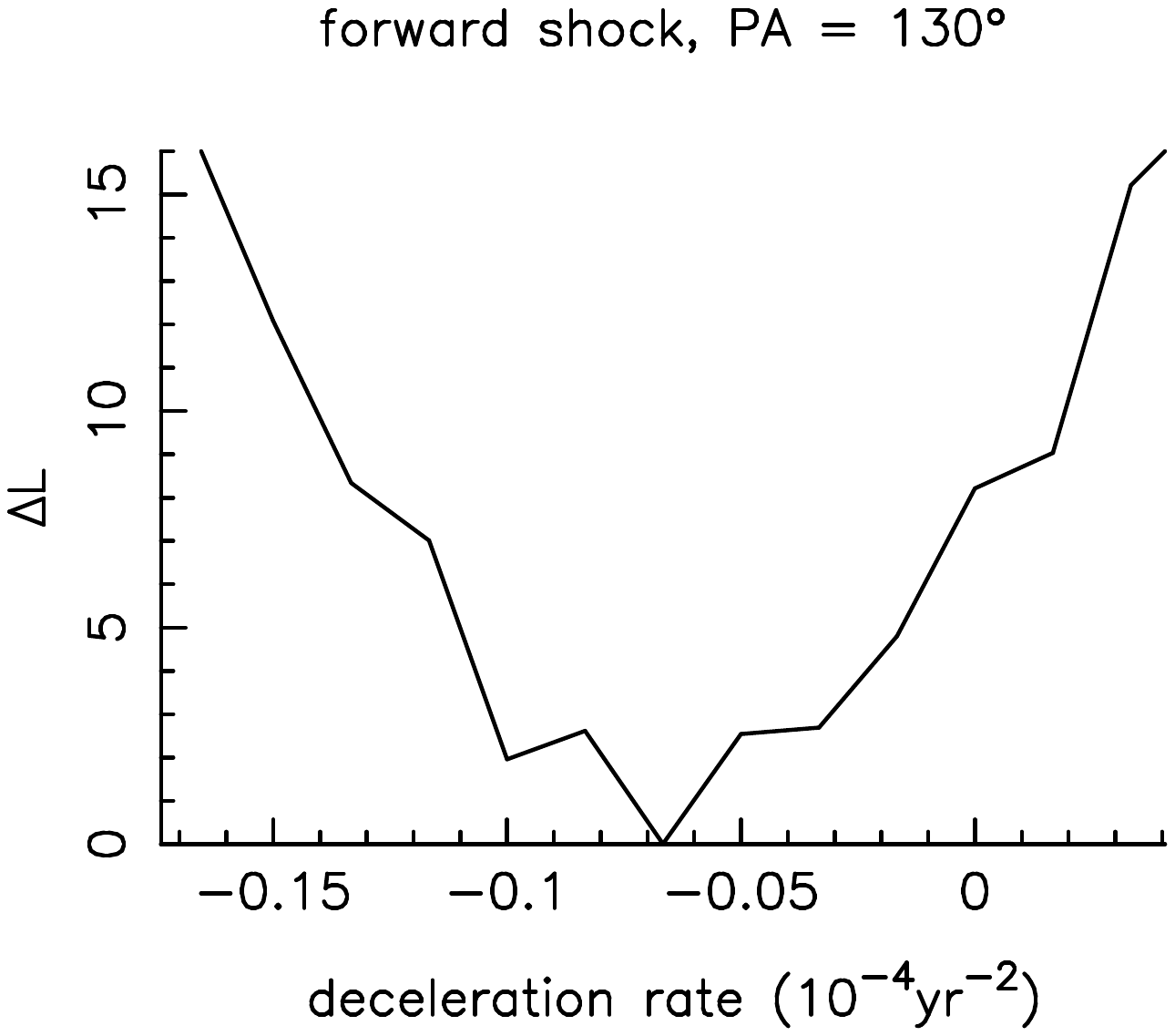}
\includegraphics[trim=50 20 120 30,clip=true,width=0.15\textwidth]{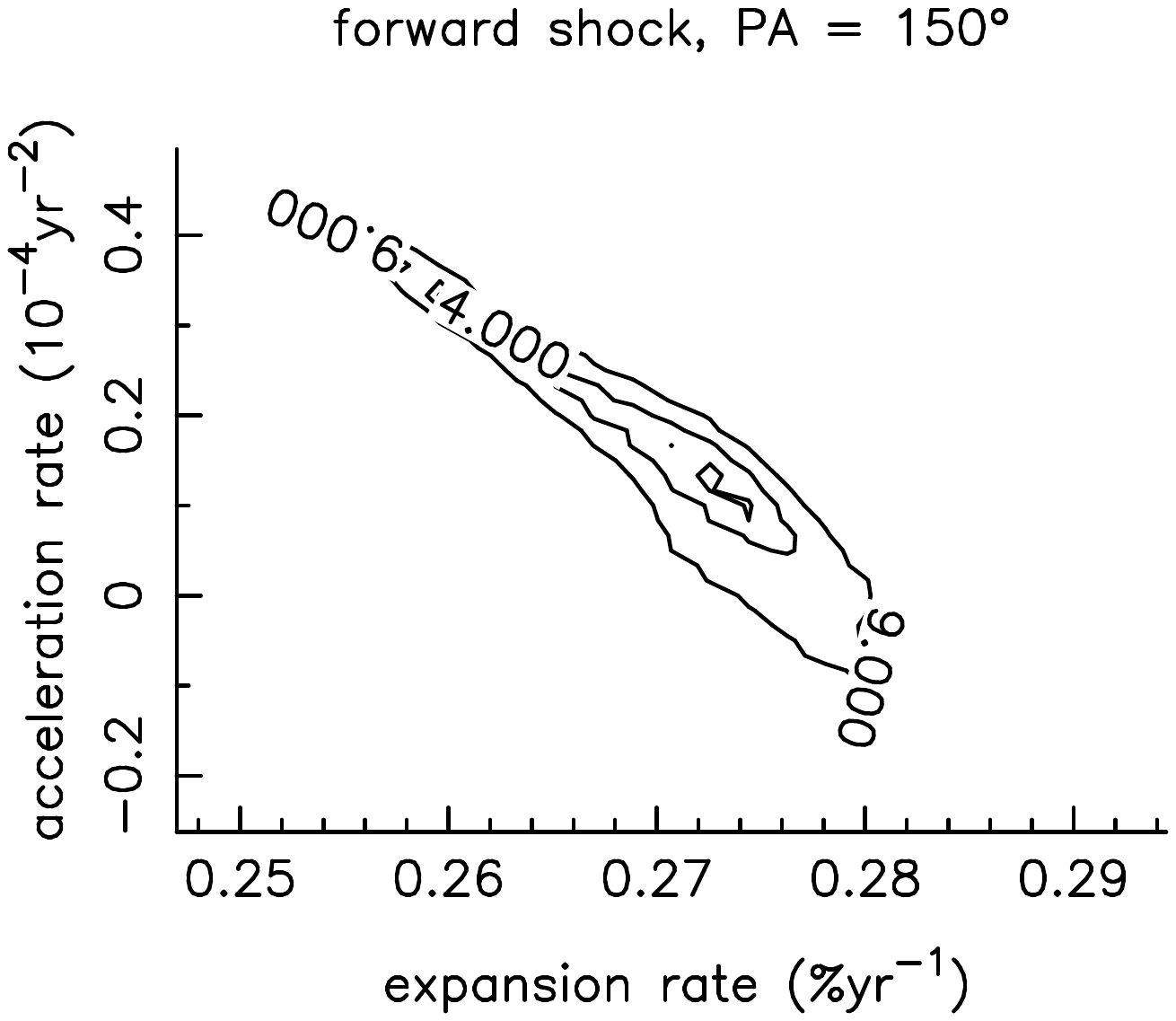}
\includegraphics[trim=50 20 120 30,clip=true,width=0.15\textwidth]{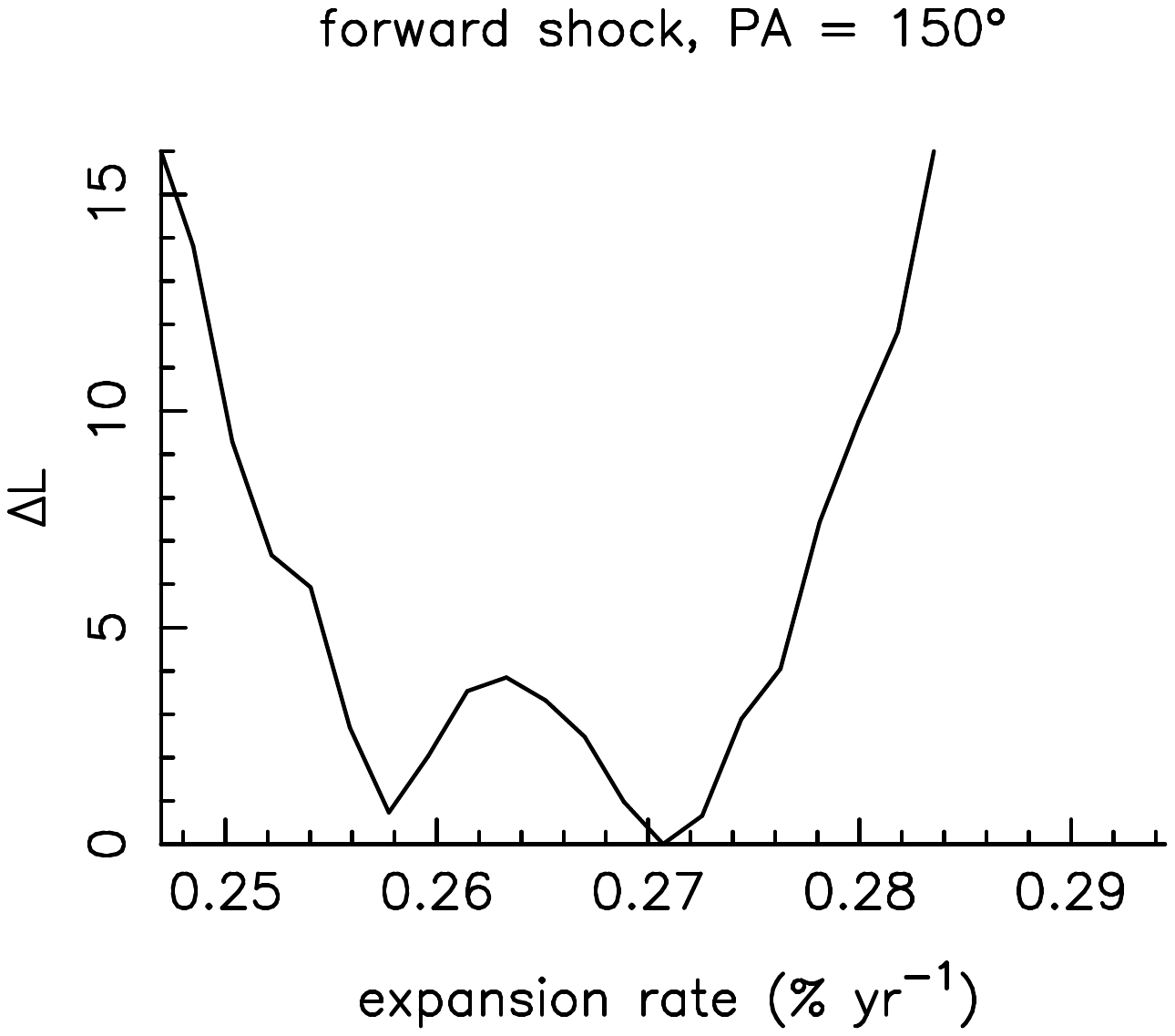}
\includegraphics[trim=50 20 120 30,clip=true,width=0.15\textwidth]{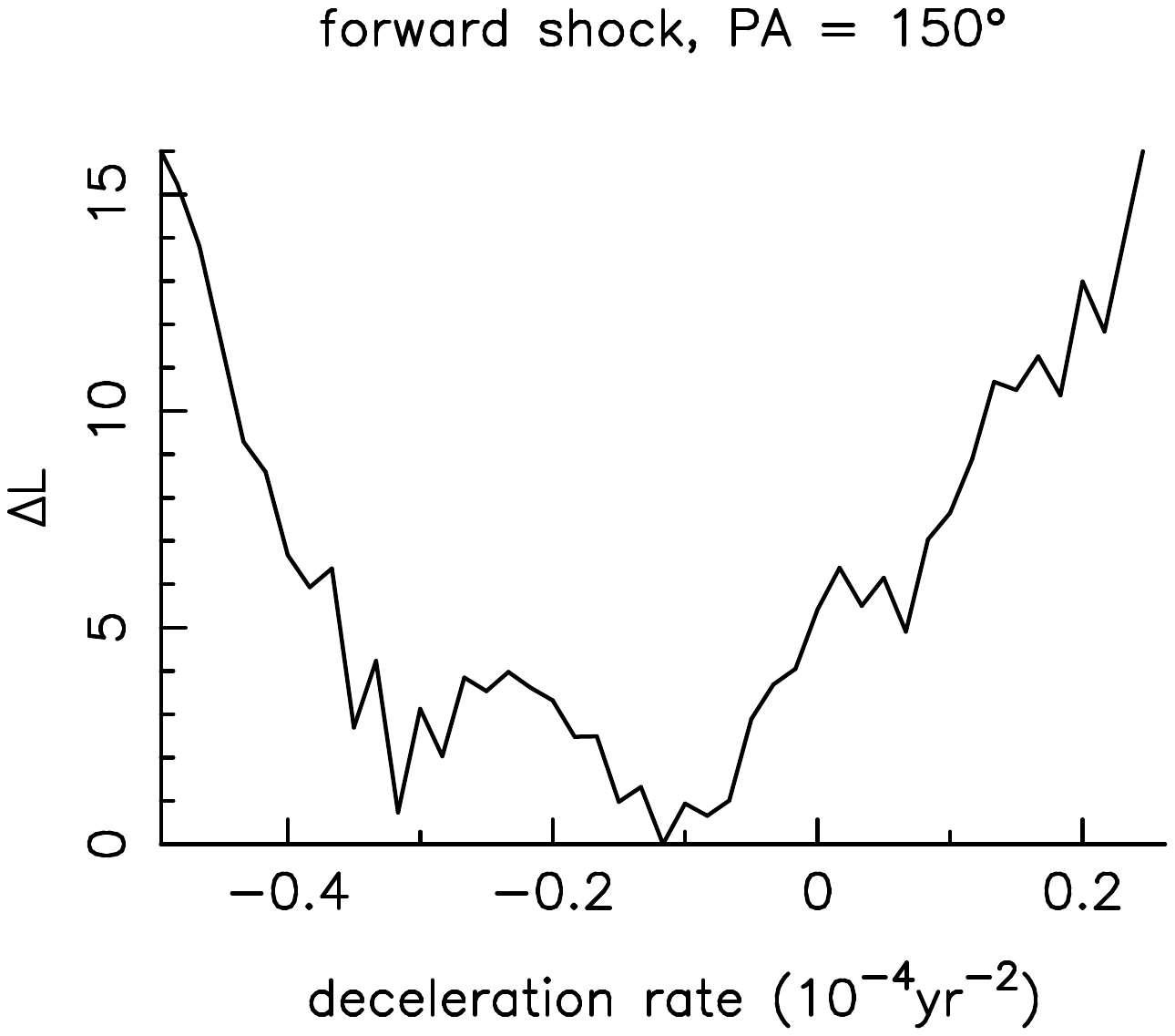}
}

\centerline{
\includegraphics[trim=50 20 120 30,clip=true,width=0.15\textwidth]{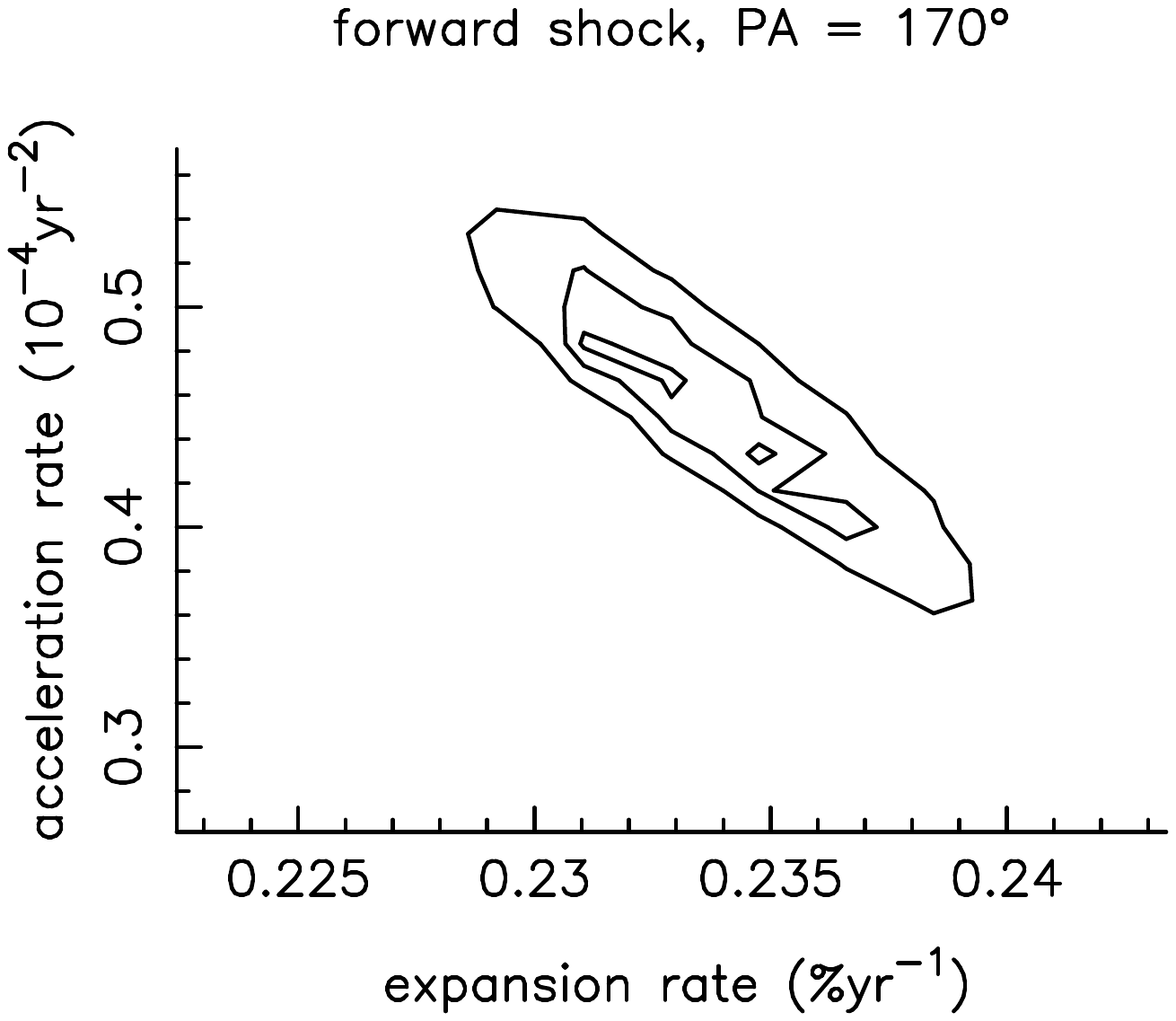}
\includegraphics[trim=50 20 120 30,clip=true,width=0.15\textwidth]{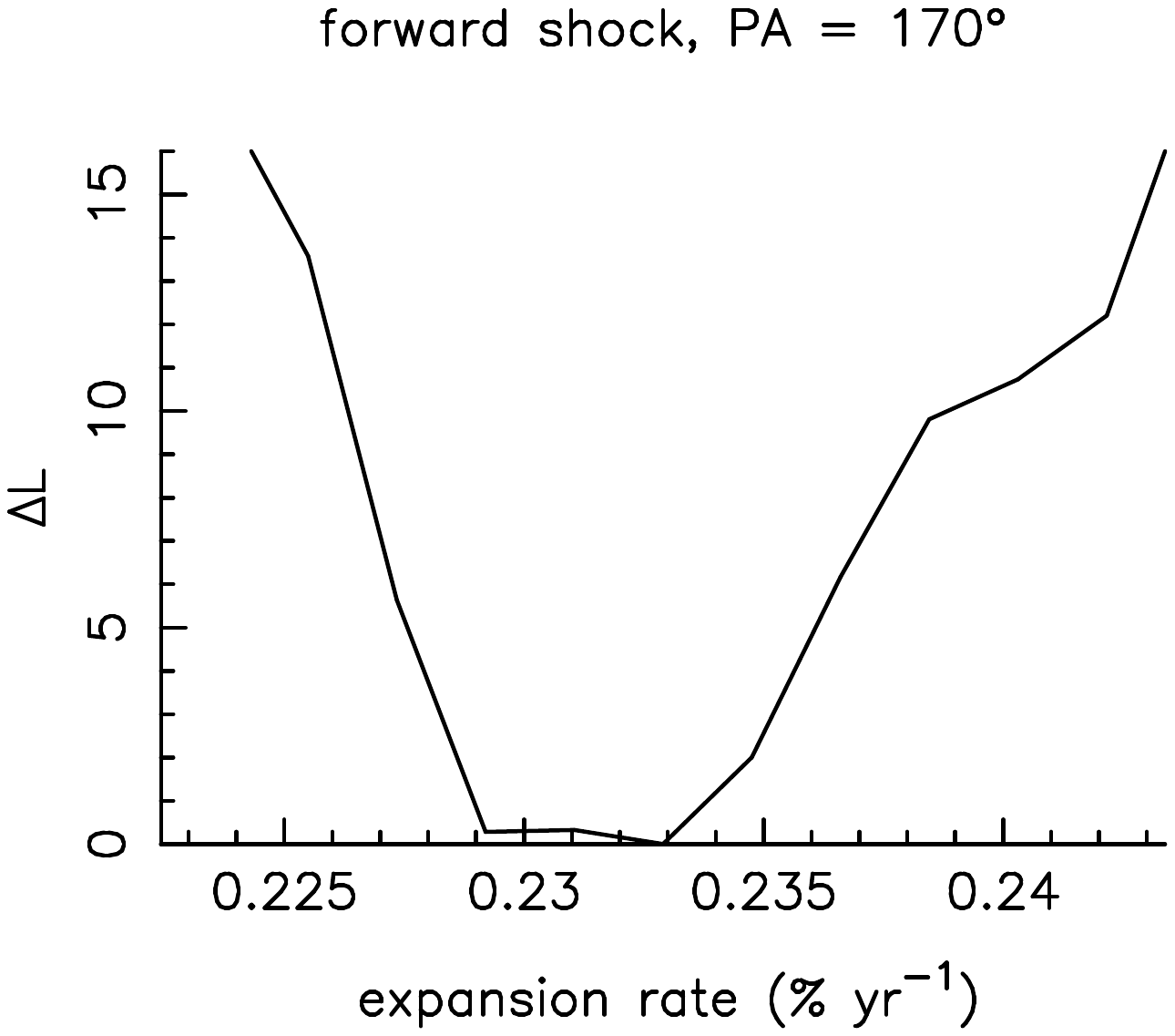}
\includegraphics[trim=50 20 120 30,clip=true,width=0.15\textwidth]{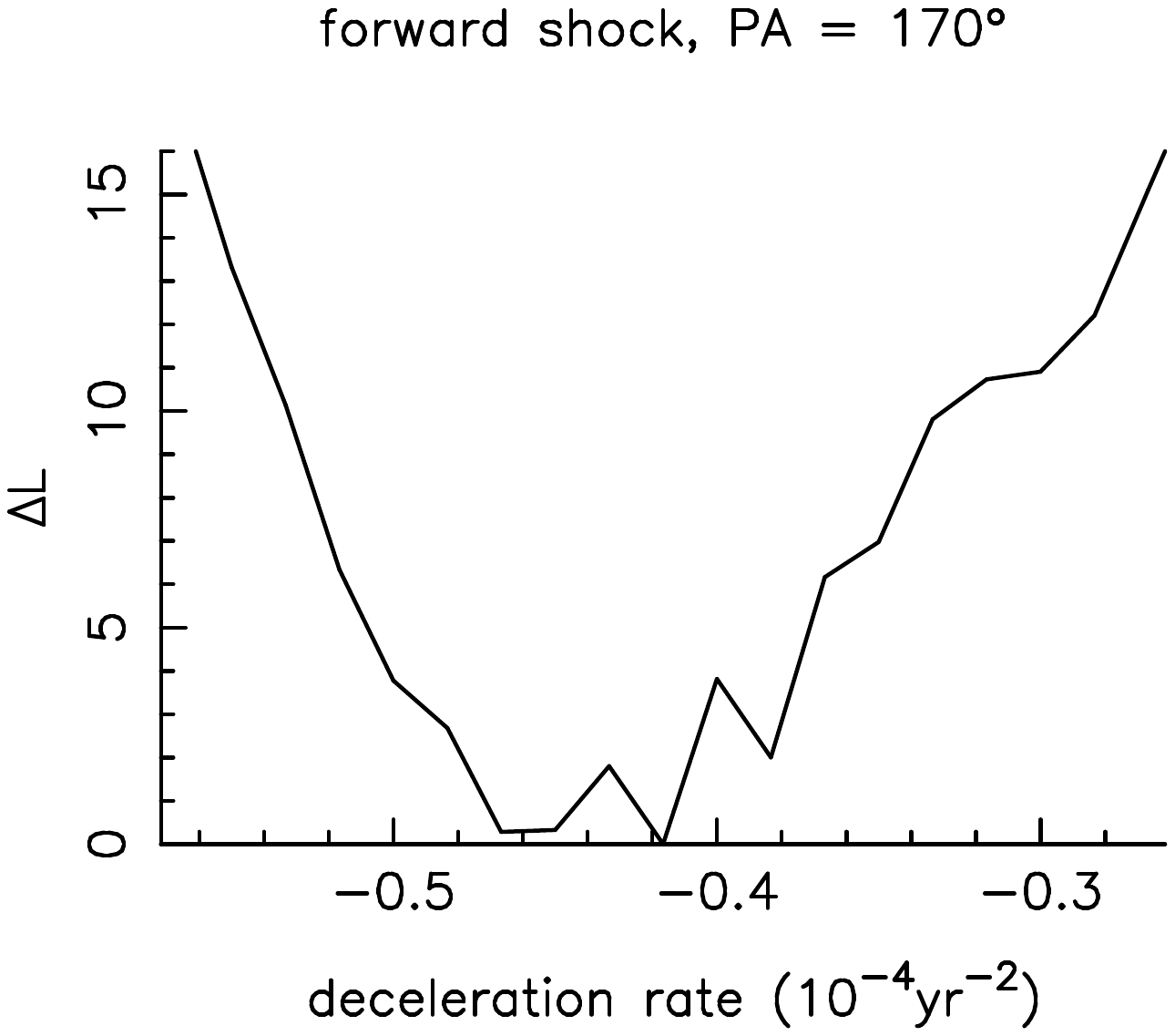}
\includegraphics[trim=50 20 120 30,clip=true,width=0.15\textwidth]{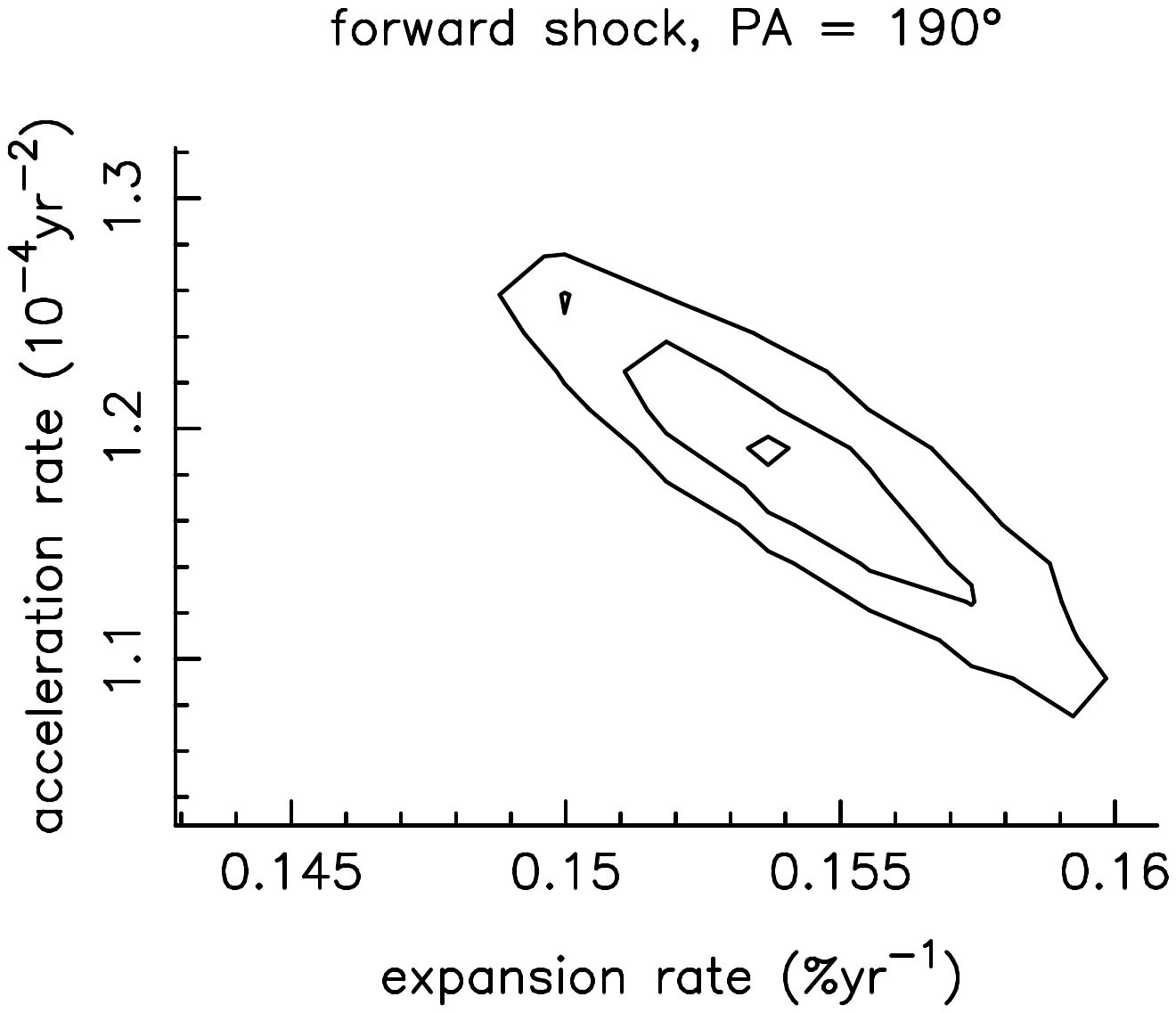}
\includegraphics[trim=50 20 120 30,clip=true,width=0.15\textwidth]{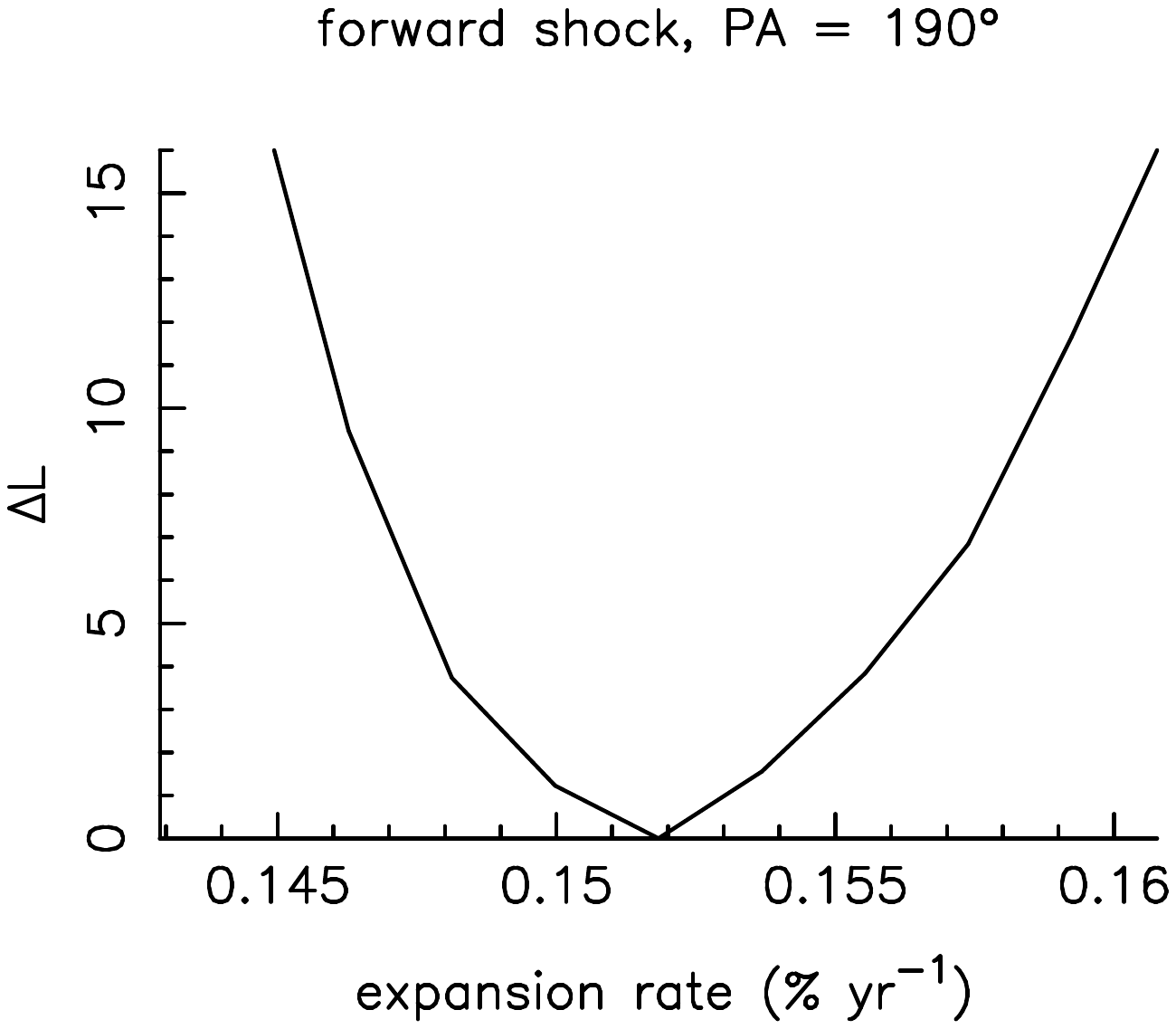}
\includegraphics[trim=50 20 120 30,clip=true,width=0.15\textwidth]{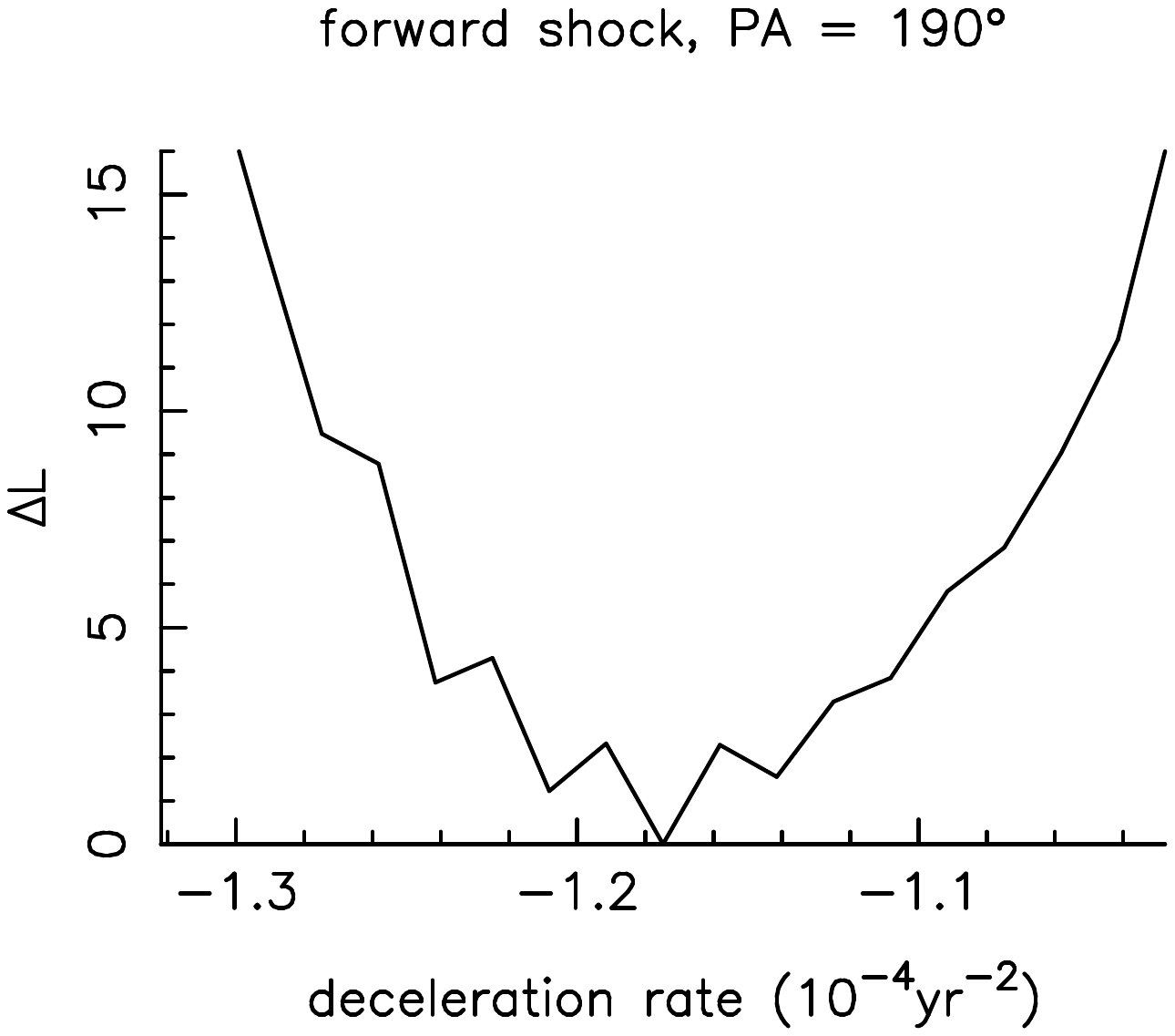}
}

\centerline{
\includegraphics[trim=50 20 120 30,clip=true,width=0.15\textwidth]{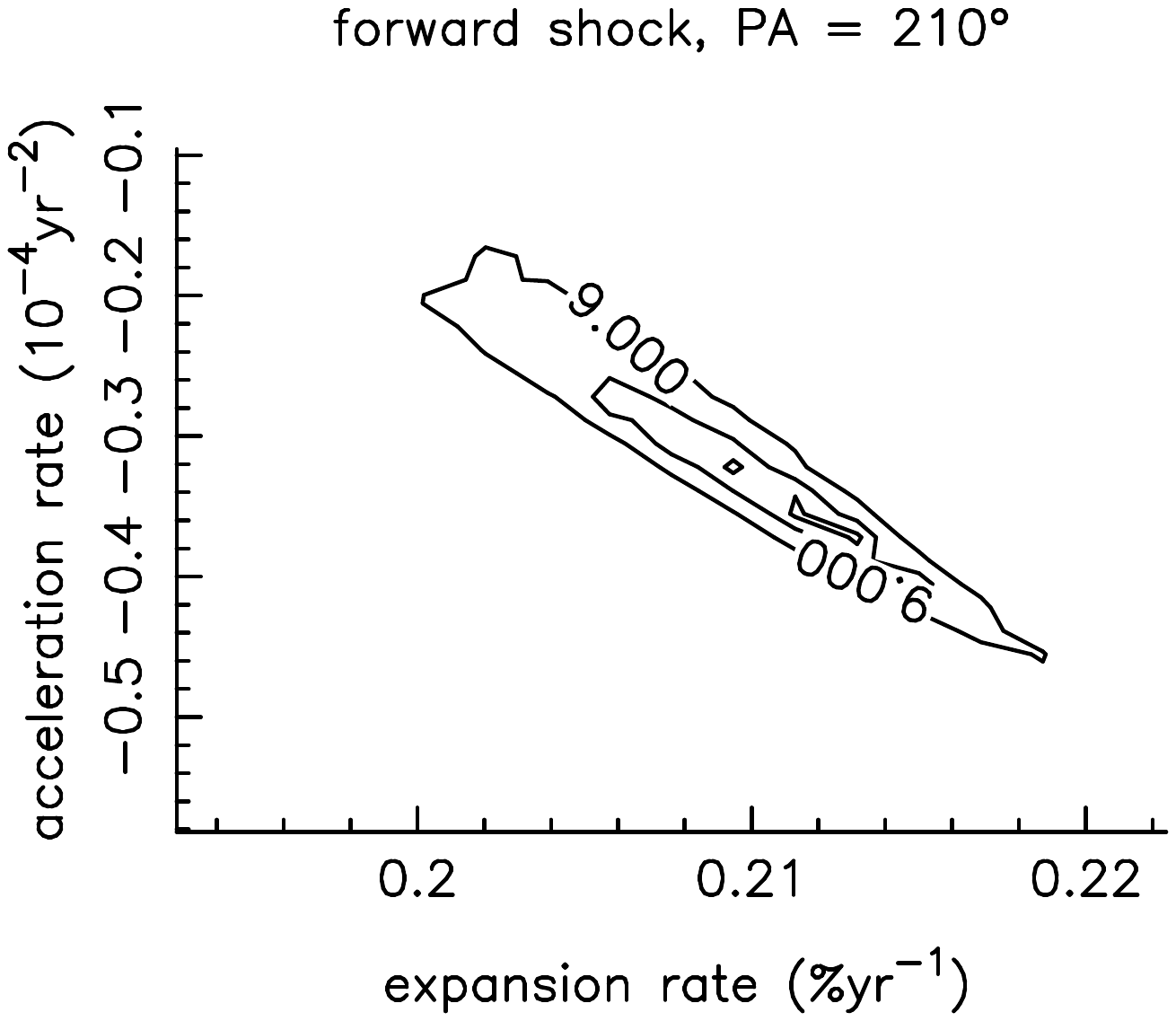}
\includegraphics[trim=50 20 120 30,clip=true,width=0.15\textwidth]{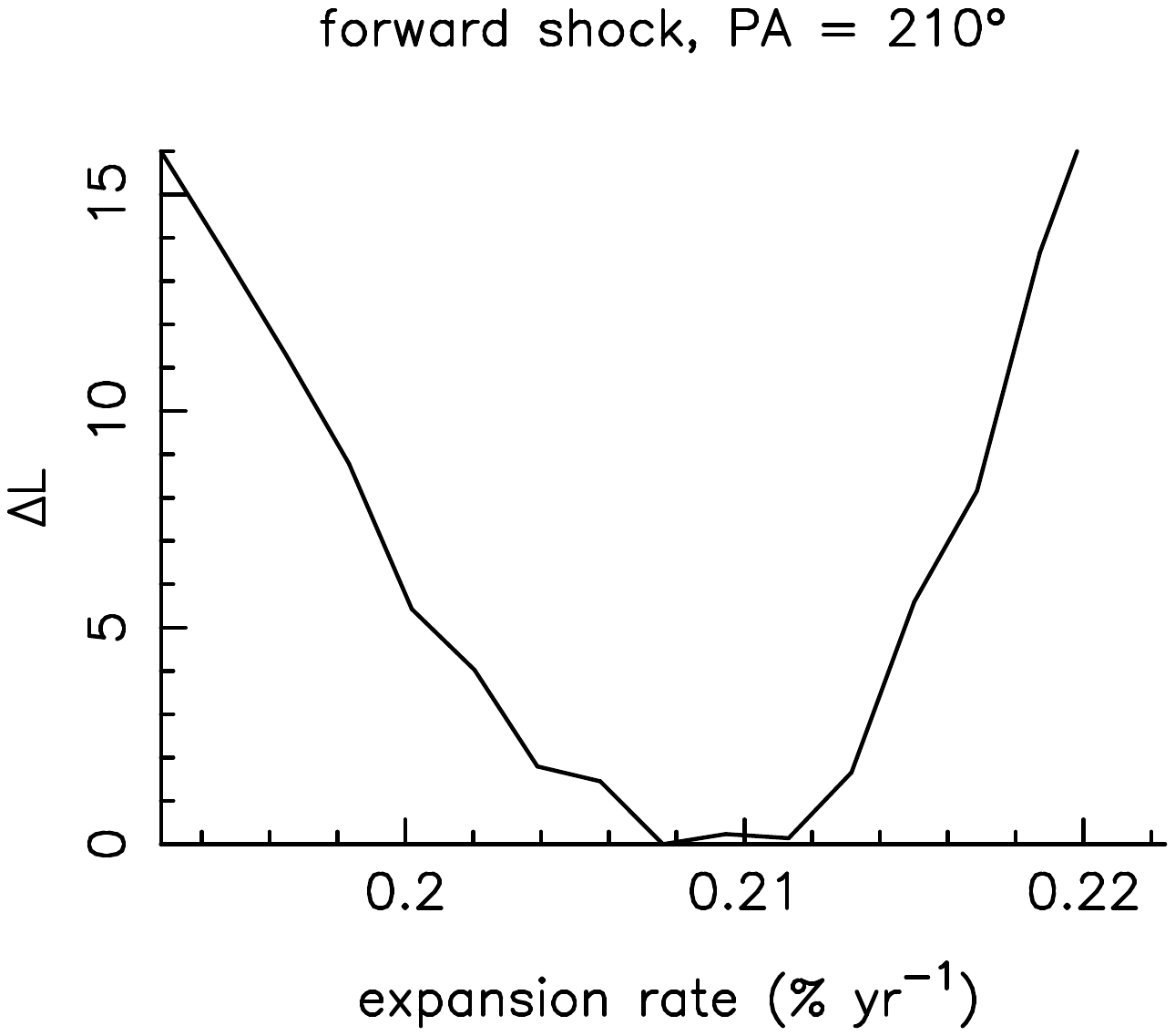}
\includegraphics[trim=50 20 120 30,clip=true,width=0.15\textwidth]{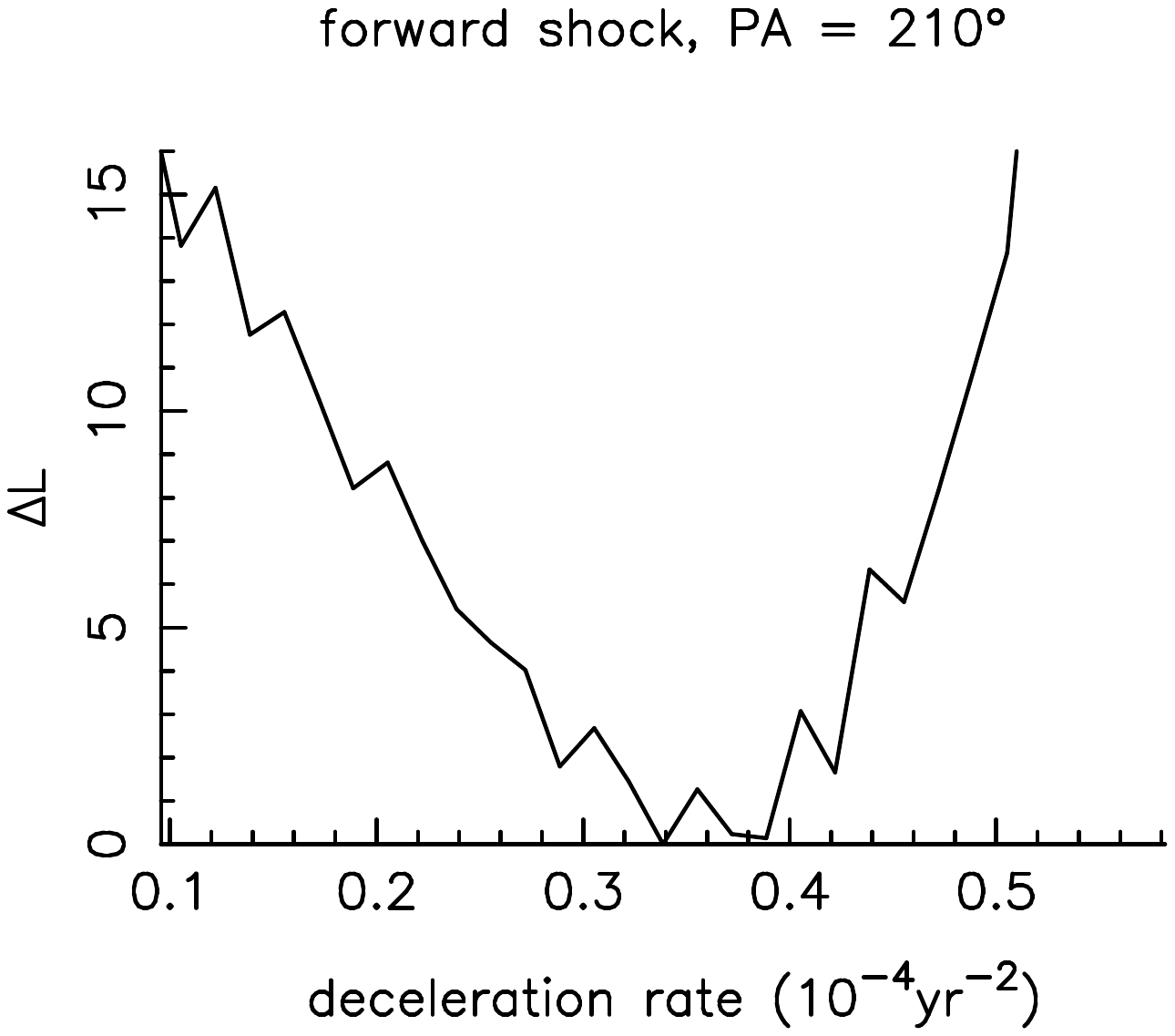}
\includegraphics[trim=50 20 120 30,clip=true,width=0.15\textwidth]{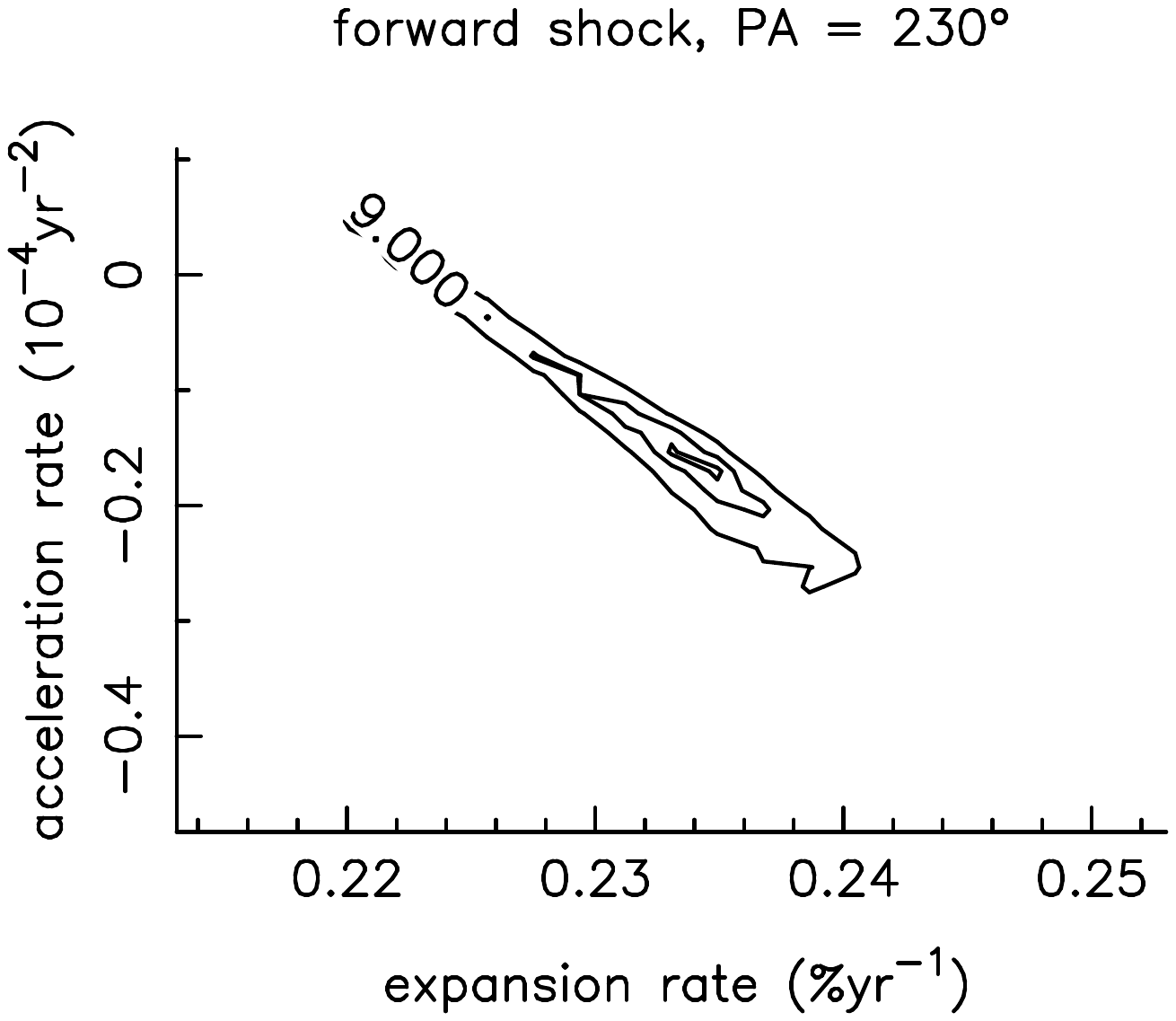}
\includegraphics[trim=50 20 120 30,clip=true,width=0.15\textwidth]{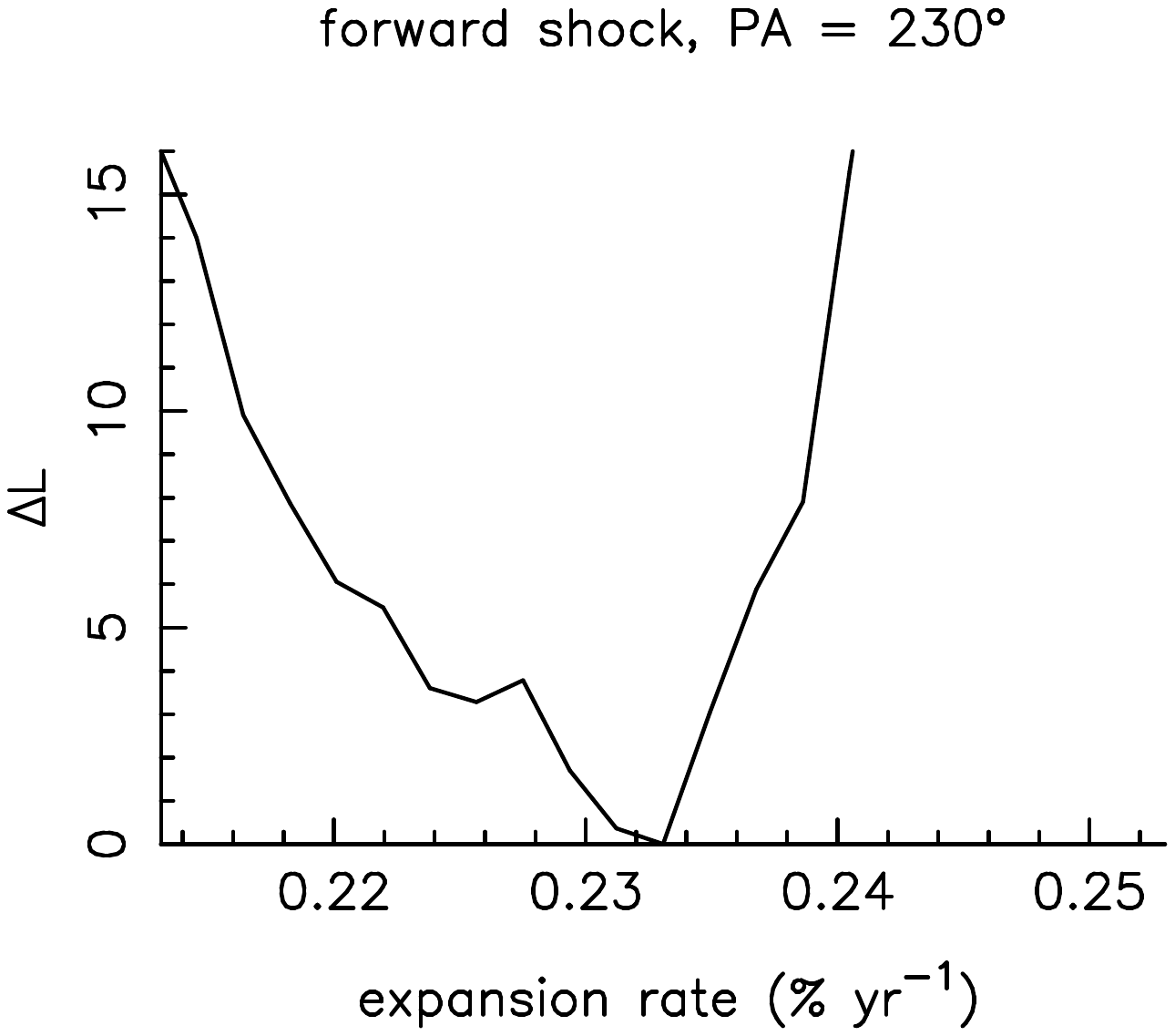}
\includegraphics[trim=50 20 120 30,clip=true,width=0.15\textwidth]{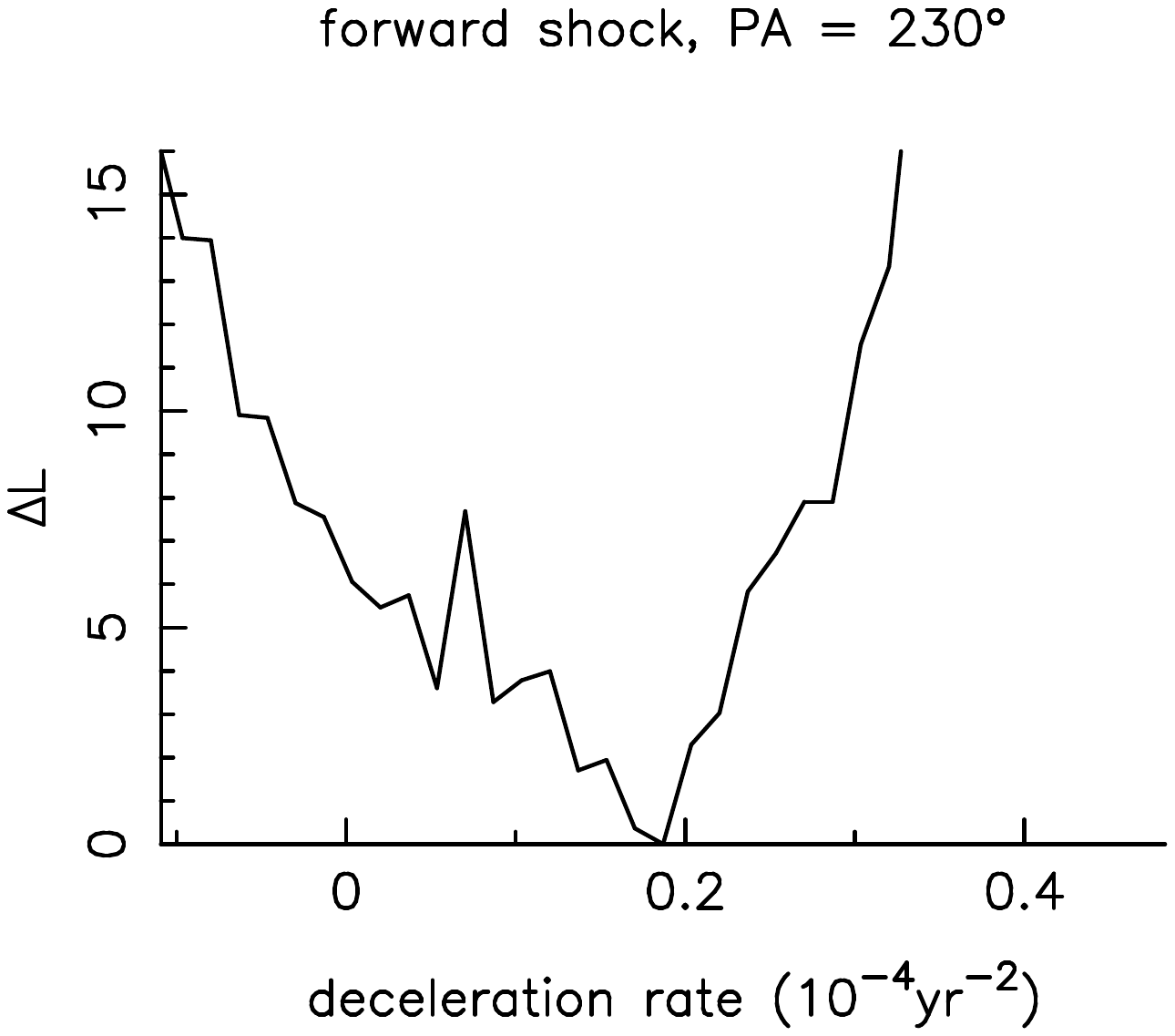}
}

\centerline{
\includegraphics[trim=50 20 120 30,clip=true,width=0.15\textwidth]{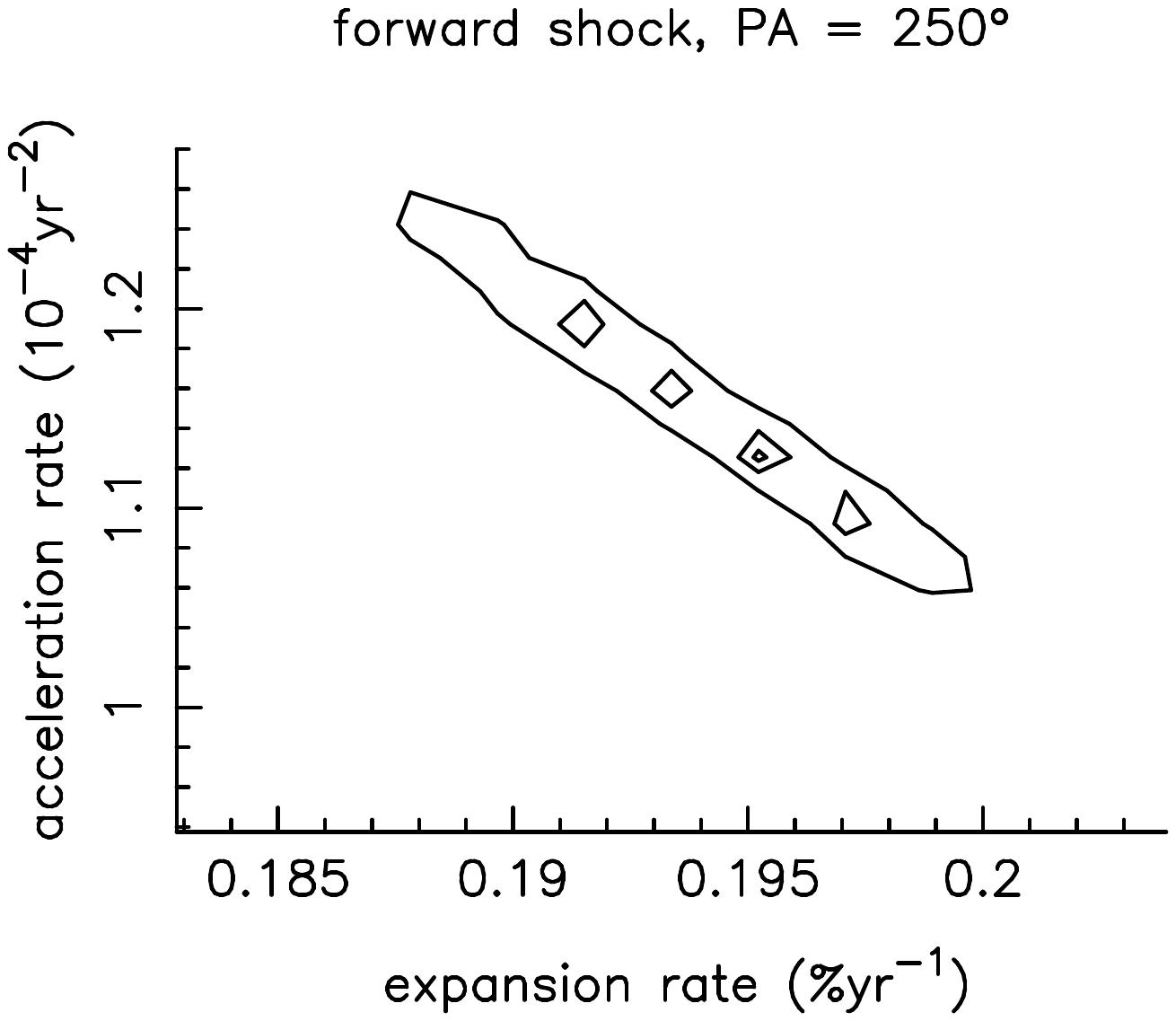}
\includegraphics[trim=50 20 120 30,clip=true,width=0.15\textwidth]{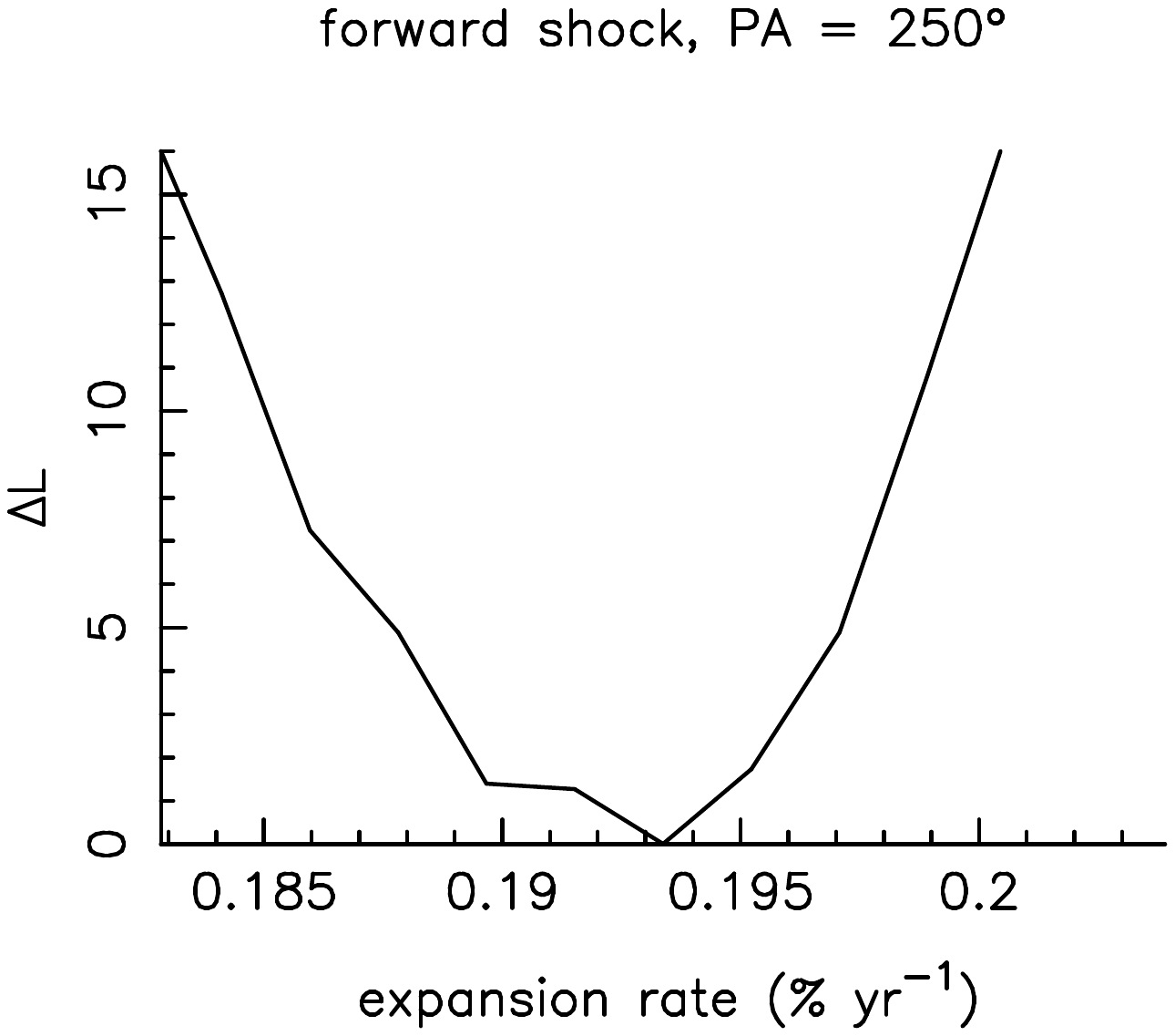}
\includegraphics[trim=50 20 120 30,clip=true,width=0.15\textwidth]{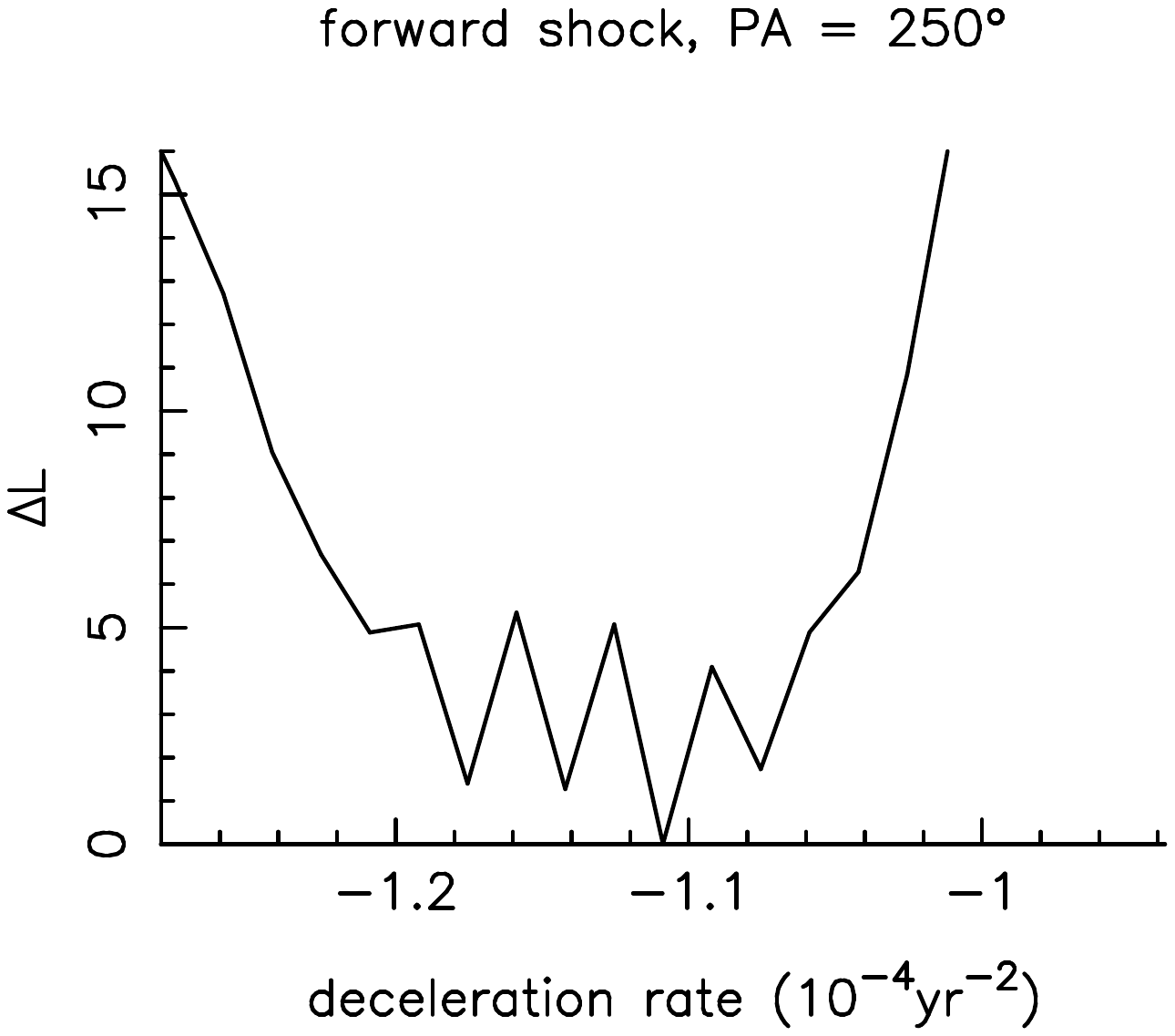}
\includegraphics[trim=50 20 120 30,clip=true,width=0.15\textwidth]{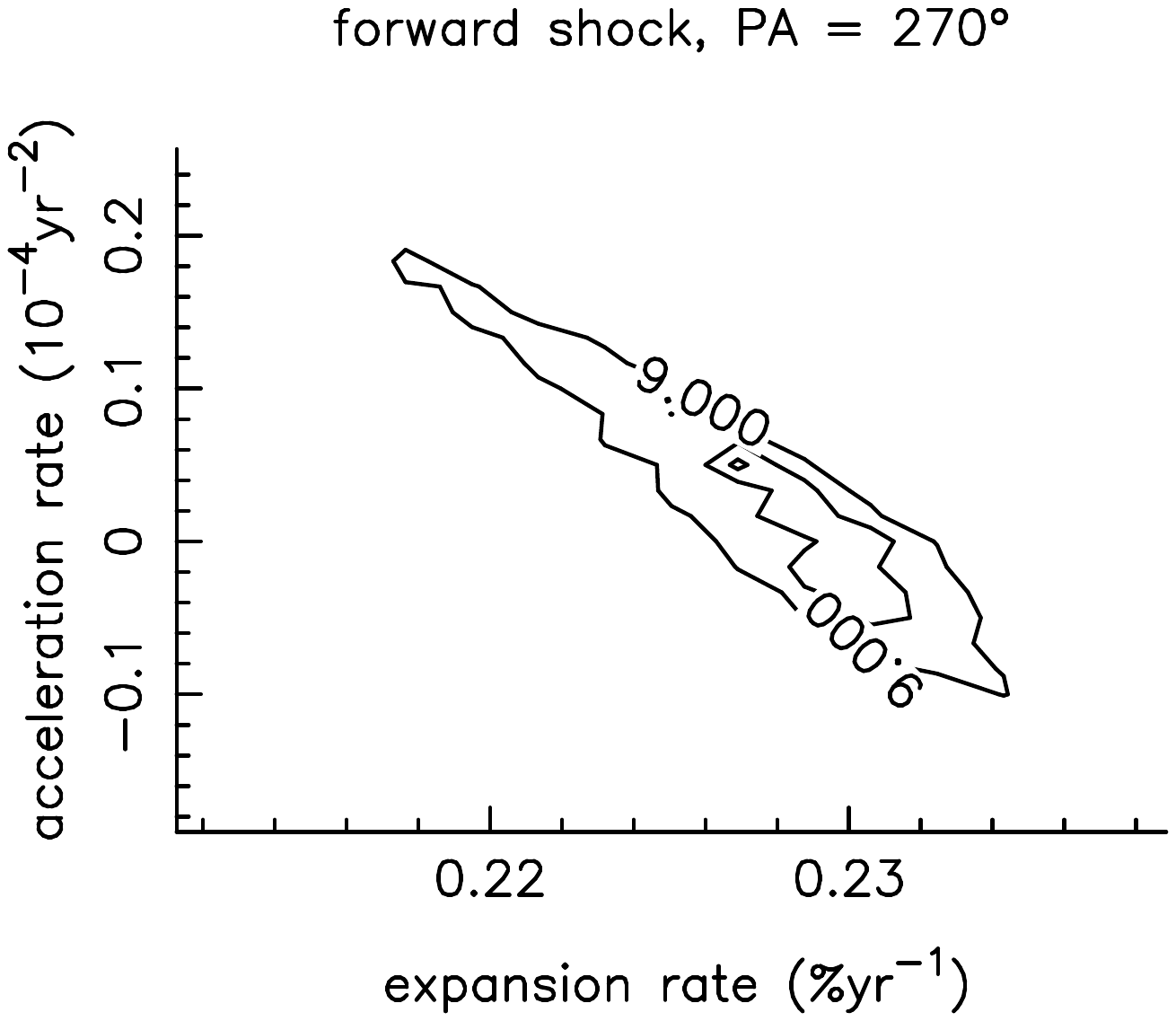}
\includegraphics[trim=50 20 120 30,clip=true,width=0.15\textwidth]{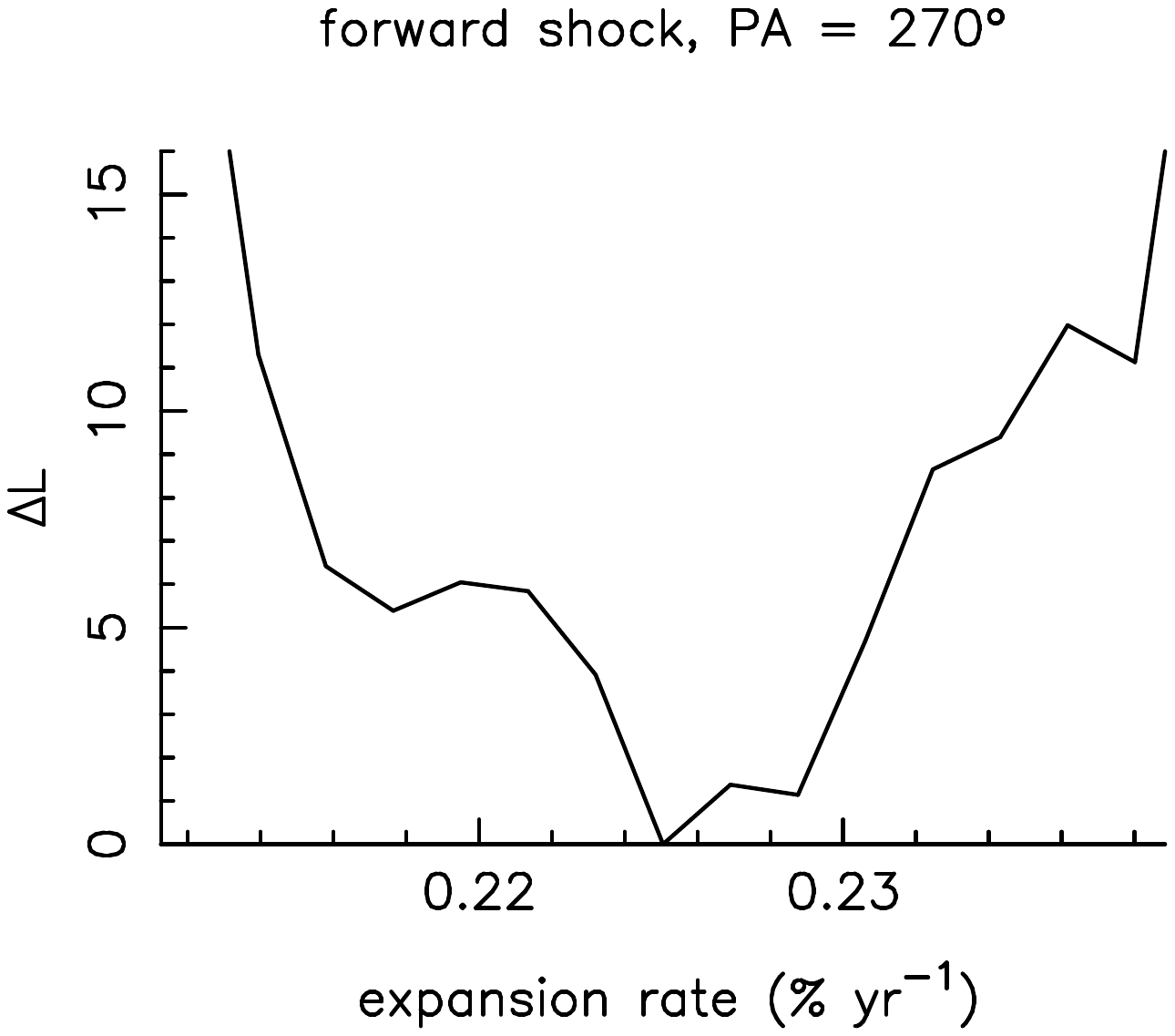}
\includegraphics[trim=50 20 120 30,clip=true,width=0.15\textwidth]{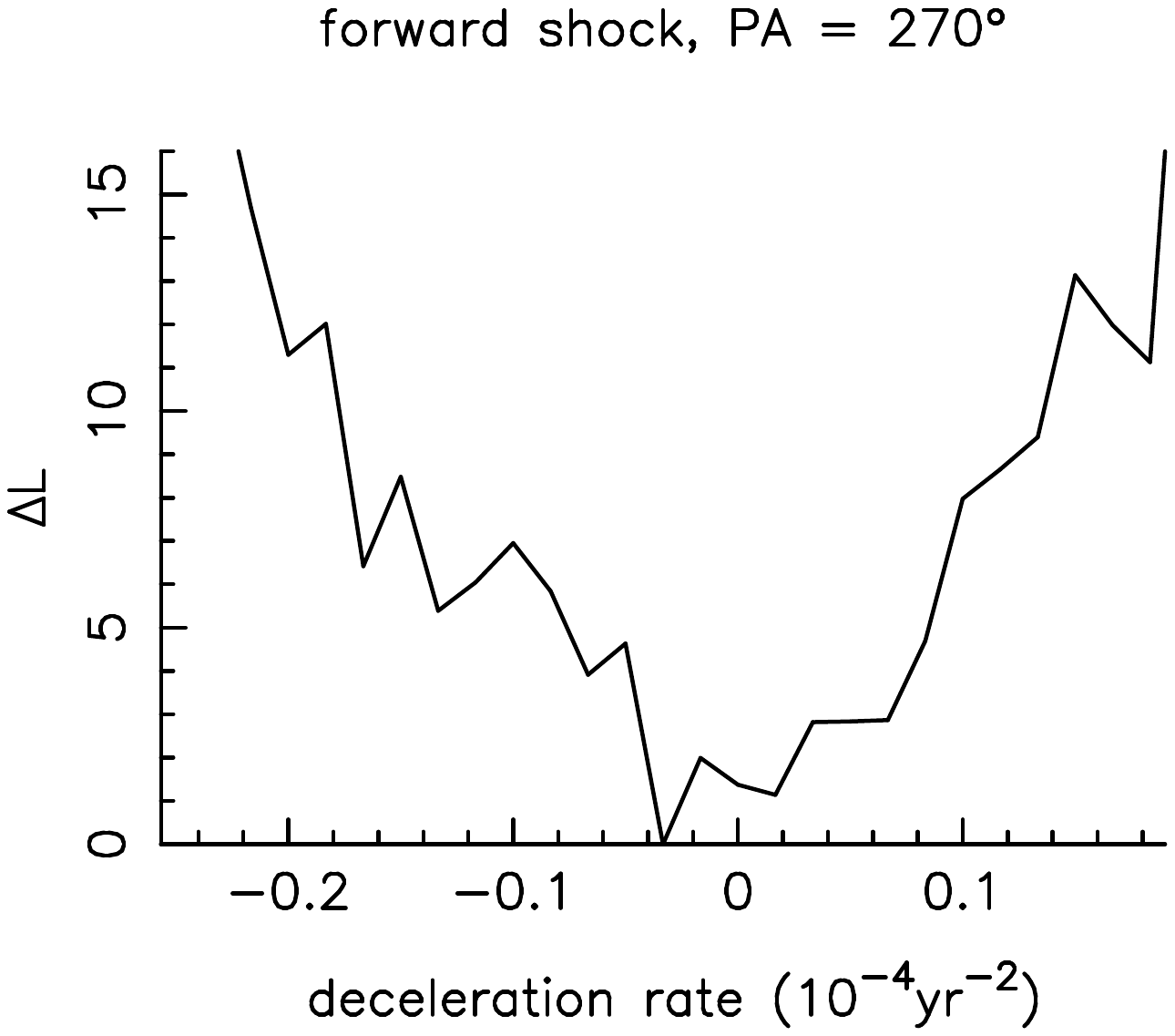}
}

\centerline{
\includegraphics[trim=50 20 120 30,clip=true,width=0.15\textwidth]{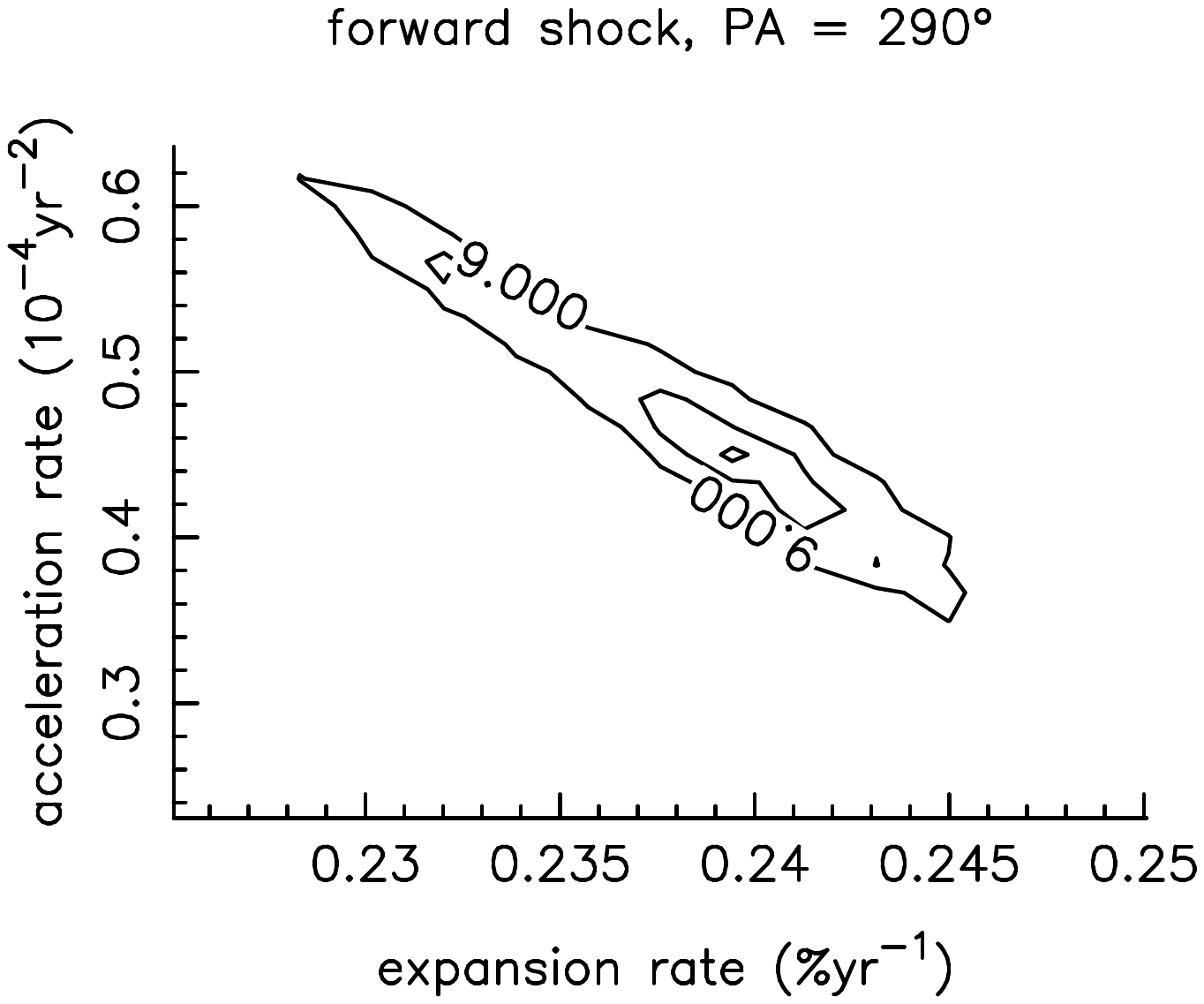}
\includegraphics[trim=50 20 120 30,clip=true,width=0.15\textwidth]{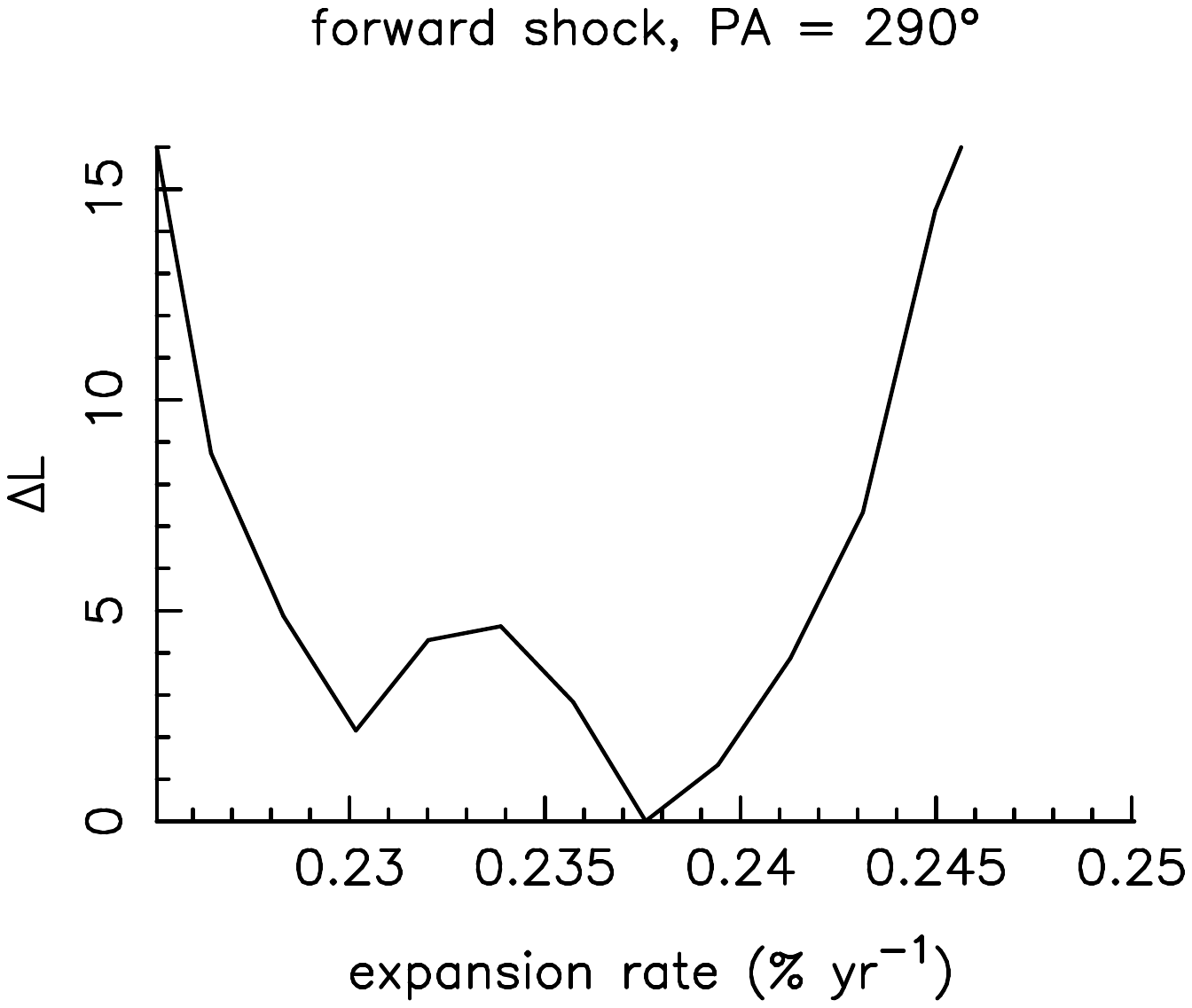}
\includegraphics[trim=50 20 120 30,clip=true,width=0.15\textwidth]{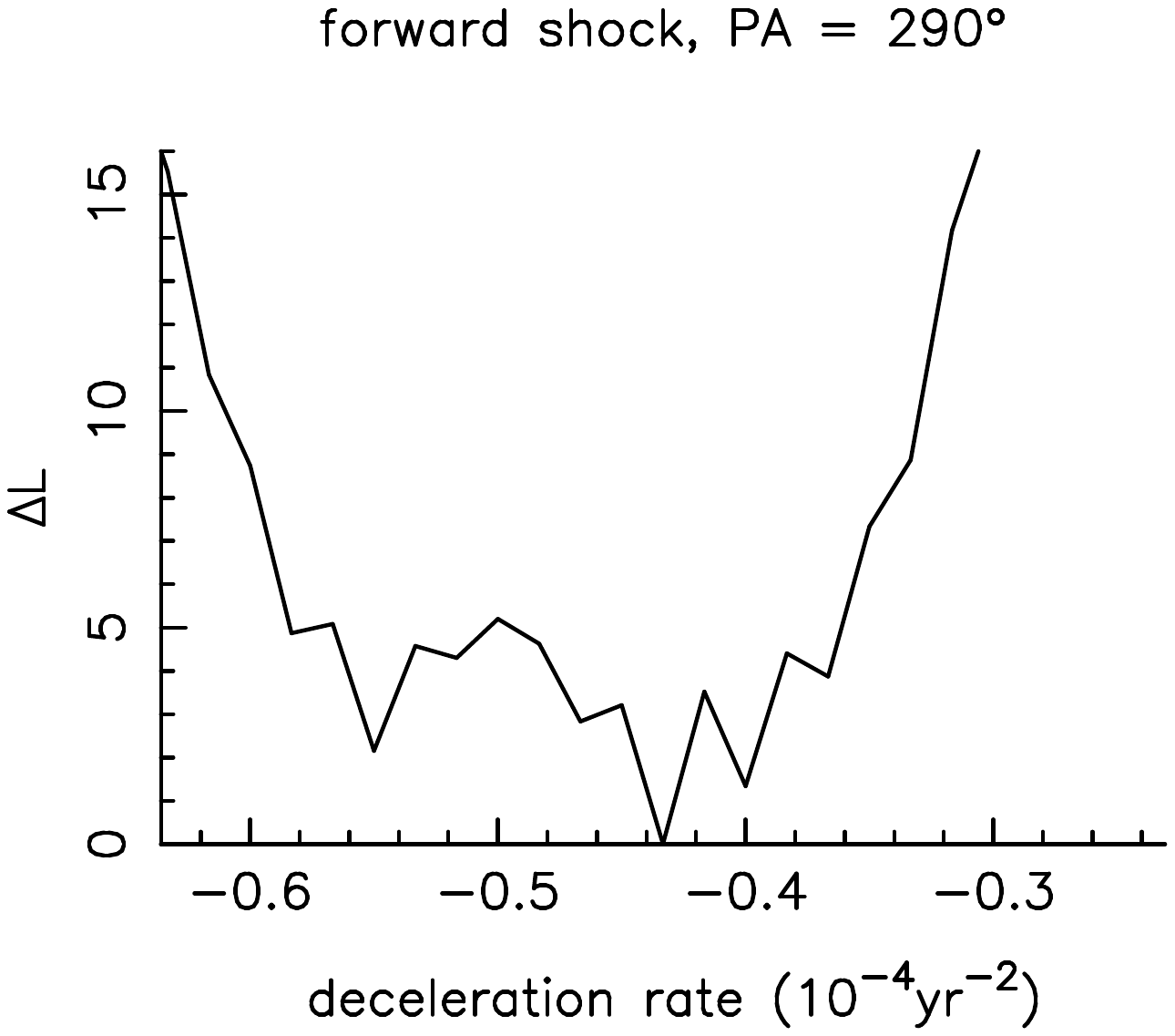}
\includegraphics[trim=50 20 120 30,clip=true,width=0.15\textwidth]{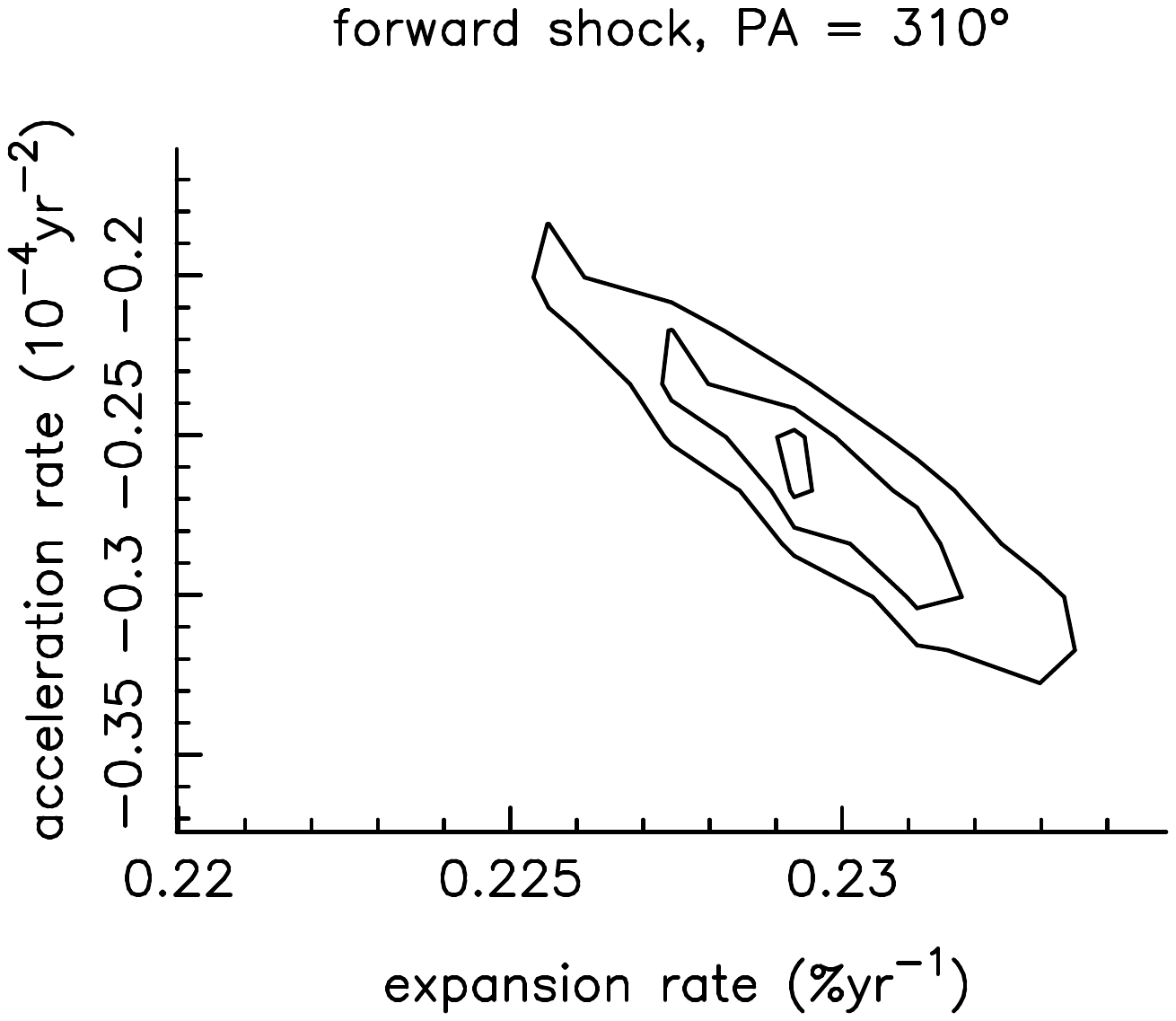}
\includegraphics[trim=50 20 120 30,clip=true,width=0.15\textwidth]{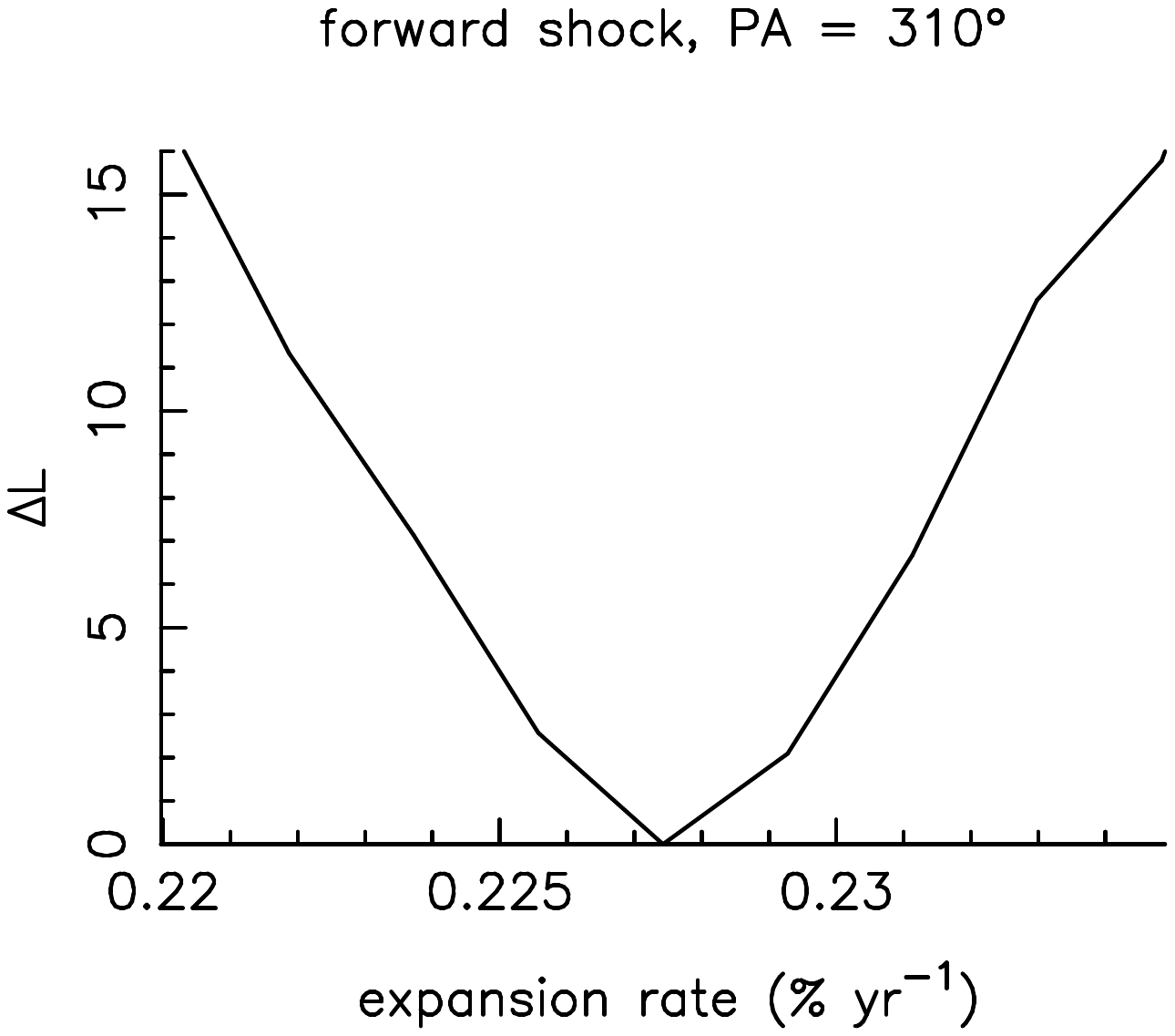}
\includegraphics[trim=50 20 120 30,clip=true,width=0.15\textwidth]{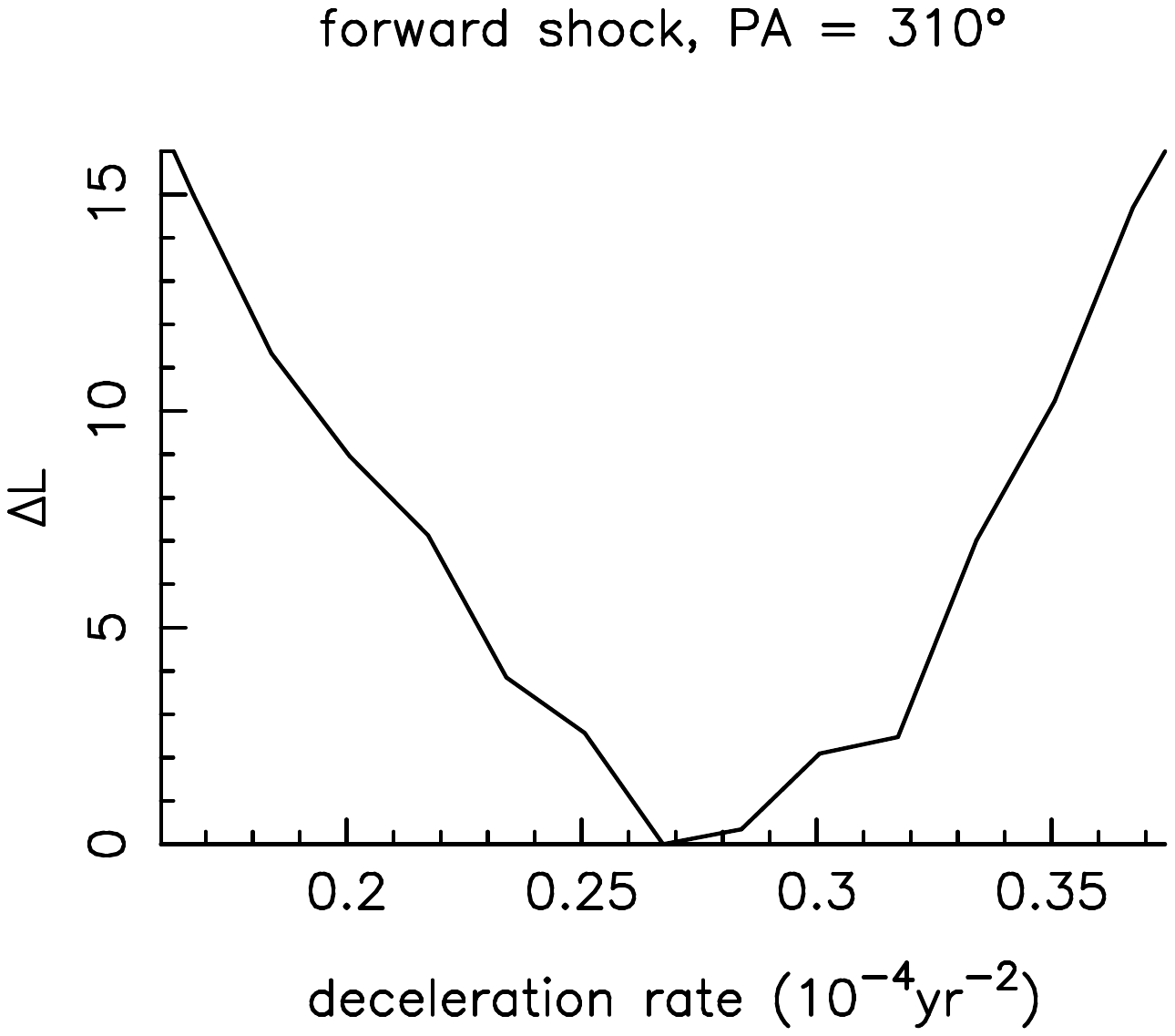}
}

\centerline{
\includegraphics[trim=50 20 120 30,clip=true,width=0.15\textwidth]{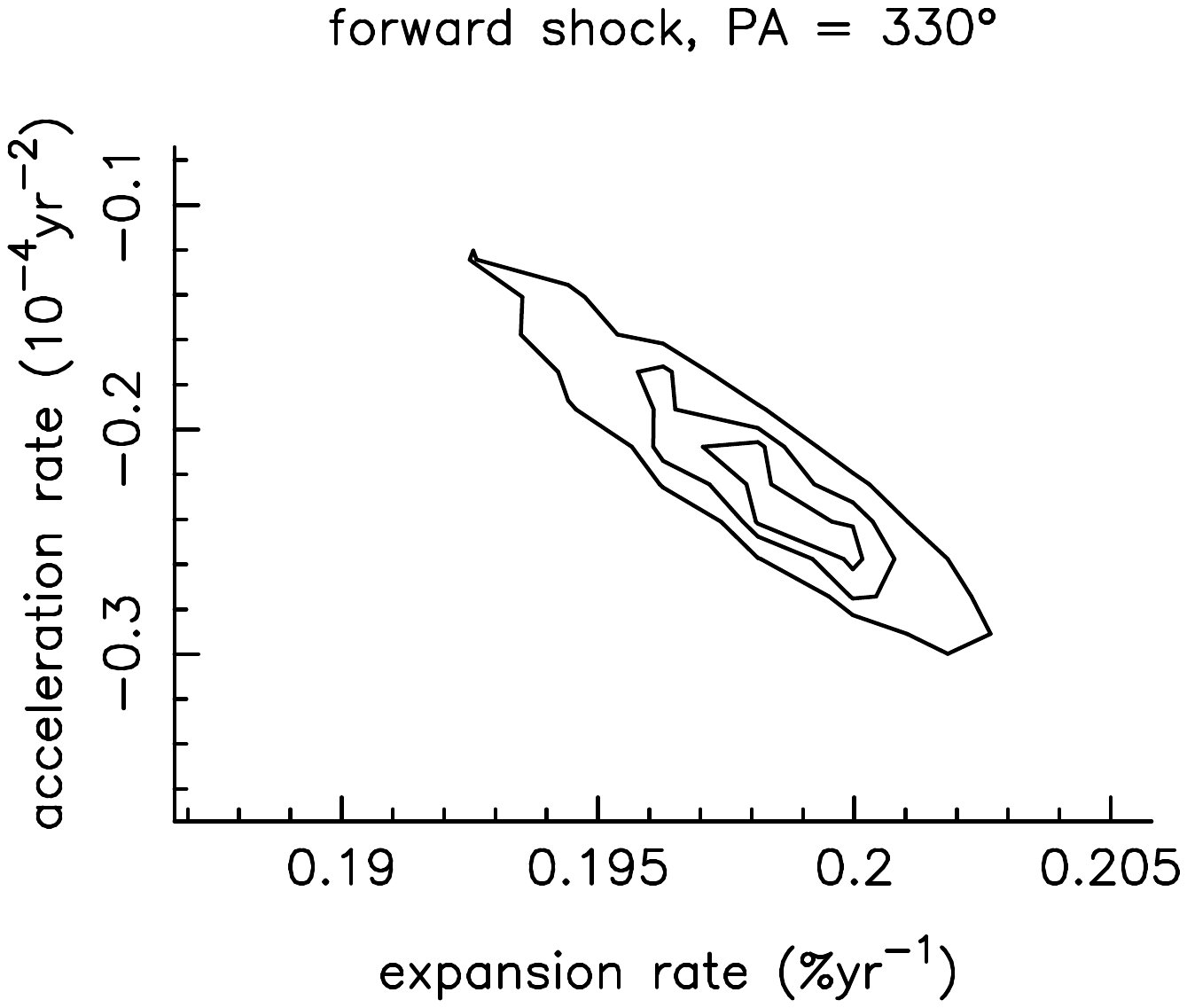}
\includegraphics[trim=50 20 120 30,clip=true,width=0.15\textwidth]{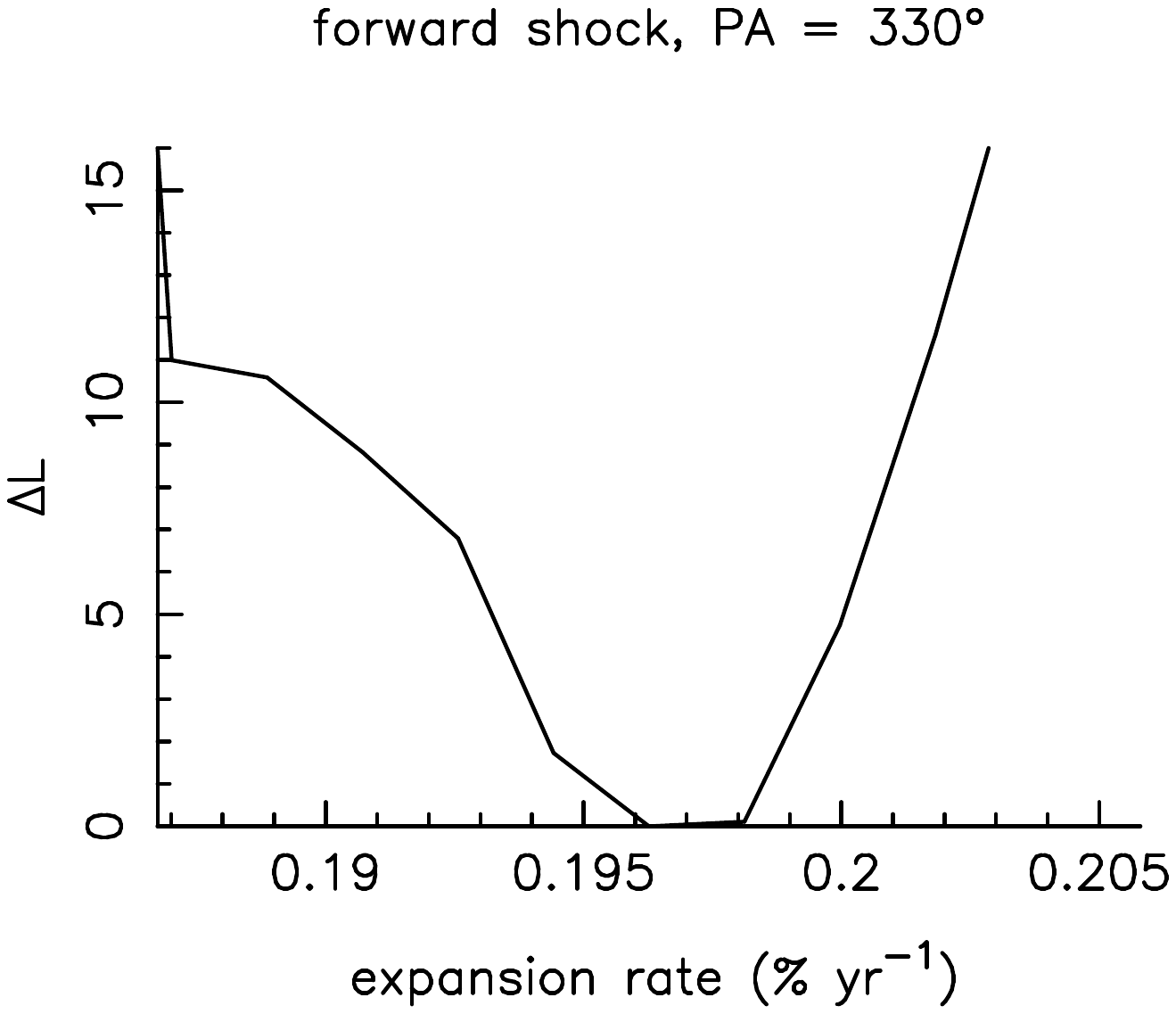}
\includegraphics[trim=50 20 120 30,clip=true,width=0.15\textwidth]{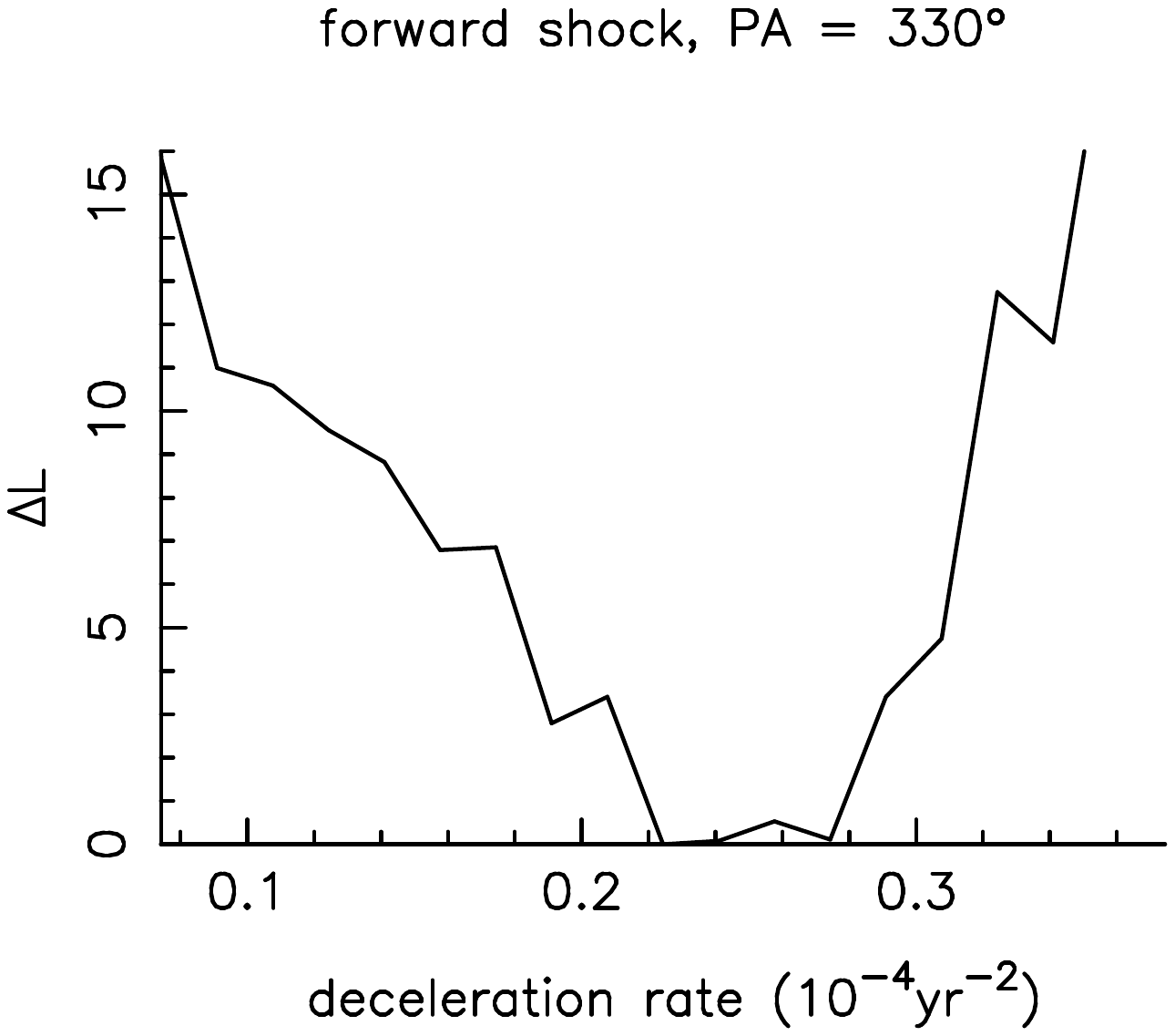}
\includegraphics[trim=50 20 120 30,clip=true,width=0.15\textwidth]{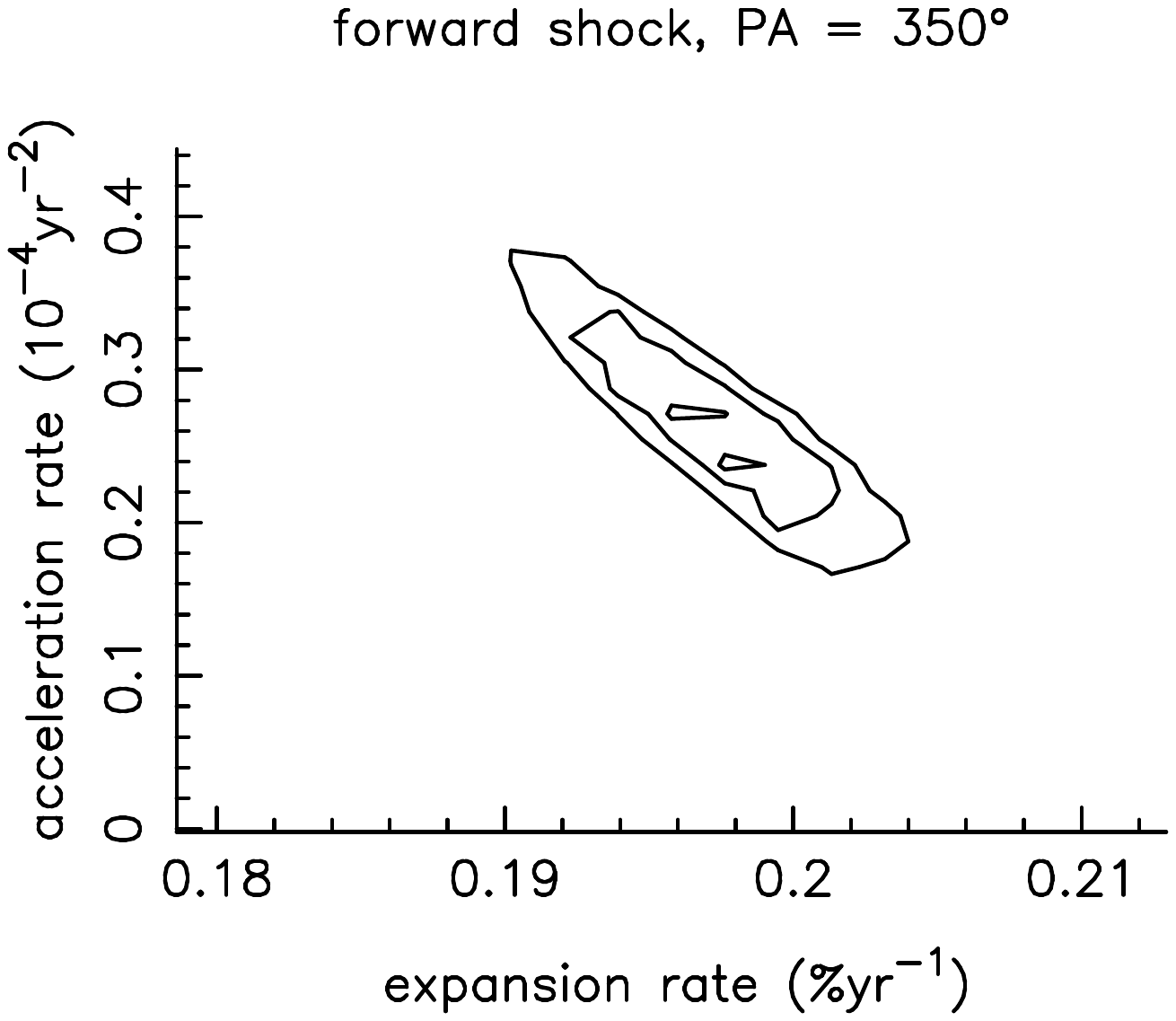}
\includegraphics[trim=50 20 120 30,clip=true,width=0.15\textwidth]{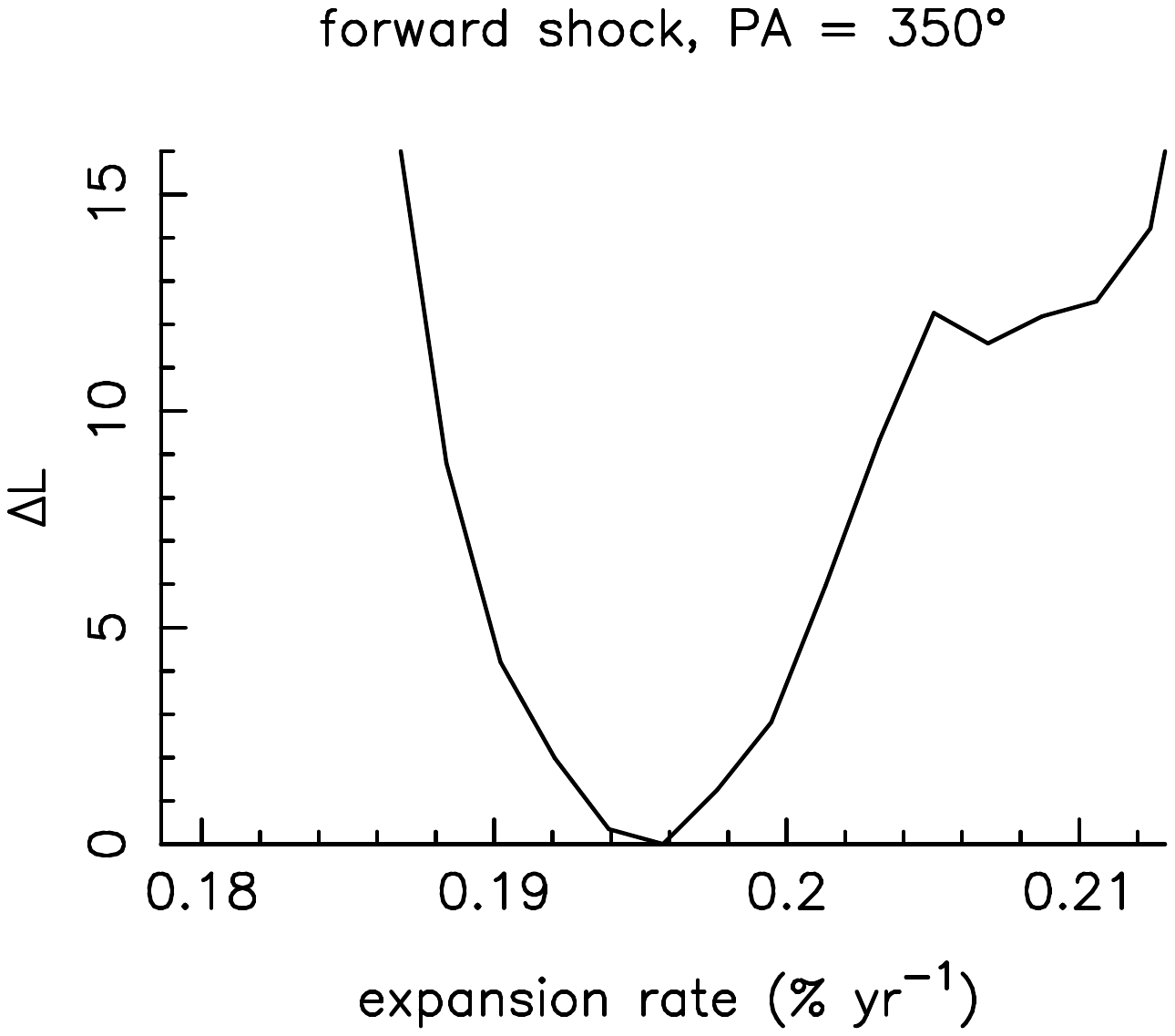}
\includegraphics[trim=50 20 120 30,clip=true,width=0.15\textwidth]{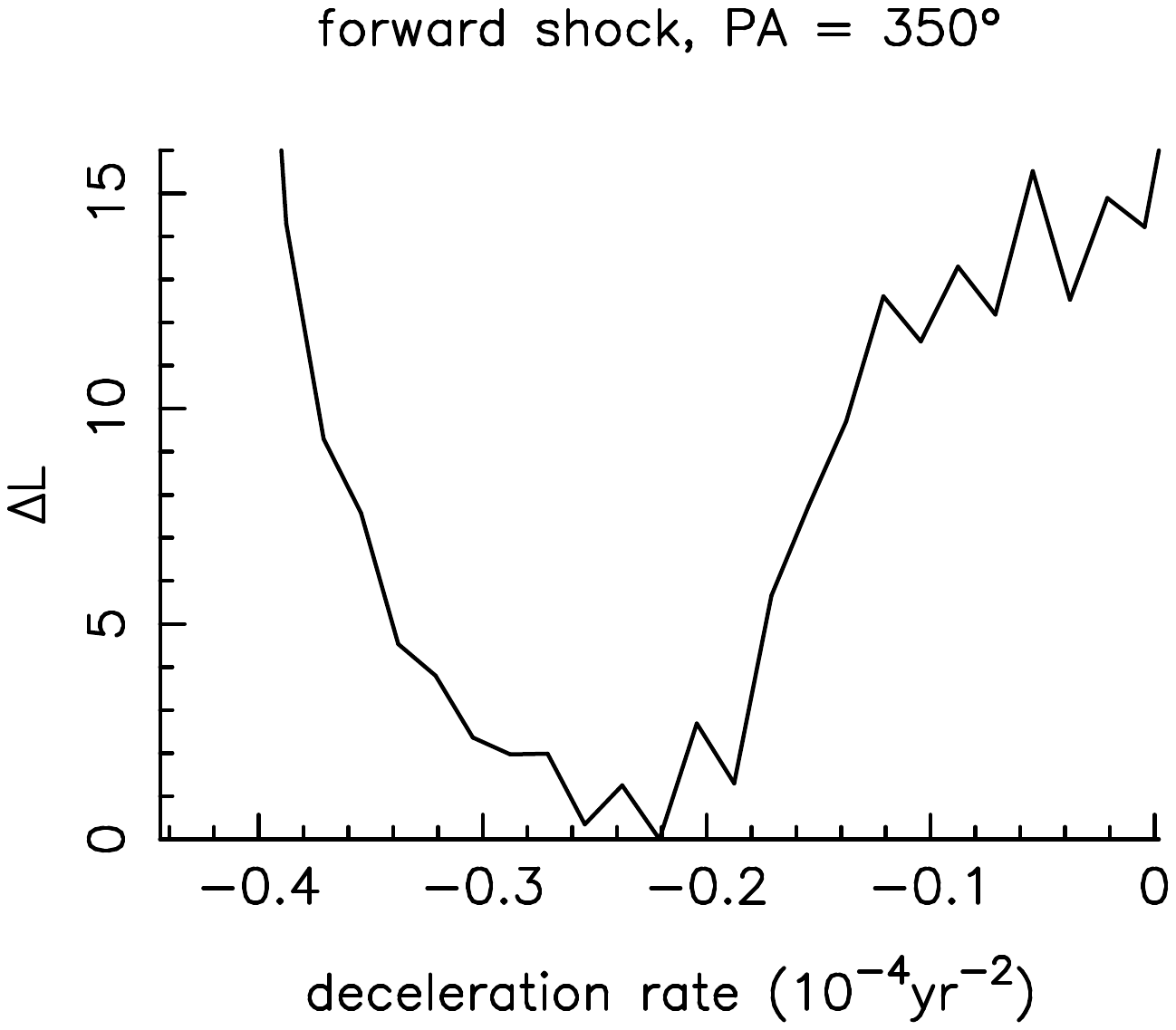}
}

\caption{Log-likelihood ($\Delta L= -2\Delta \ln \mathcal{L}$) distributions relative to the minimum values given as a contour plot with levels $\Delta L=1,4,9$ (corresponding to $1\sigma,2\sigma$,3$\sigma$ confidence regions) and the distribution of $\Delta L$ as function of $a$ and the deceleration parameter $-b$, marginalised over the other parameter.
\label{fig:logLfs}
}
\end{figure*}

\begin{figure*}
\centerline{
\includegraphics[trim=50 20 120 30,clip=true,width=0.15\textwidth]{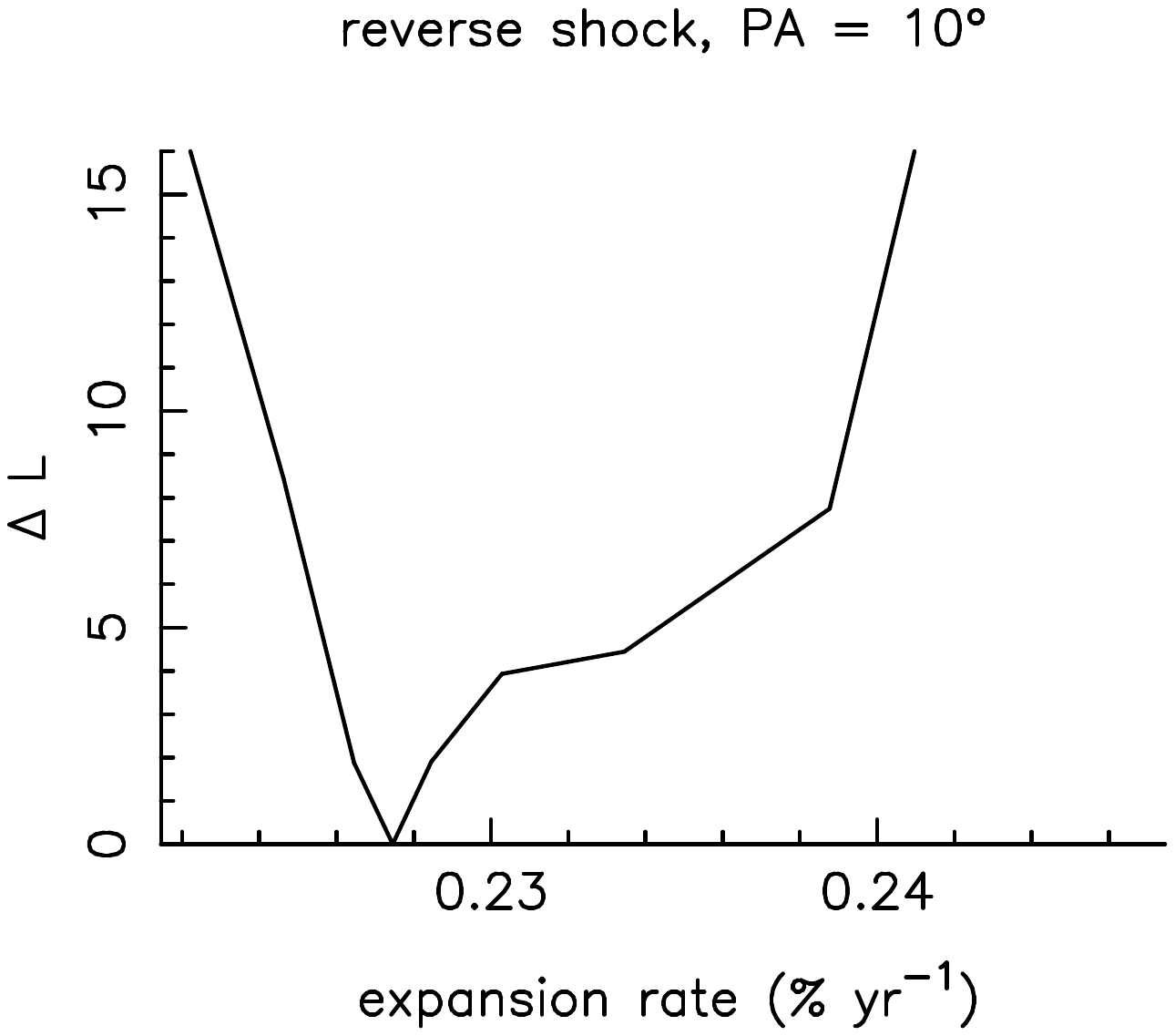}
\includegraphics[trim=50 20 120 30,clip=true,width=0.15\textwidth]{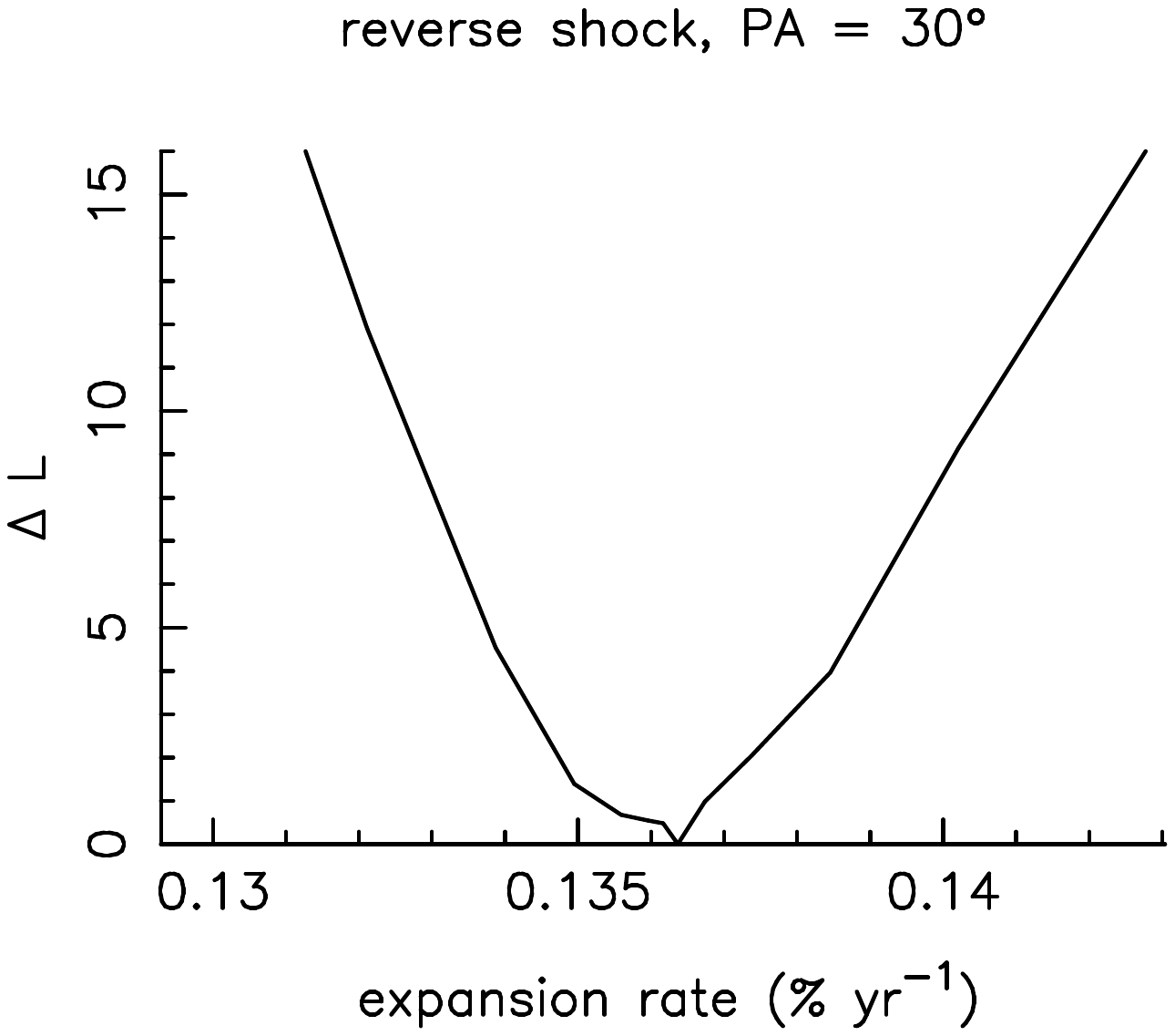}
\includegraphics[trim=50 20 120 30,clip=true,width=0.15\textwidth]{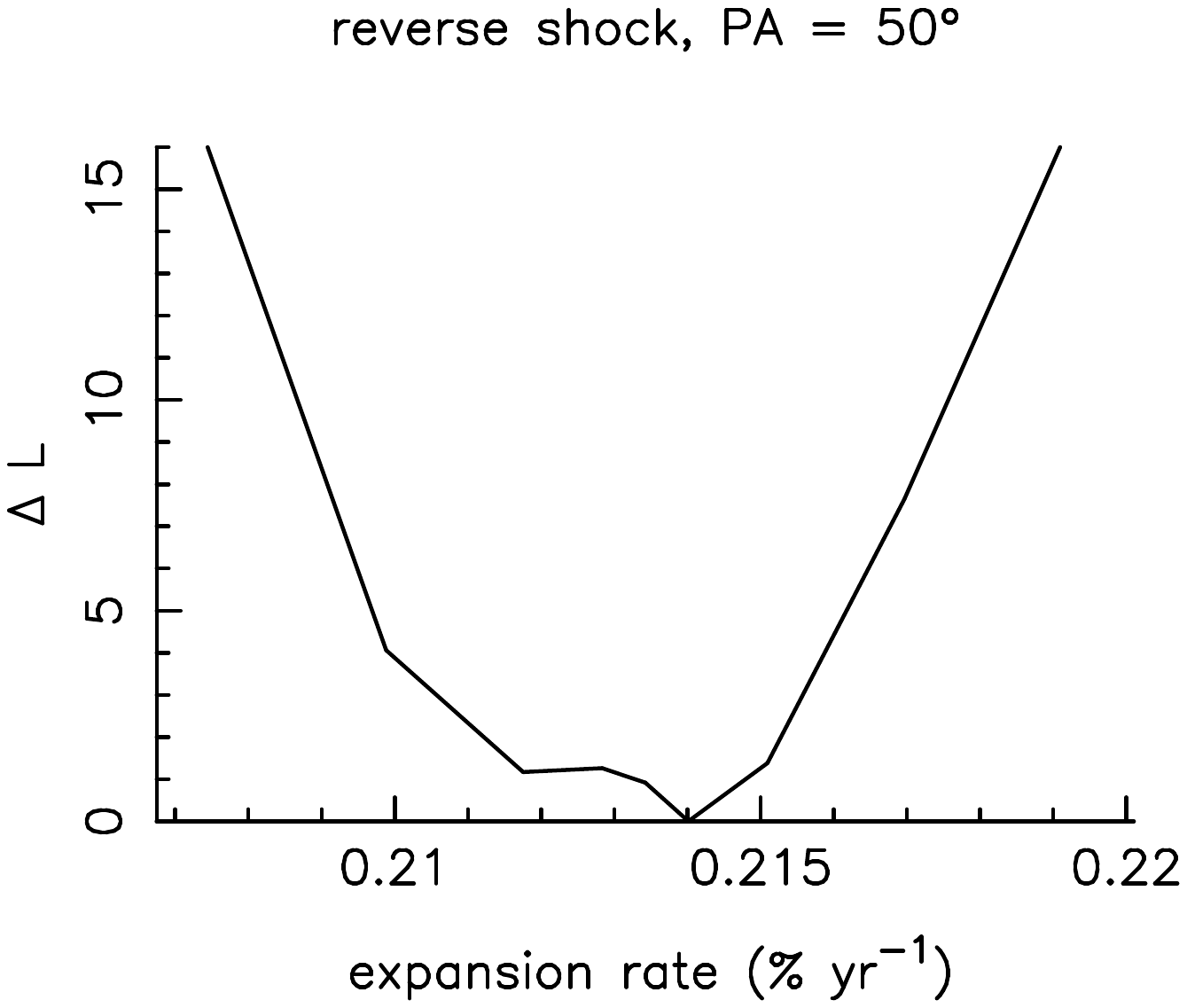}
\includegraphics[trim=50 20 120 30,clip=true,width=0.15\textwidth]{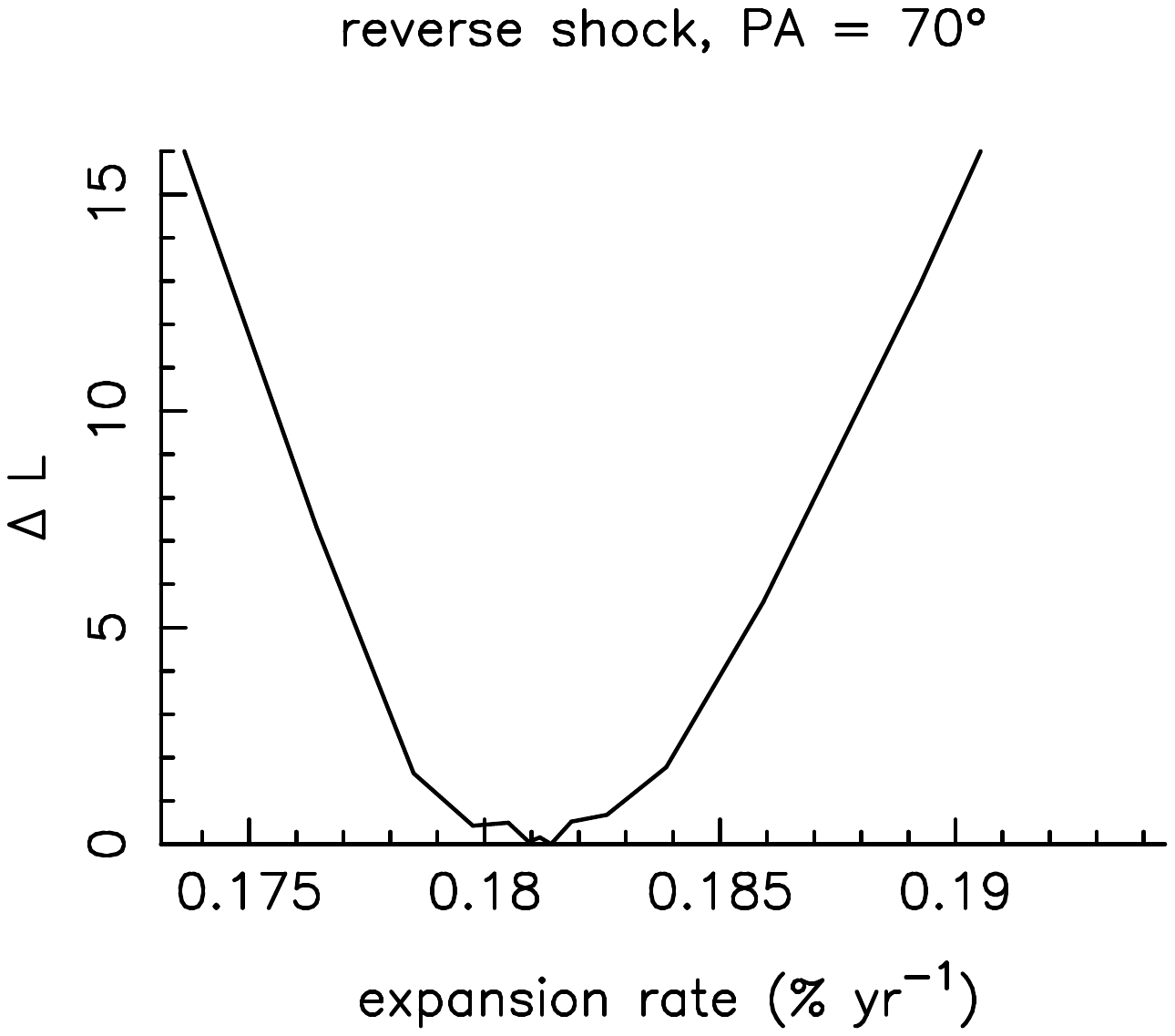}
\includegraphics[trim=50 20 120 30,clip=true,width=0.15\textwidth]{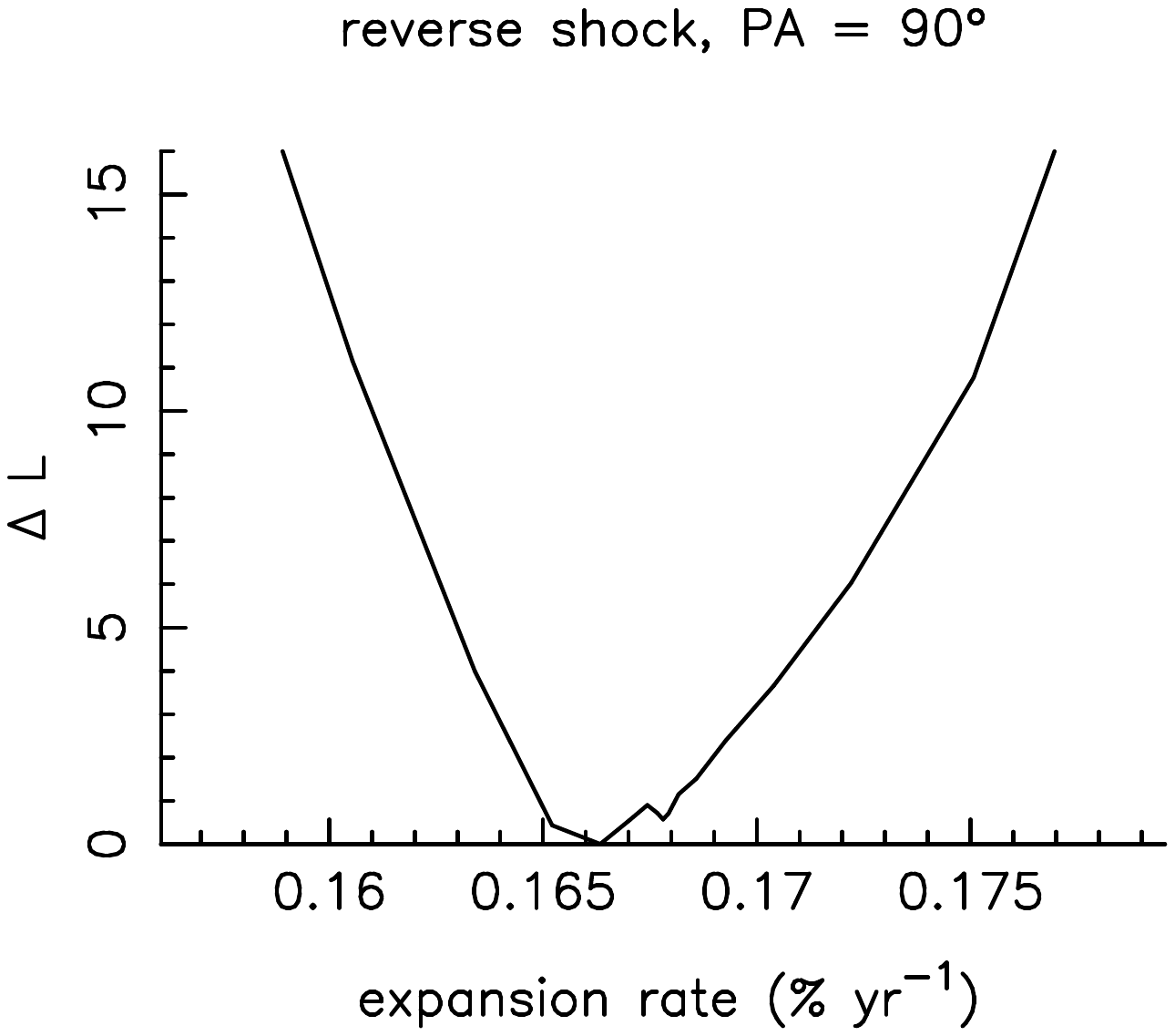}
\includegraphics[trim=50 20 120 30,clip=true,width=0.15\textwidth]{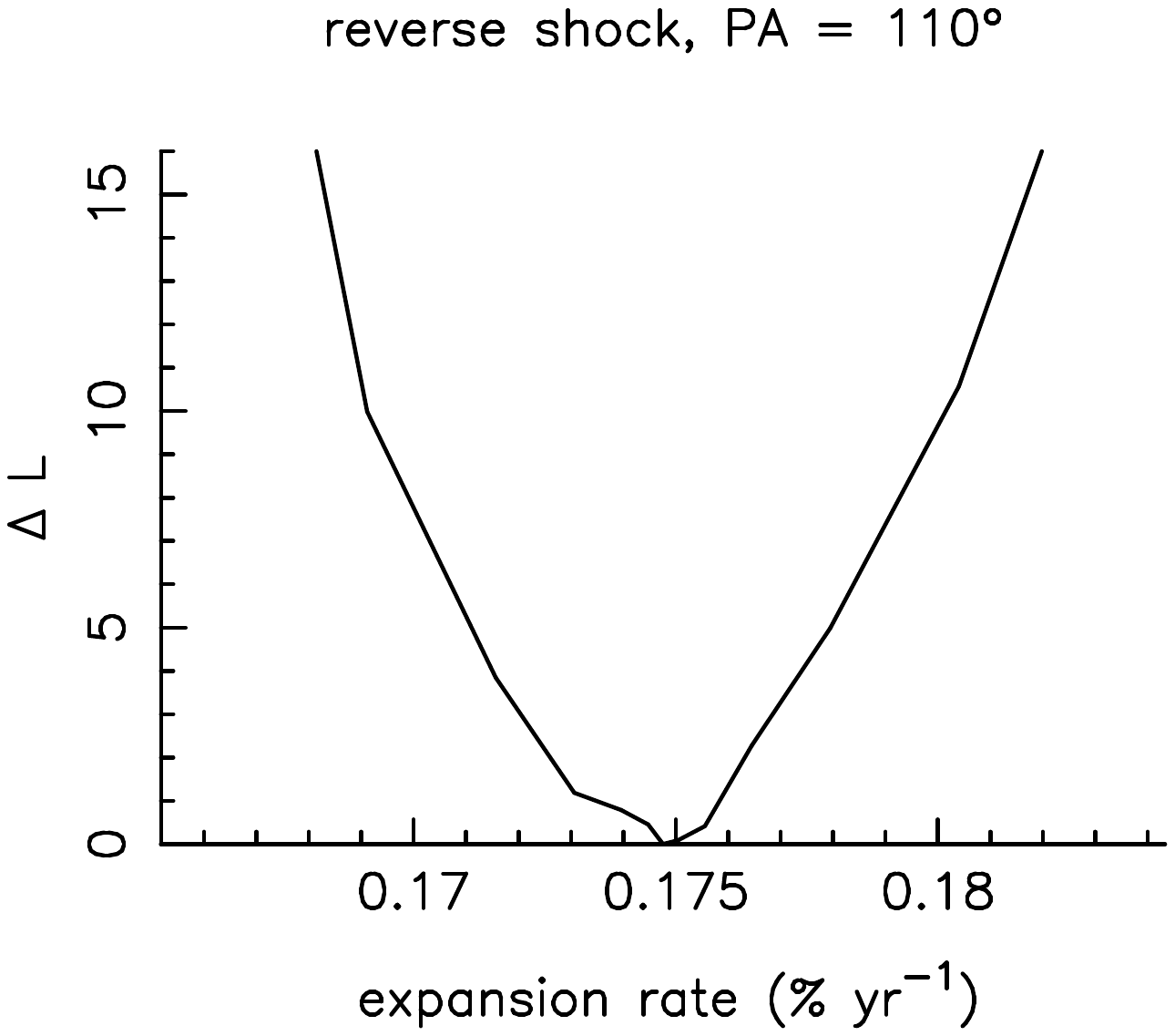}
}
\centerline{
\includegraphics[trim=50 20 120 30,clip=true,width=0.15\textwidth]{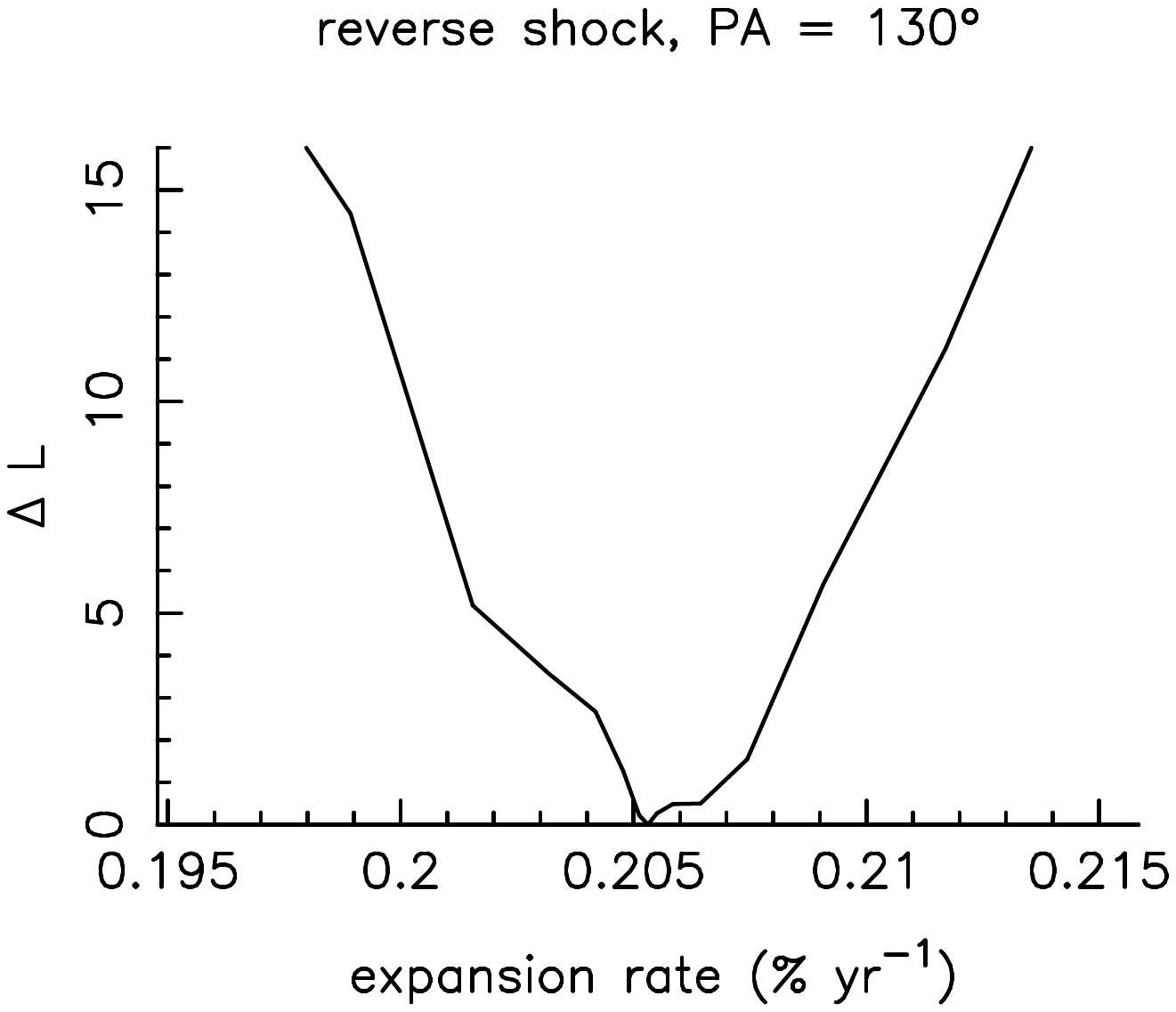}
\includegraphics[trim=50 20 120 30,clip=true,width=0.15\textwidth]{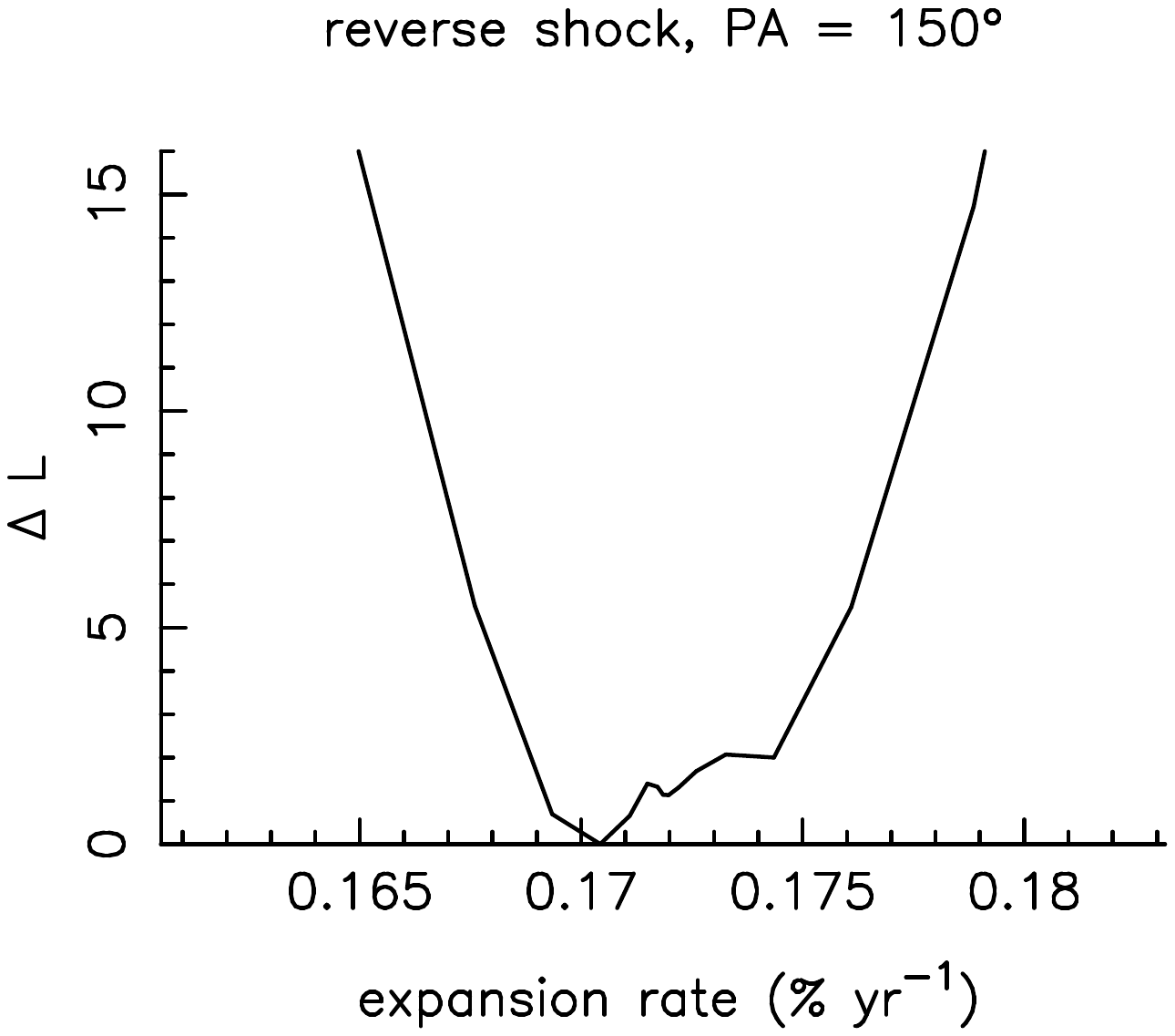}
\includegraphics[trim=50 20 120 30,clip=true,width=0.15\textwidth]{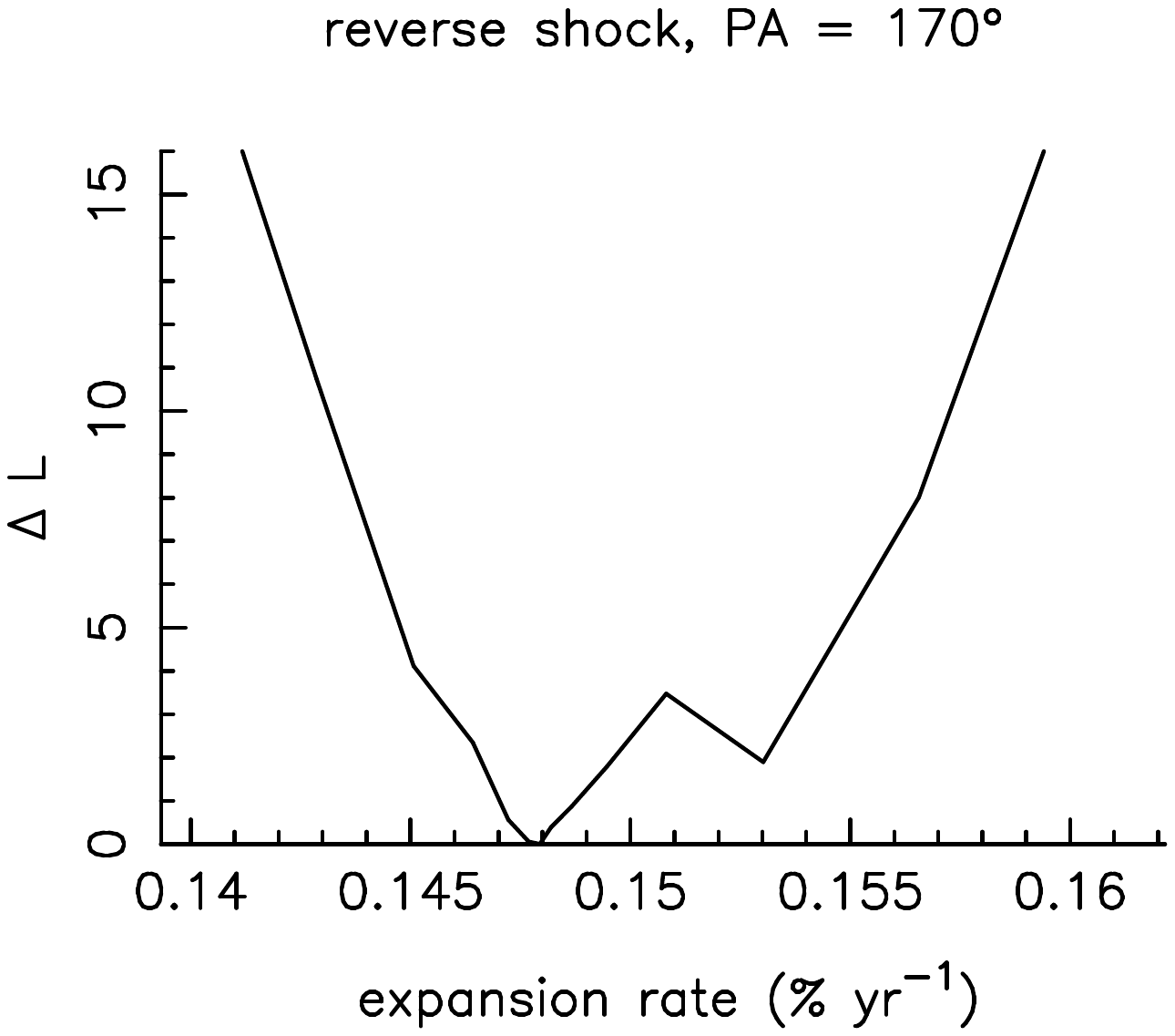}
\includegraphics[trim=50 20 120 30,clip=true,width=0.15\textwidth]{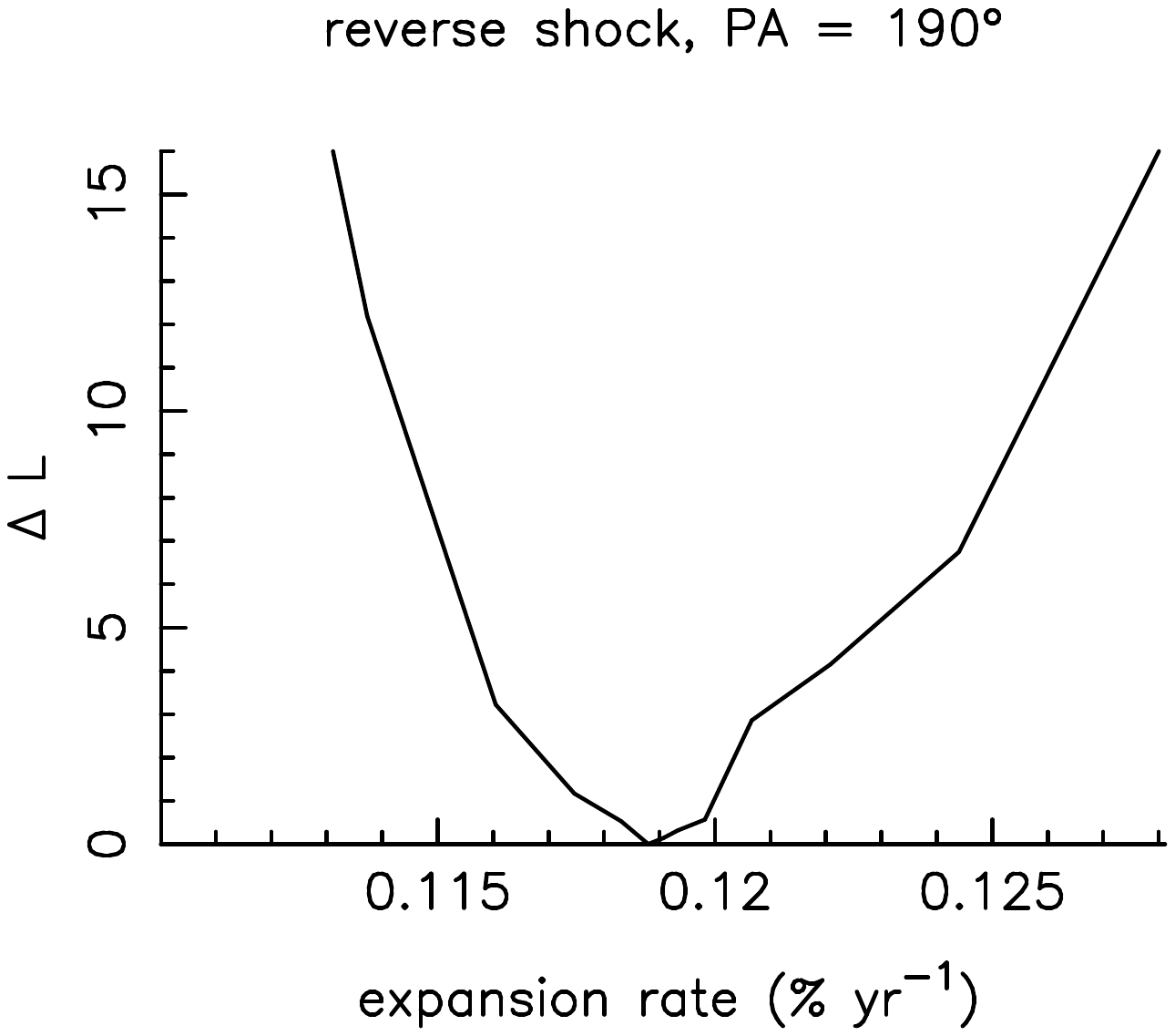}
\includegraphics[trim=50 20 120 30,clip=true,width=0.15\textwidth]{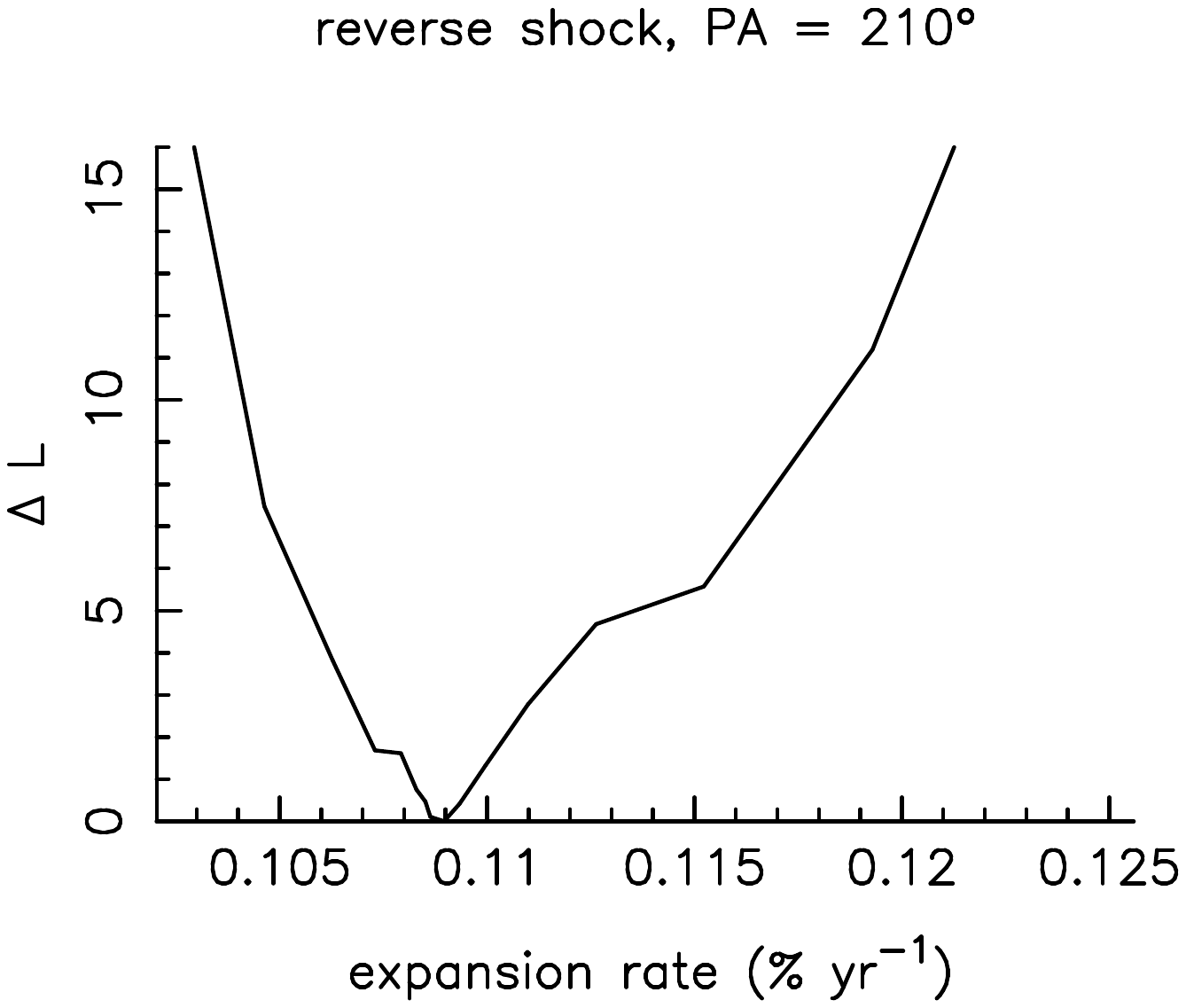}
\includegraphics[trim=50 20 120 30,clip=true,width=0.15\textwidth]{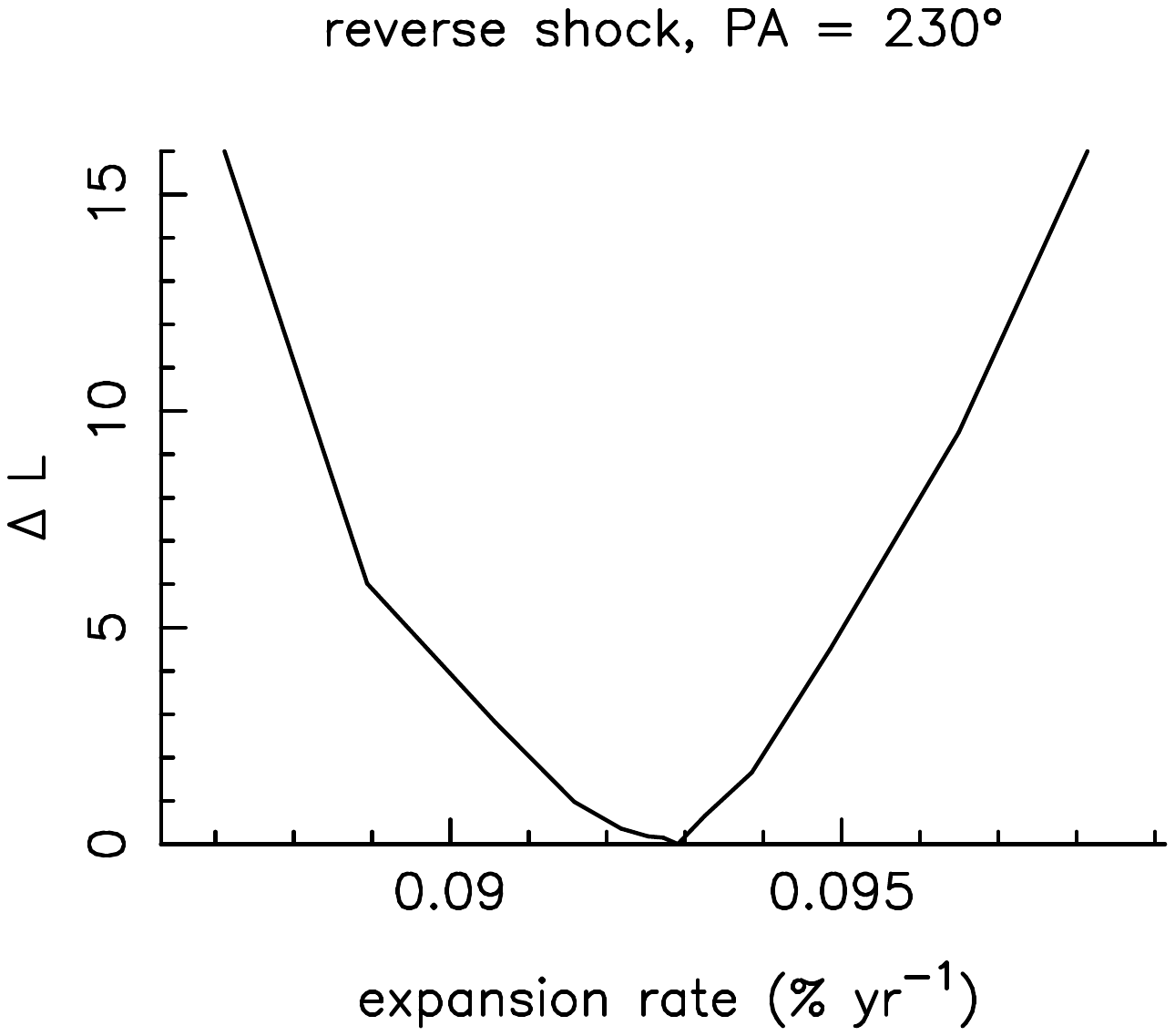}
}
\centerline{
\includegraphics[trim=50 20 120 30,clip=true,width=0.15\textwidth]{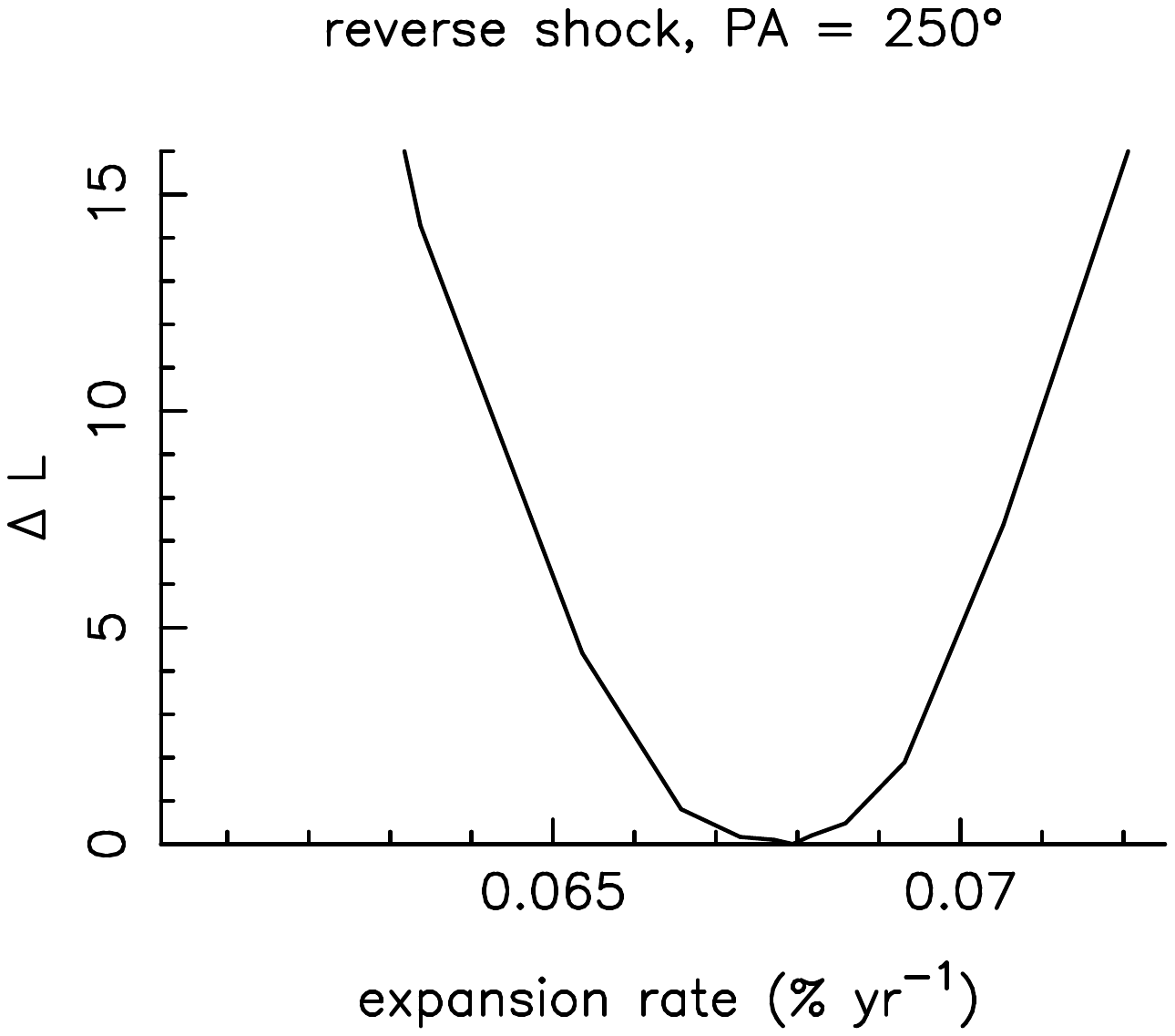}
\includegraphics[trim=50 20 120 30,clip=true,width=0.15\textwidth]{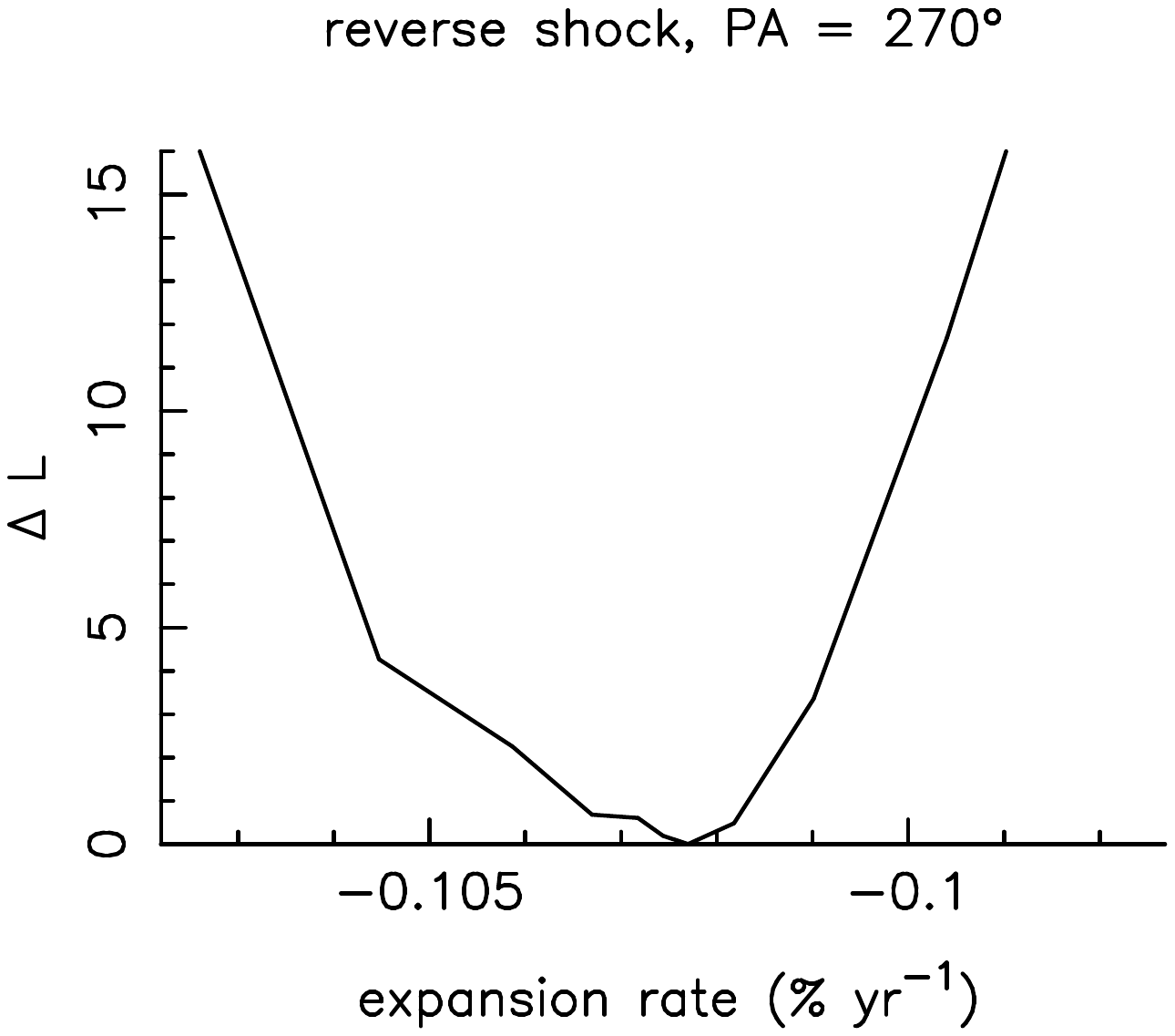}
\includegraphics[trim=50 20 120 30,clip=true,width=0.15\textwidth]{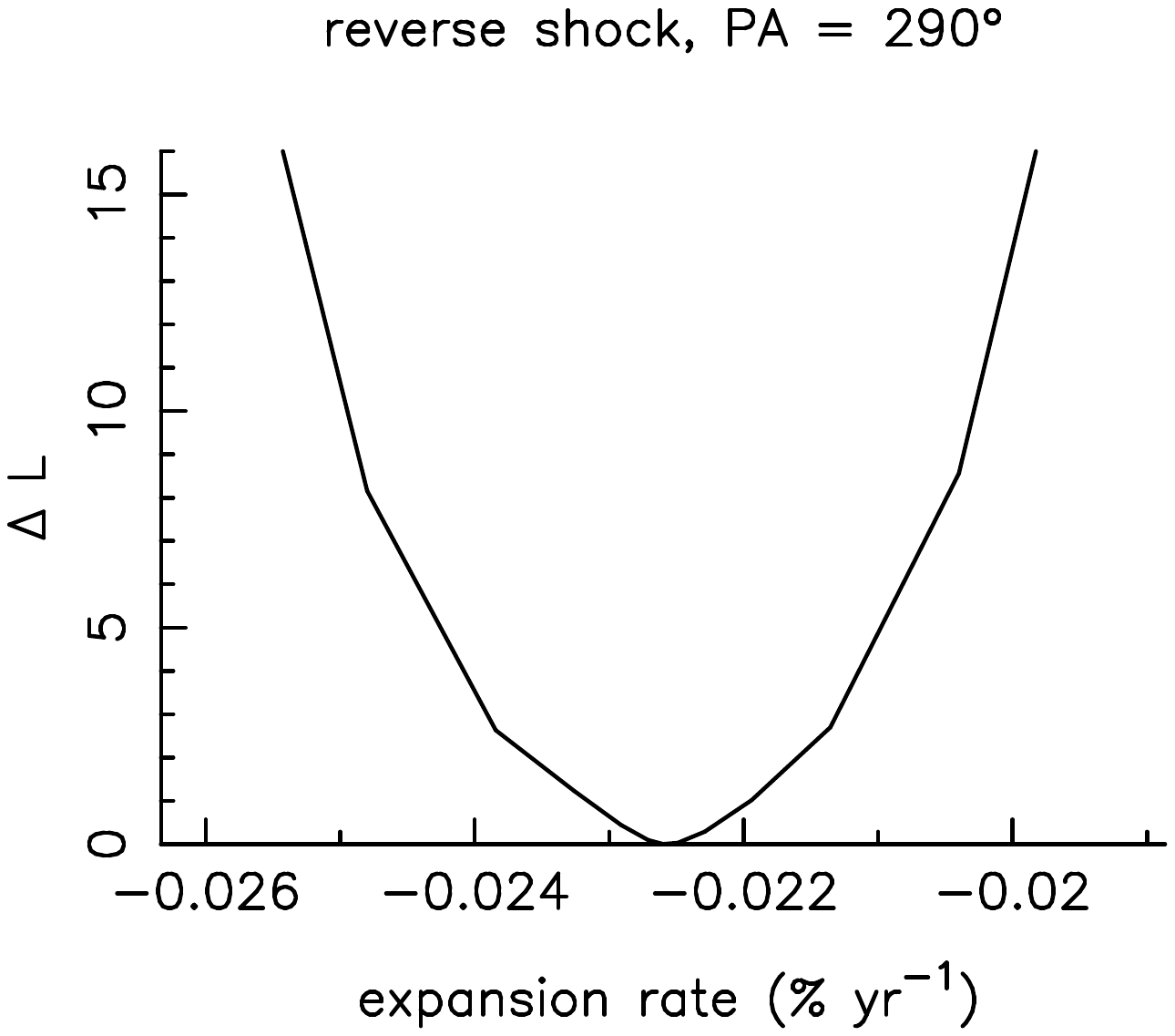}
\includegraphics[trim=50 20 120 30,clip=true,width=0.15\textwidth]{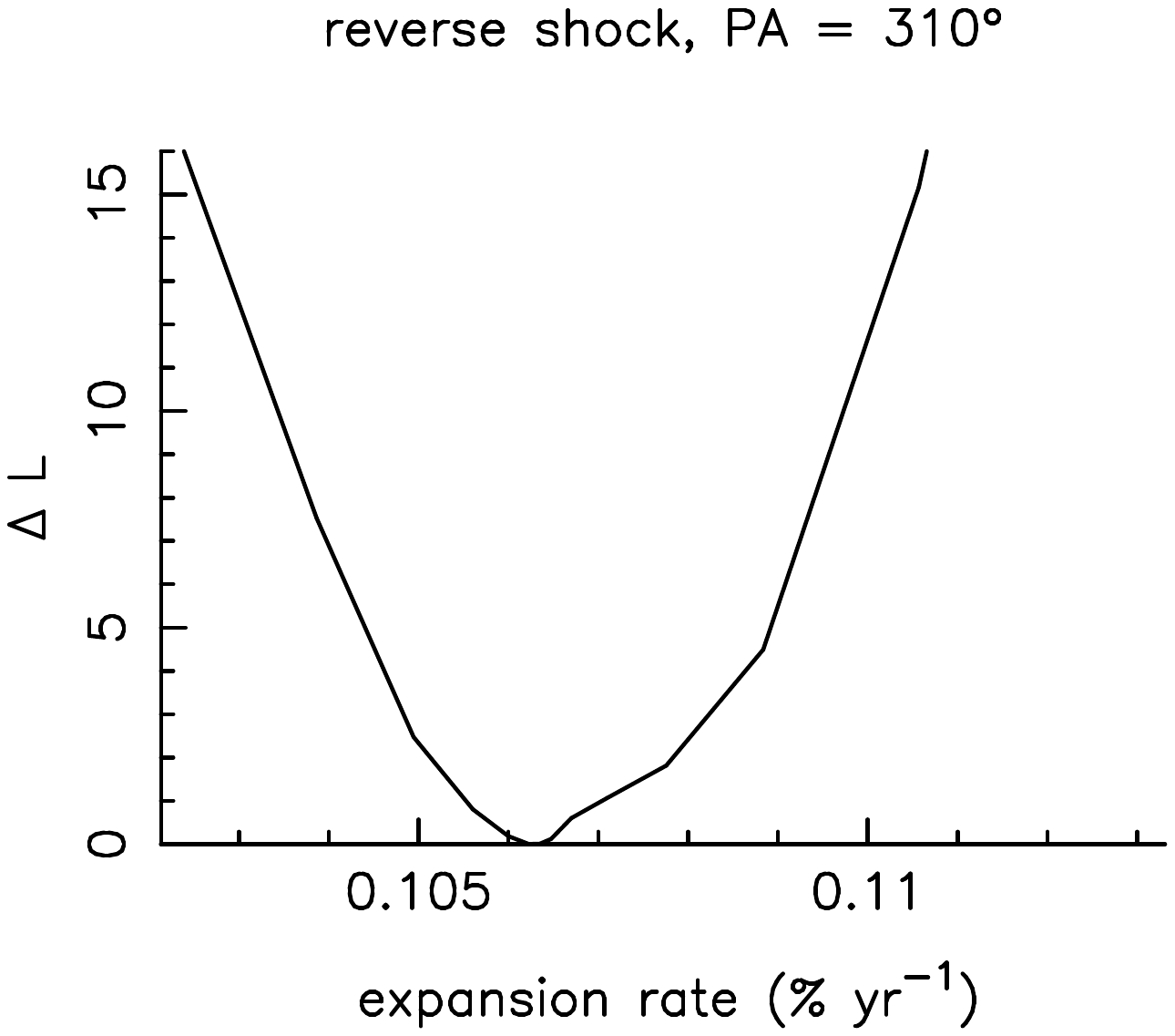}
\includegraphics[trim=50 20 120 30,clip=true,width=0.15\textwidth]{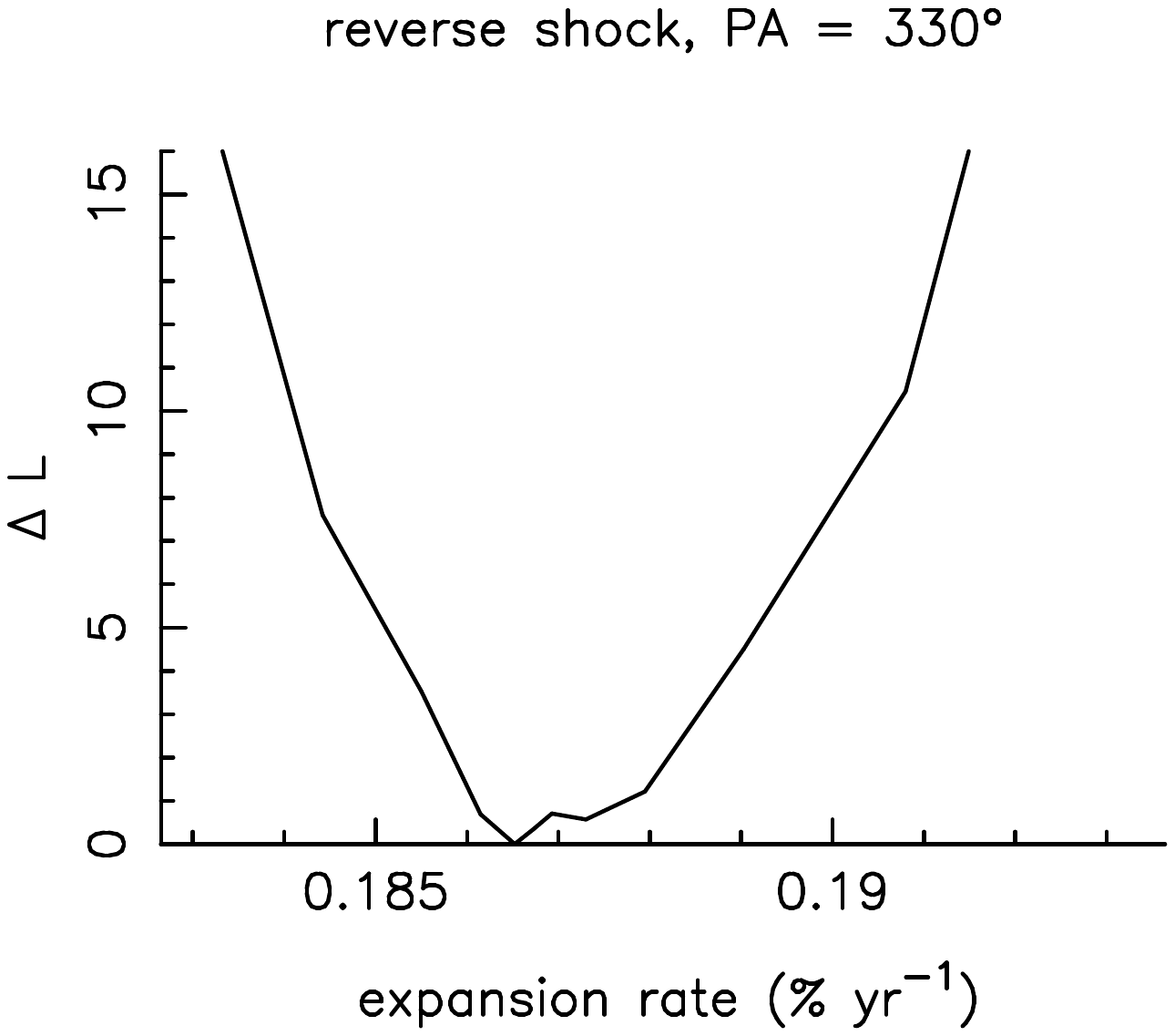}
\includegraphics[trim=50 20 120 30,clip=true,width=0.15\textwidth]{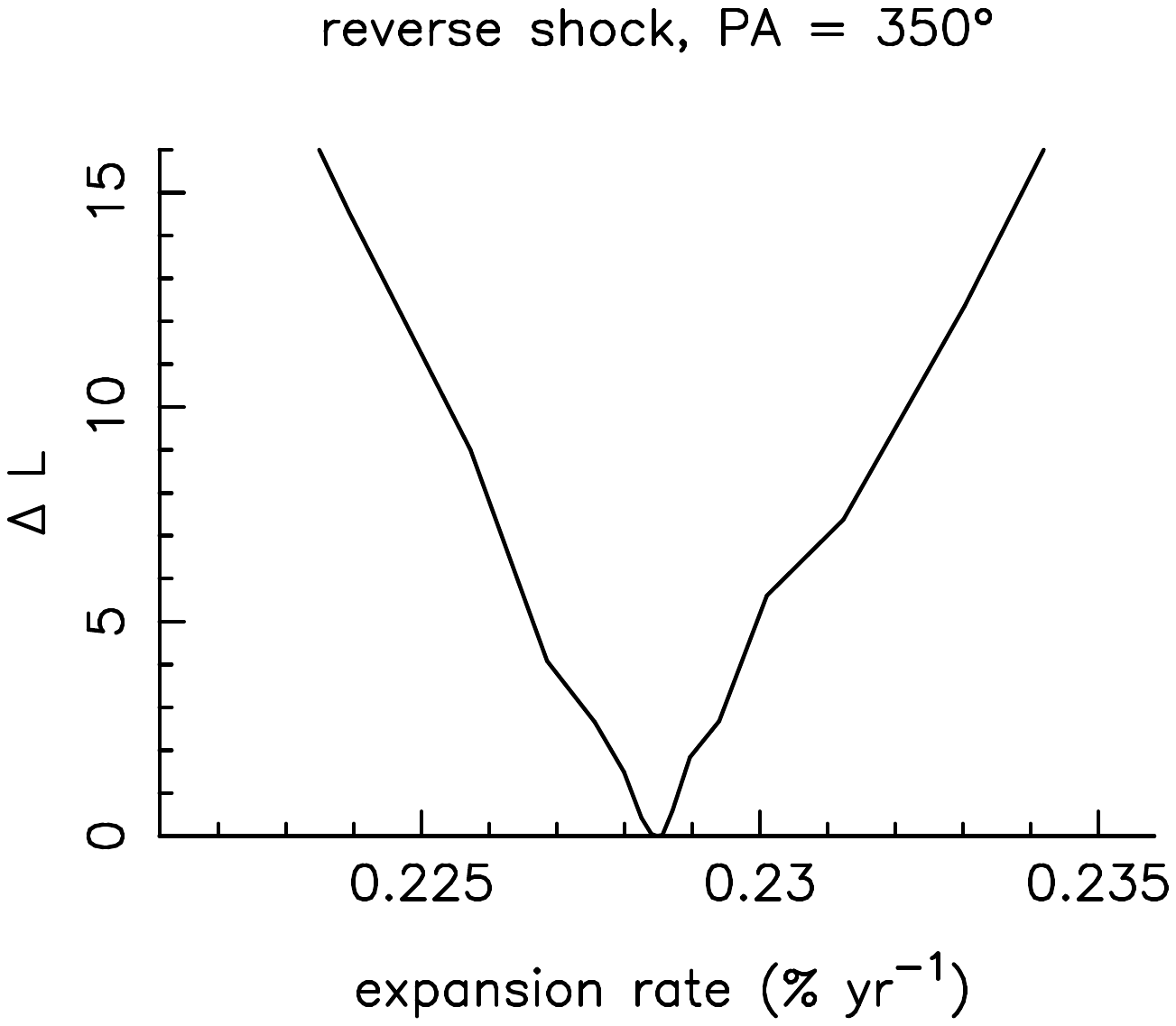}
}
\caption{Log-likelihood ($\Delta L= -2\Delta \ln \mathcal{L}$) distributions  relative to the minimum values for the reverse shock regions as a function of the expansion rate $a$.
\label{fig:logLrs}
}
\end{figure*}

In Figure~\ref{fig:logLfs} and \ref{fig:logLrs} shows the log-likelihood profiles, $\Delta L$, for respectively the forward and reverse shock regions.
For the reverse shock $L$ only depends on the expansion rate $a$. For the forward shock a two-dimensional fit was employed, fitting
both the expansion and acceleration rates $a$ and $b$, and both the two-dimensional likelihood contours and the likelihood curves as a function of $a$ and $b$
are shown, marginalised over respectively $b$ and $a$. The curves for $b$ appear rather ragged, which is caused by the finite stepsize in $b$.

\begin{figure*}
    \centerline{
     \includegraphics[trim=70 90 80 263,clip=true,width=0.25\textwidth]{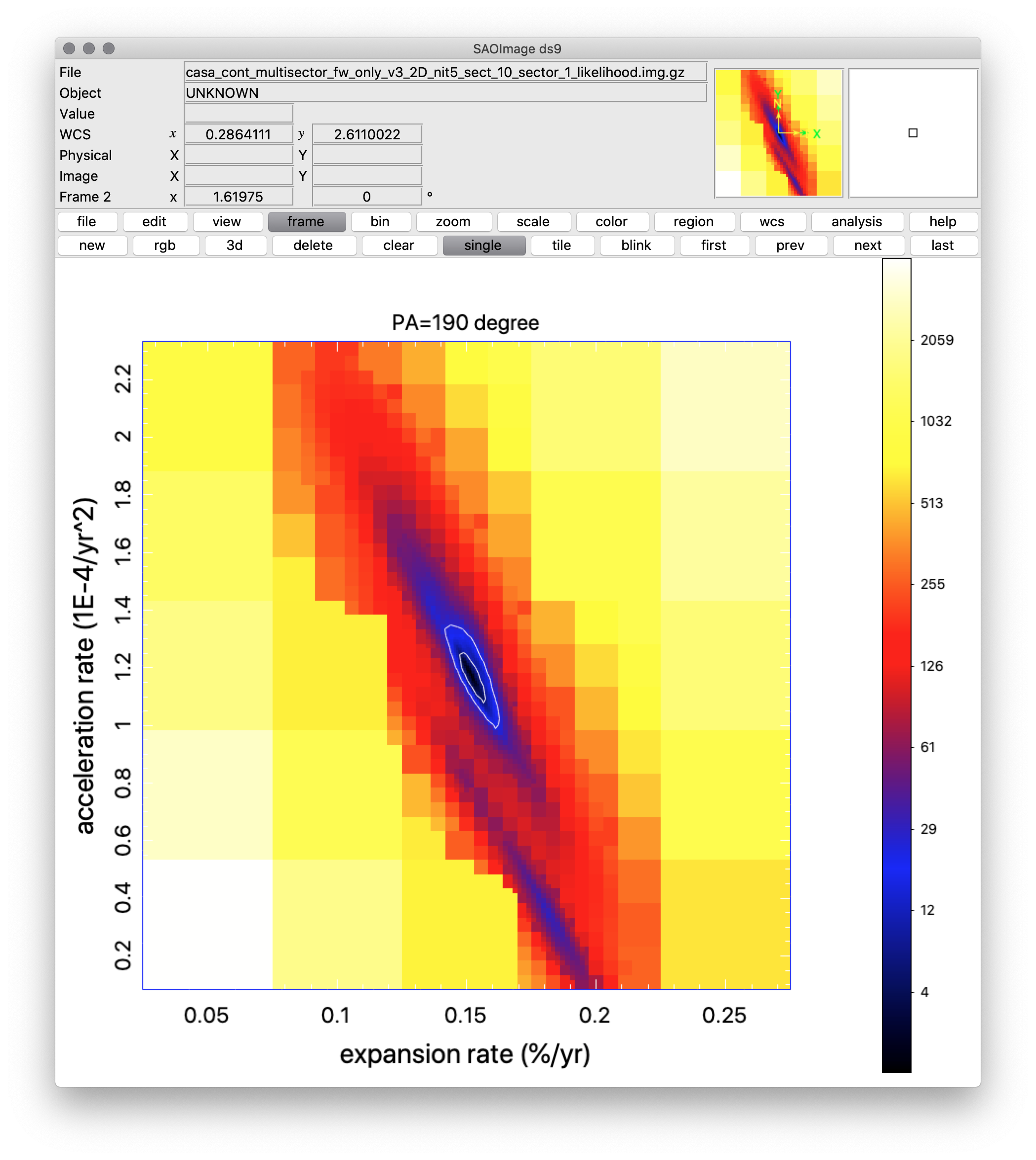}
    \includegraphics[trim=70 90 80 263,clip=true,width=0.25\textwidth]{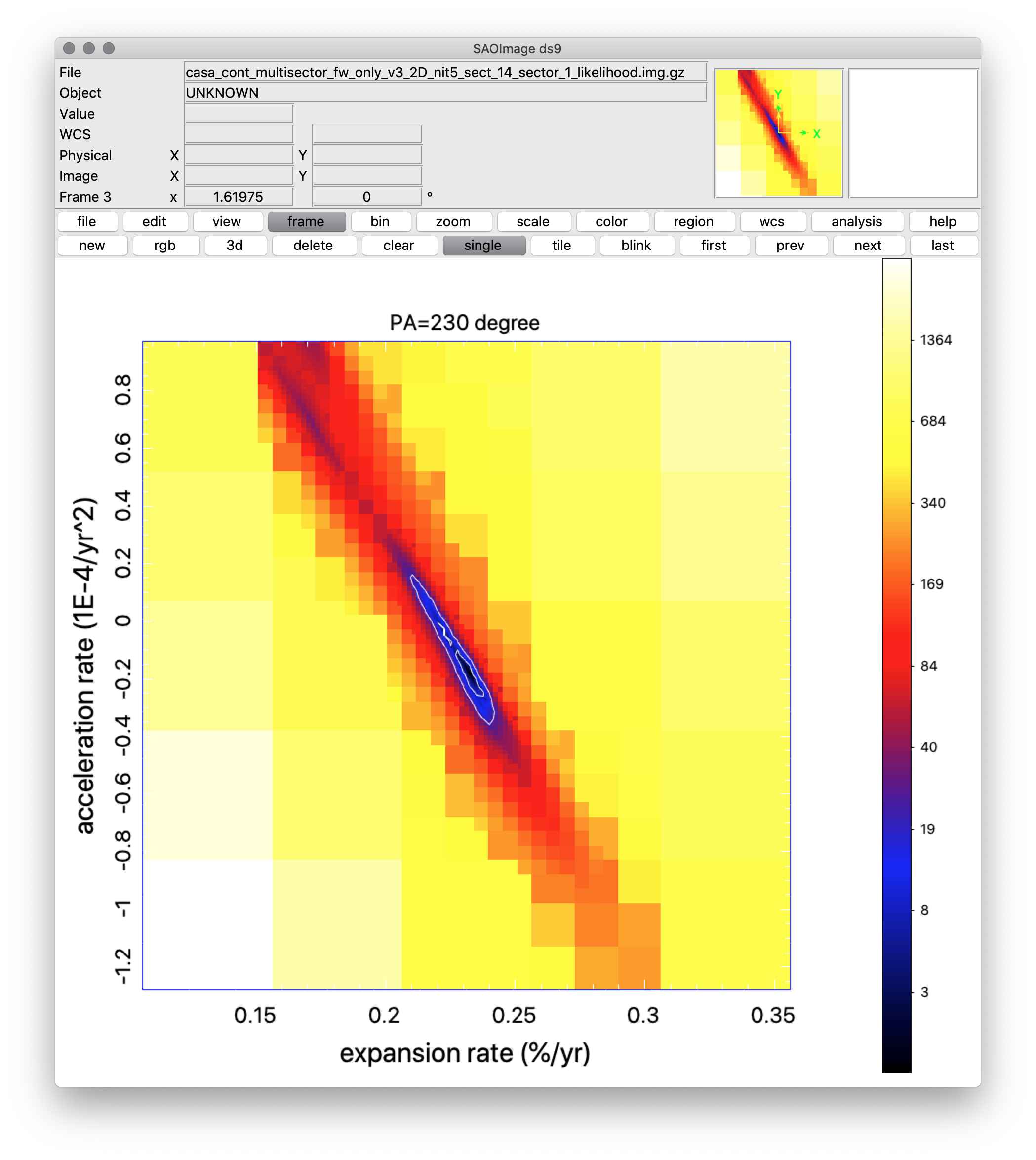}
       \includegraphics[trim=70 90 80 263,clip=true,width=0.25\textwidth]{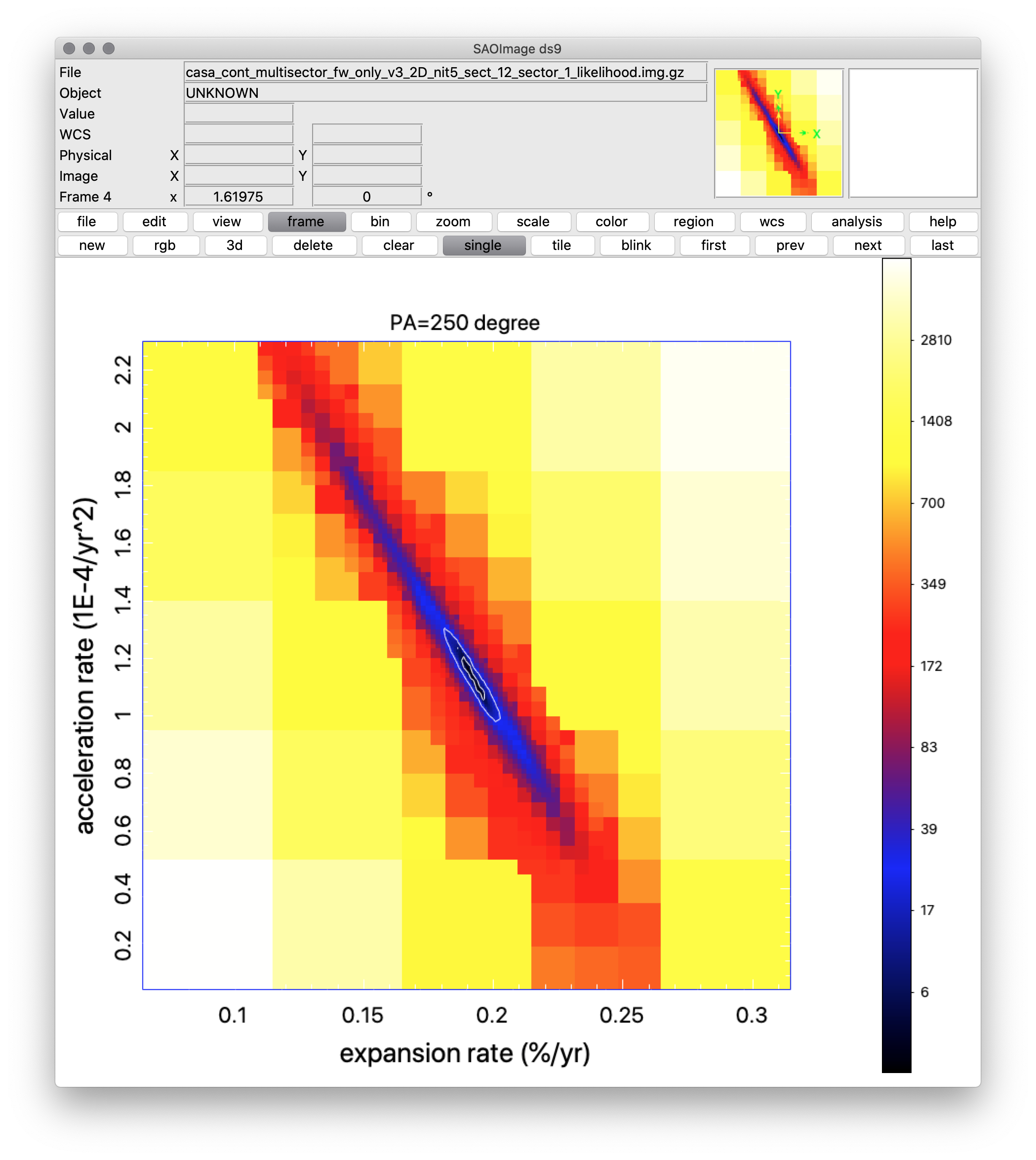}
    \includegraphics[trim=70 90 80 263,clip=true,width=0.25\textwidth]{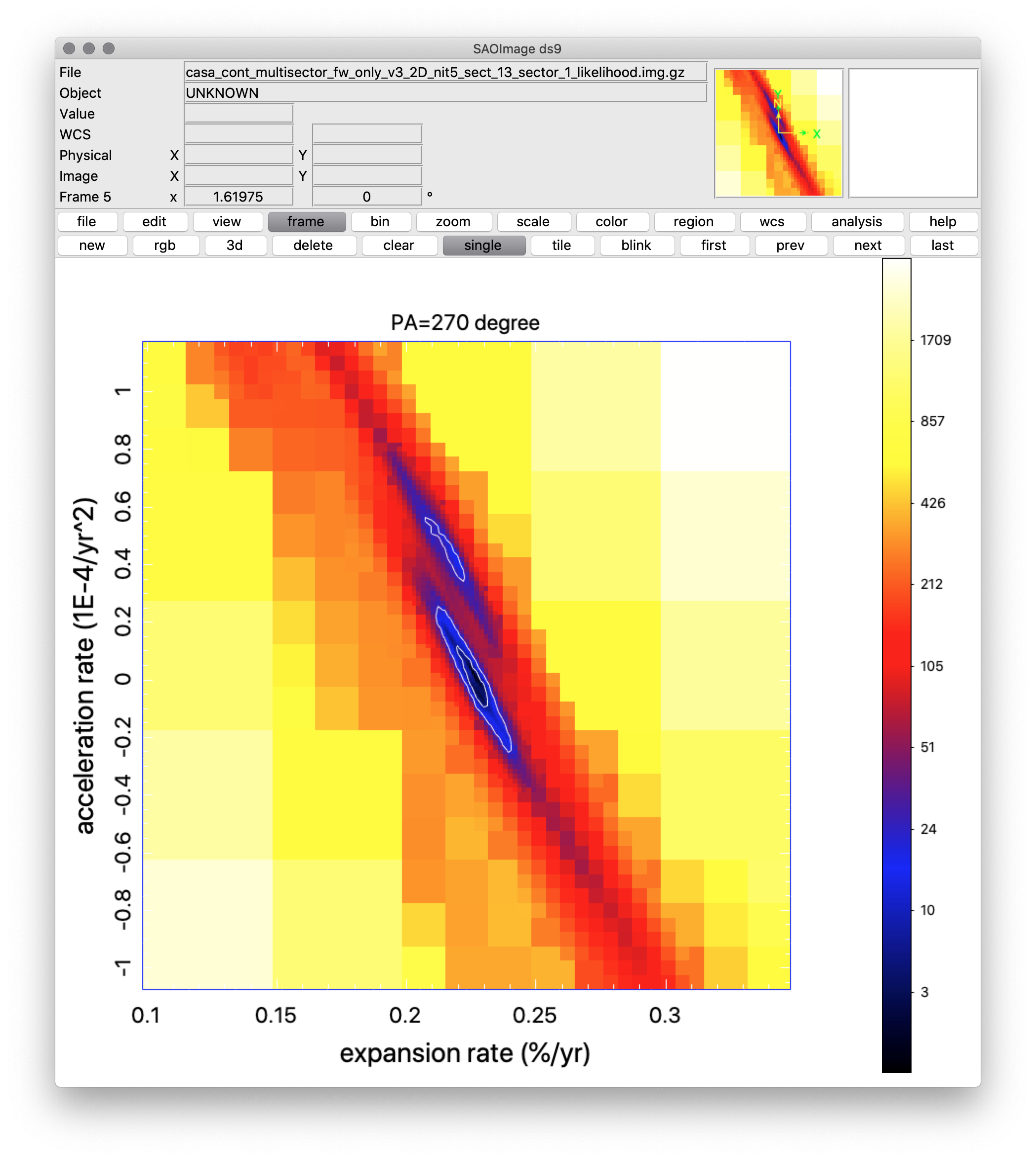}    
    }
  \caption{\label{fig:likmaps_hires}
    Likelihood maps, i.e. a map of  $\Delta L$ as function of
    $a$ (horizontal) and $b$ (vertical)
    for four different sectors associated with  forward shock region.
    The maps were made at a higher resolution in $a$ and $b$ than for the
    general fits.
    The sectors with $PA=190^\circ,230^\circ, 270^\circ$were chosen to show the presence of some interesting
    substructure, with subminima that may indicate filaments deviating from the best-fit solutions.
    $PA=250^\circ$ provides a more
    regular example.
    The contours are for $\Delta L=9,25$, corresponding to 3$\sigma$ and 5$\sigma$ confidence
    regions. 
      }
\end{figure*}

In Figure~\ref{fig:likmaps_hires} color coded log-likelihood maps--$\Delta L$ as function of $a$ and $b$---are shown, which were calculated at a smaller stepsize
in $a$ and $b$. Log-likelihood maps of those forward shock regions are shown that appear to have substructure in the  maps, possibly caused by filaments with different $a$ and $b$ colocated
within the same region. The map with $PA=250^\circ$ is shown as a counter example, displaying a rather smooth likelihood map.

\begin{table*}
\centering
\caption{Expansion measurements for the forward shock region, but here without fitting for deceleration ($b=0$).
 \label{tab:fs_alternative}}
\begin{tabular}{rccccr}\hline\hline\noalign{\smallskip}
PA & $f$ & $\tau_{\rm exp}$ & $m$ & $v_{\rm s,obs}$ & $L$\\ 
{[$^{\circ}$]} & {[\% yr$^{-1}$]} & {[yr]} & & {[km\,s$^{-1}$]}\\\noalign{\smallskip}\hline\noalign{\smallskip}
10 & 0.1797 $\pm$ 0.0008 & $556.4^{+2.4}_{-2.4}$ & $0.5985\pm 0.0025$ & 4780 $\pm$ 20 & 126866.1\\
30 & 0.2378 $\pm$ 0.0001 & $420.5^{+0.2}_{-0.2}$ & $0.7919\pm 0.0004$ & 6324 $\pm$ 3 & 102435.1\\
50 & 0.1930 $\pm$ 0.0004 & $518.2^{+1.0}_{-1.0}$ & $0.6426\pm 0.0013$ & 5132 $\pm$ 10 & 165858.4\\
70 & 0.1998 $\pm$ 0.0004 & $500.5^{+1.1}_{-1.1}$ & $0.6653\pm 0.0014$ & 5313 $\pm$ 11 & 162904.8\\
90 & 0.1813 $\pm$ 0.0006 & $551.7^{+1.7}_{-1.7}$ & $0.6036\pm 0.0019$ & 4821 $\pm$ 15 & 60876.9\\
110 & 0.2026 $\pm$ 0.0034 & $493.6^{+8.3}_{-8.1}$ & $0.6747\pm 0.0112$ & 5388 $\pm$ 89 & 95754.8\\
130 & 0.2426 $\pm$ 0.0003 & $412.2^{+0.5}_{-0.5}$ & $0.8079\pm 0.0010$ & 6452 $\pm$ 8 & 114888.0\\
150 & 0.2754 $\pm$ 0.0022 & $363.1^{+2.9}_{-2.8}$ & $0.9170\pm 0.0072$ & 7323 $\pm$ 58 & 64908.0\\
170 & 0.2578 $\pm$ 0.0004 & $387.9^{+0.6}_{-0.6}$ & $0.8585\pm 0.0014$ & 6856 $\pm$ 11 & 52894.9\\
190 & 0.2013 $\pm$ 0.0030 & $496.7^{+7.5}_{-7.3}$ & $0.6704\pm 0.0100$ & 5354 $\pm$ 80 & 63857.5\\
210 & 0.1900 $\pm$ 0.0028 & $526.4^{+8.0}_{-7.8}$ & $0.6326\pm 0.0095$ & 5052 $\pm$ 76 & 67674.0\\
230 & 0.2202 $\pm$ 0.0006 & $454.1^{+1.3}_{-1.3}$ & $0.7333\pm 0.0020$ & 5857 $\pm$ 16 & 53435.7\\
250 & 0.2541 $\pm$ 0.0003 & $393.6^{+0.4}_{-0.4}$ & $0.8460\pm 0.0009$ & 6756 $\pm$ 7 & 97440.0\\
270 & 0.2271 $\pm$ 0.0008 & $440.3^{+1.5}_{-1.5}$ & $0.7564\pm 0.0026$ & 6040 $\pm$ 21 & 114109.1\\
290 & 0.2648 $\pm$ 0.0008 & $377.7^{+1.1}_{-1.1}$ & $0.8816\pm 0.0026$ & 7041 $\pm$ 21 & 104314.2\\
310 & 0.2141 $\pm$ 0.0003 & $467.0^{+0.7}_{-0.7}$ & $0.7131\pm 0.0010$ & 5695 $\pm$ 8 & 113587.3\\
330 & 0.1834 $\pm$ 0.0006 & $545.3^{+1.9}_{-1.9}$ & $0.6106\pm 0.0021$ & 4876 $\pm$ 17 & 156120.8\\
350 & 0.2125 $\pm$ 0.0003 & $470.7^{+0.6}_{-0.6}$ & $0.7075\pm 0.0008$ & 5650 $\pm$ 7 & 142833.5\\
\noalign{\smallskip}\hline\noalign{\smallskip}
Mean & 0.219 $\pm$ 0.030 & & 0.728 $\pm$ 0.099 & 5817 $\pm$ 787 \\
\noalign{\smallskip}\hline
\end{tabular}
\end{table*}

\begin{figure*}
\centerline{
\includegraphics[trim=50 50 100 50,clip=true,width=0.5\textwidth]{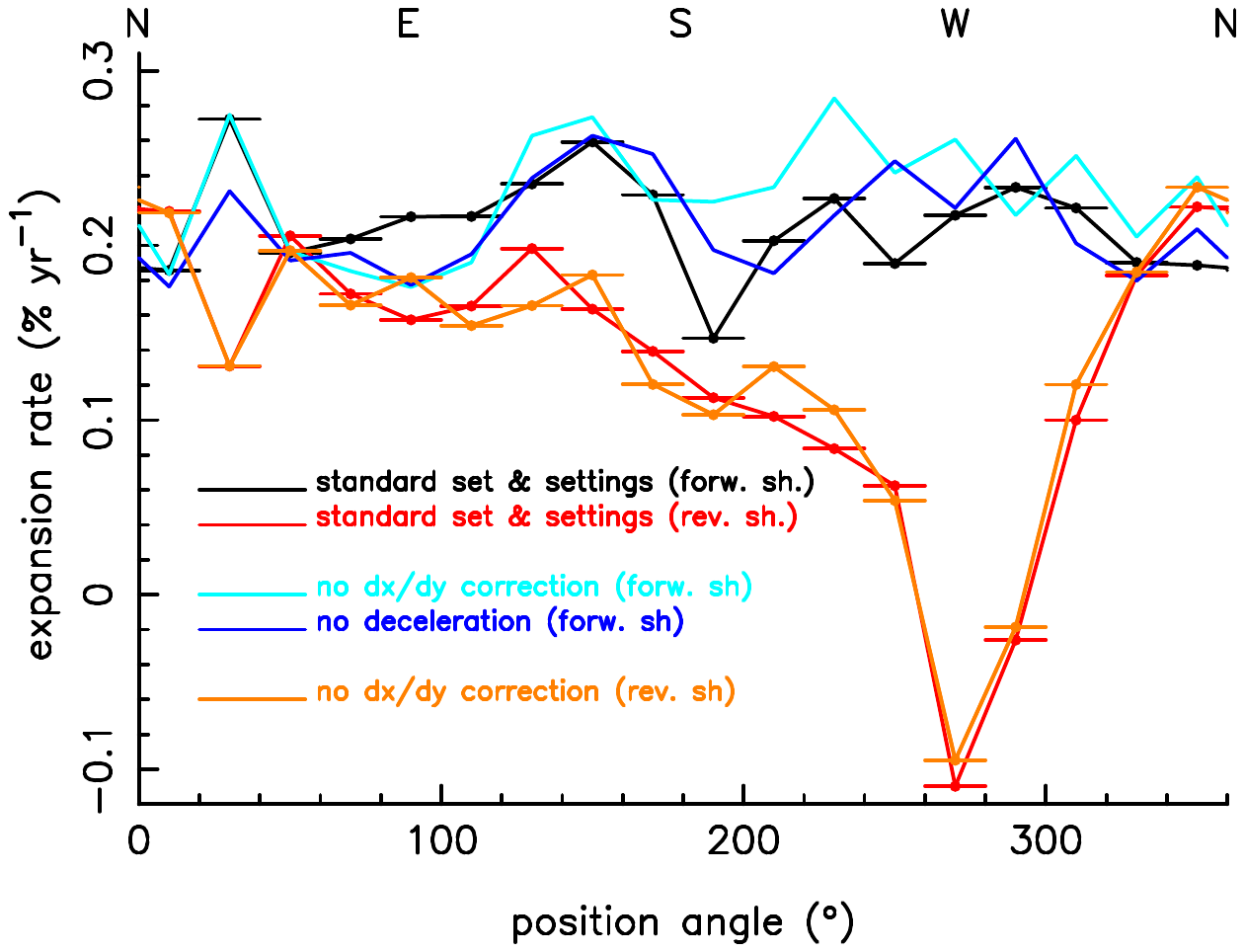}
\includegraphics[trim=50 50 100 100,clip=true,width=0.5\textwidth]{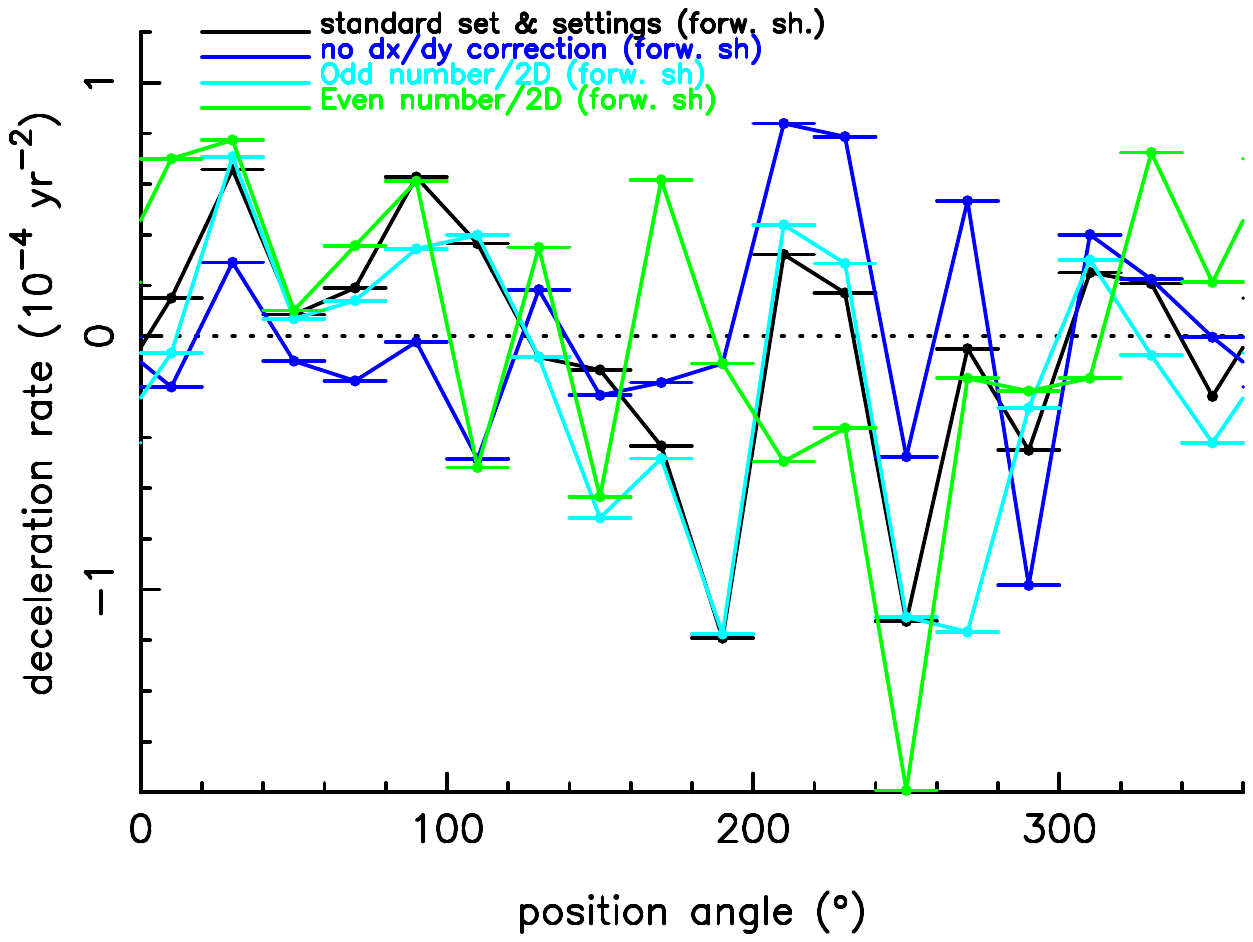}
}
  \caption{
    \label{fig:comparison}
Left:
    The expansion rate measurements of the forward and reverse shock regions
    employing different settings.
    The black (forward shock) and red (reverse shock) lines correspond to the measurements 
    discussed in the main text, and shown in Fig.~\ref{fig:expansion}.
    The cyan line shows the best fit expansion rates if no correction to the invidual pointings is applied.
    The blue line shows the solution if no there is no deceleration measurement performed, i.e. setting $b=0$.
    Right:
    The measured deceleration parameter for the forward shock region shown for the optimal fit (black line),
    a solution with $\Delta x=0, \Delta y=0$ for all epochs, and for two subsets of the  observations, namely the odd
    and even numbered observations.
  }
\end{figure*}

\section{The robustness of the best fit expansion and deceleration}

For our best fit solutions, presented in the main text, we made a correction for image misalignments (pointing errors) as explained in Section~\ref{sec:pointing}, with the values for
$\Delta x, \Delta y$ listed in Table~\ref{tab:dxdy}. Moreover,
we fitted for the acceleration/deceleration (parameter $b$) for the forward shock. In Figure~\ref{fig:comparison} we show the effects of assuming no pointing misalignments
(for all epoch $\Delta x=0, \Delta y=0$) and for not fitting for $b$, but setting it to $b=0$.  For the expansion parameters  there are clear differences in measured expansion parameters,
but the general trends are not affected, indicating that the results are robust.  For the forward shock differences in measured $a$ are below 0.04\%yr$^{-1}$, and for the reverse shock
region the deviation on $a$ are even below 0.02\%yr$^{-1}$. 

The difference in total $L$ for all 18 forward shock regions between the best fit value and one for which $\Delta x=0, \Delta y=0$ is $\Delta L=3803$, for
34 extra degree of freedom (corresponding to $\Delta x$, $\Delta y$ for 17 epochs). The difference in log-likelihood if one does not fit for $b$, but set the values for all sectors to $b=0$, is
$\Delta L=1156$, for 18 degrees of freedom.

Clearly, solving for both $\Delta x,\Delta y$ and $b$ leads to a much better fit, and the preferred values for the expansion are those listed in Table~\ref{tab:forward} and Table~\ref{tab:reverse}.
For comparison we do list the forward shock expansion rates for $b=0$ in Table~\ref{tab:fs_alternative}.

\end{document}